\documentclass[longauth]{aa}

\usepackage{graphicx}
\usepackage{natbib}
\usepackage{scalerel}
\usepackage{enumitem}
\usepackage{multirow}
\usepackage{url}
\usepackage{glossaries}
\usepackage{cancel}
\usepackage{float}

\newacronym{dlnn}{DLNN}{Deep Learning Neural Network}
\newacronym{GBDT}{GBDT}{Gradient-Boosted Decision-Trees}
\newacronym{ml}{ML}{machine learning}
\newacronym{ms}{SFMS}{star-forming main-sequence}
\newacronym{snr}{S/N}{signal-to-noise ratio}
\newacronym{sed}{SED}{spectral energy distribution}
\newacronym{odr}{ODR}{orthogonal distance regression}
\newacronym{pp}{PPs}{physical parameters}
\newacronym{UV}{UV}{ultraviolet}
\newacronym{NIR}{NIR}{near-infrared}
\newacronym{EDF}{EDF}{Euclid Deep Fields}
\newacronym{EWS}{EWS}{Euclid Wide Survey}
\newacronym{CSMR}{CSMR}{\texttt{CatBoost} single-model regressor}
\newacronym{CCR}{CCR}{\texttt{CatBoost} chained regressors}

\usepackage[table]{xcolor}

\usepackage{txfonts}
\usepackage[pdfencoding=auto,psdextra]{hyperref}
\hypersetup{colorlinks=true, linkcolor=blue, filecolor=magenta, urlcolor=blue, citecolor=blue}
\makeatletter
\renewcommand*\aa@pageof{, page \thepage{} of \pageref*{LastPage}}
\makeatother

\newcommand{\nnpz}{\texttt{nnpz}}
\newcommand{\phosphoros}{\texttt{Phosphoros}}
\newcommand{\catboost}{\texttt{CatBoost}}
\newcommand{\sfr}{\logten\,\mathrm{SFR}}

\newcommand{\Mstarwun}{\logten(M_{\star}/M_\odot)}
\newcommand{\sfrwun}{\logten\,(\mathrm{SFR}/M_\odot\,\mathrm{yr}^{-1})}

\usepackage[utf8]{inputenc}
\usepackage{bm}

\usepackage[switch, modulo]{lineno}

\usepackage{euclid}
\usepackage{txfonts}

\usepackage{cleveref}
\usepackage{orcidlink}

\begin{document} 

    \title{\Euclid\/ preparation. LI.} \subtitle{Forecasting the recovery of galaxy physical properties and their relations with template-fitting and machine-learning methods.} 

    \newcommand{\orcid}[1]{} 
    
    \author{Euclid Collaboration: A.~Enia\orcid{0000-0002-0200-2857}\thanks{\email{andrea.enia@unibo.it}}\inst{\ref{aff1},\ref{aff2}}
    \and M.~Bolzonella\orcid{0000-0003-3278-4607}\inst{\ref{aff2}}
    \and L.~Pozzetti\orcid{0000-0001-7085-0412}\inst{\ref{aff2}}
    \and A.~Humphrey\inst{\ref{aff3},\ref{aff4}}
    \and P.~A.~C.~Cunha\orcid{0000-0002-9454-859X}\inst{\ref{aff5},\ref{aff3}}
    \and W.~G.~Hartley\inst{\ref{aff6}}
    \and F.~Dubath\orcid{0000-0002-6533-2810}\inst{\ref{aff6}}
    \and S.~Paltani\orcid{0000-0002-8108-9179}\inst{\ref{aff6}}
    \and X.~Lopez~Lopez\inst{\ref{aff1},\ref{aff2}}
    \and S.~Quai\orcid{0000-0002-0449-8163}\inst{\ref{aff1},\ref{aff2}}
    \and S.~Bardelli\orcid{0000-0002-8900-0298}\inst{\ref{aff2}}
    \and L.~Bisigello\orcid{0000-0003-0492-4924}\inst{\ref{aff7},\ref{aff8}}
    \and S.~Cavuoti\orcid{0000-0002-3787-4196}\inst{\ref{aff9},\ref{aff10}}
    \and G.~De~Lucia\orcid{0000-0002-6220-9104}\inst{\ref{aff11}}
    \and M.~Ginolfi\orcid{0000-0002-9122-1700}\inst{\ref{aff12},\ref{aff13}}
    \and A.~Grazian\orcid{0000-0002-5688-0663}\inst{\ref{aff14}}
    \and M.~Siudek\orcid{0000-0002-2949-2155}\inst{\ref{aff15},\ref{aff16}}
    \and C.~Tortora\orcid{0000-0001-7958-6531}\inst{\ref{aff9}}
    \and G.~Zamorani\orcid{0000-0002-2318-301X}\inst{\ref{aff2}}
    \and N.~Aghanim\inst{\ref{aff17}}
    \and B.~Altieri\orcid{0000-0003-3936-0284}\inst{\ref{aff18}}
    \and A.~Amara\inst{\ref{aff19}}
    \and S.~Andreon\orcid{0000-0002-2041-8784}\inst{\ref{aff20}}
    \and N.~Auricchio\orcid{0000-0003-4444-8651}\inst{\ref{aff2}}
    \and C.~Baccigalupi\orcid{0000-0002-8211-1630}\inst{\ref{aff21},\ref{aff11},\ref{aff22},\ref{aff23}}
    \and M.~Baldi\orcid{0000-0003-4145-1943}\inst{\ref{aff24},\ref{aff2},\ref{aff25}}
    \and R.~Bender\orcid{0000-0001-7179-0626}\inst{\ref{aff26},\ref{aff27}}
    \and C.~Bodendorf\inst{\ref{aff26}}
    \and D.~Bonino\orcid{0000-0002-3336-9977}\inst{\ref{aff28}}
    \and E.~Branchini\orcid{0000-0002-0808-6908}\inst{\ref{aff29},\ref{aff30},\ref{aff20}}
    \and M.~Brescia\orcid{0000-0001-9506-5680}\inst{\ref{aff31},\ref{aff9},\ref{aff10}}
    \and J.~Brinchmann\orcid{0000-0003-4359-8797}\inst{\ref{aff3}}
    \and S.~Camera\orcid{0000-0003-3399-3574}\inst{\ref{aff32},\ref{aff33},\ref{aff28}}
    \and V.~Capobianco\orcid{0000-0002-3309-7692}\inst{\ref{aff28}}
    \and C.~Carbone\orcid{0000-0003-0125-3563}\inst{\ref{aff34}}
    \and J.~Carretero\orcid{0000-0002-3130-0204}\inst{\ref{aff35},\ref{aff36}}
    \and S.~Casas\orcid{0000-0002-4751-5138}\inst{\ref{aff37}}
    \and F.~J.~Castander\orcid{0000-0001-7316-4573}\inst{\ref{aff16},\ref{aff38}}
    \and M.~Castellano\orcid{0000-0001-9875-8263}\inst{\ref{aff39}}
    \and G.~Castignani\orcid{0000-0001-6831-0687}\inst{\ref{aff2}}
    \and A.~Cimatti\inst{\ref{aff40}}
    \and C.~Colodro-Conde\inst{\ref{aff41}}
    \and G.~Congedo\orcid{0000-0003-2508-0046}\inst{\ref{aff42}}
    \and C.~J.~Conselice\orcid{0000-0003-1949-7638}\inst{\ref{aff43}}
    \and L.~Conversi\orcid{0000-0002-6710-8476}\inst{\ref{aff44},\ref{aff18}}
    \and Y.~Copin\orcid{0000-0002-5317-7518}\inst{\ref{aff45}}
    \and L.~Corcione\orcid{0000-0002-6497-5881}\inst{\ref{aff28}}
    \and F.~Courbin\orcid{0000-0003-0758-6510}\inst{\ref{aff46}}
    \and H.~M.~Courtois\orcid{0000-0003-0509-1776}\inst{\ref{aff47}}
    \and A.~Da~Silva\orcid{0000-0002-6385-1609}\inst{\ref{aff48},\ref{aff49}}
    \and H.~Degaudenzi\orcid{0000-0002-5887-6799}\inst{\ref{aff6}}
    \and A.~M.~Di~Giorgio\orcid{0000-0002-4767-2360}\inst{\ref{aff50}}
    \and J.~Dinis\orcid{0000-0001-5075-1601}\inst{\ref{aff48},\ref{aff49}}
    \and X.~Dupac\inst{\ref{aff18}}
    \and S.~Dusini\orcid{0000-0002-1128-0664}\inst{\ref{aff51}}
    \and M.~Fabricius\orcid{0000-0002-7025-6058}\inst{\ref{aff26},\ref{aff27}}
    \and M.~Farina\orcid{0000-0002-3089-7846}\inst{\ref{aff50}}
    \and S.~Farrens\orcid{0000-0002-9594-9387}\inst{\ref{aff52}}
    \and S.~Ferriol\inst{\ref{aff45}}
    \and P.~Fosalba\orcid{0000-0002-1510-5214}\inst{\ref{aff38},\ref{aff53}}
    \and S.~Fotopoulou\orcid{0000-0002-9686-254X}\inst{\ref{aff54}}
    \and M.~Frailis\orcid{0000-0002-7400-2135}\inst{\ref{aff11}}
    \and E.~Franceschi\orcid{0000-0002-0585-6591}\inst{\ref{aff2}}
    \and M.~Fumana\orcid{0000-0001-6787-5950}\inst{\ref{aff34}}
    \and S.~Galeotta\orcid{0000-0002-3748-5115}\inst{\ref{aff11}}
    \and B.~Gillis\orcid{0000-0002-4478-1270}\inst{\ref{aff42}}
    \and C.~Giocoli\orcid{0000-0002-9590-7961}\inst{\ref{aff2},\ref{aff55}}
    \and F.~Grupp\inst{\ref{aff26},\ref{aff27}}
    \and S.~V.~H.~Haugan\orcid{0000-0001-9648-7260}\inst{\ref{aff56}}
    \and W.~Holmes\inst{\ref{aff57}}
    \and I.~Hook\orcid{0000-0002-2960-978X}\inst{\ref{aff58}}
    \and F.~Hormuth\inst{\ref{aff59}}
    \and A.~Hornstrup\orcid{0000-0002-3363-0936}\inst{\ref{aff60},\ref{aff61}}
    \and K.~Jahnke\orcid{0000-0003-3804-2137}\inst{\ref{aff62}}
    \and B.~Joachimi\orcid{0000-0001-7494-1303}\inst{\ref{aff63}}
    \and E.~Keih\"anen\orcid{0000-0003-1804-7715}\inst{\ref{aff64}}
    \and S.~Kermiche\orcid{0000-0002-0302-5735}\inst{\ref{aff65}}
    \and A.~Kiessling\orcid{0000-0002-2590-1273}\inst{\ref{aff57}}
    \and B.~Kubik\orcid{0009-0006-5823-4880}\inst{\ref{aff45}}
    \and M.~K\"ummel\orcid{0000-0003-2791-2117}\inst{\ref{aff27}}
    \and M.~Kunz\orcid{0000-0002-3052-7394}\inst{\ref{aff66}}
    \and H.~Kurki-Suonio\orcid{0000-0002-4618-3063}\inst{\ref{aff67},\ref{aff68}}
    \and S.~Ligori\orcid{0000-0003-4172-4606}\inst{\ref{aff28}}
    \and P.~B.~Lilje\orcid{0000-0003-4324-7794}\inst{\ref{aff56}}
    \and V.~Lindholm\orcid{0000-0003-2317-5471}\inst{\ref{aff67},\ref{aff68}}
    \and I.~Lloro\inst{\ref{aff69}}
    \and E.~Maiorano\orcid{0000-0003-2593-4355}\inst{\ref{aff2}}
    \and O.~Mansutti\orcid{0000-0001-5758-4658}\inst{\ref{aff11}}
    \and O.~Marggraf\orcid{0000-0001-7242-3852}\inst{\ref{aff70}}
    \and K.~Markovic\orcid{0000-0001-6764-073X}\inst{\ref{aff57}}
    \and M.~Martinelli\orcid{0000-0002-6943-7732}\inst{\ref{aff39},\ref{aff71}}
    \and N.~Martinet\orcid{0000-0003-2786-7790}\inst{\ref{aff72}}
    \and F.~Marulli\orcid{0000-0002-8850-0303}\inst{\ref{aff1},\ref{aff2},\ref{aff25}}
    \and R.~Massey\orcid{0000-0002-6085-3780}\inst{\ref{aff73}}
    \and H.~J.~McCracken\orcid{0000-0002-9489-7765}\inst{\ref{aff74}}
    \and E.~Medinaceli\orcid{0000-0002-4040-7783}\inst{\ref{aff2}}
    \and S.~Mei\orcid{0000-0002-2849-559X}\inst{\ref{aff75}}
    \and M.~Melchior\inst{\ref{aff76}}
    \and Y.~Mellier\inst{\ref{aff77},\ref{aff74}}
    \and M.~Meneghetti\orcid{0000-0003-1225-7084}\inst{\ref{aff2},\ref{aff25}}
    \and E.~Merlin\orcid{0000-0001-6870-8900}\inst{\ref{aff39}}
    \and G.~Meylan\inst{\ref{aff46}}
    \and M.~Moresco\orcid{0000-0002-7616-7136}\inst{\ref{aff1},\ref{aff2}}
    \and L.~Moscardini\orcid{0000-0002-3473-6716}\inst{\ref{aff1},\ref{aff2},\ref{aff25}}
    \and E.~Munari\orcid{0000-0002-1751-5946}\inst{\ref{aff11},\ref{aff21}}
    \and C.~Neissner\orcid{0000-0001-8524-4968}\inst{\ref{aff78},\ref{aff36}}
    \and S.-M.~Niemi\inst{\ref{aff79}}
    \and J.~W.~Nightingale\orcid{0000-0002-8987-7401}\inst{\ref{aff80},\ref{aff73}}
    \and C.~Padilla\orcid{0000-0001-7951-0166}\inst{\ref{aff78}}
    \and F.~Pasian\orcid{0000-0002-4869-3227}\inst{\ref{aff11}}
    \and K.~Pedersen\inst{\ref{aff81}}
    \and V.~Pettorino\inst{\ref{aff79}}
    \and G.~Polenta\orcid{0000-0003-4067-9196}\inst{\ref{aff82}}
    \and M.~Poncet\inst{\ref{aff83}}
    \and L.~A.~Popa\inst{\ref{aff84}}
    \and F.~Raison\orcid{0000-0002-7819-6918}\inst{\ref{aff26}}
    \and R.~Rebolo\inst{\ref{aff41},\ref{aff85}}
    \and A.~Renzi\orcid{0000-0001-9856-1970}\inst{\ref{aff8},\ref{aff51}}
    \and J.~Rhodes\inst{\ref{aff57}}
    \and G.~Riccio\inst{\ref{aff9}}
    \and E.~Romelli\orcid{0000-0003-3069-9222}\inst{\ref{aff11}}
    \and M.~Roncarelli\orcid{0000-0001-9587-7822}\inst{\ref{aff2}}
    \and E.~Rossetti\orcid{0000-0003-0238-4047}\inst{\ref{aff24}}
    \and R.~Saglia\orcid{0000-0003-0378-7032}\inst{\ref{aff27},\ref{aff26}}
    \and Z.~Sakr\orcid{0000-0002-4823-3757}\inst{\ref{aff86},\ref{aff87},\ref{aff88}}
    \and D.~Sapone\orcid{0000-0001-7089-4503}\inst{\ref{aff89}}
    \and P.~Schneider\orcid{0000-0001-8561-2679}\inst{\ref{aff70}}
    \and T.~Schrabback\orcid{0000-0002-6987-7834}\inst{\ref{aff90}}
    \and M.~Scodeggio\inst{\ref{aff34}}
    \and A.~Secroun\orcid{0000-0003-0505-3710}\inst{\ref{aff65}}
    \and E.~Sefusatti\orcid{0000-0003-0473-1567}\inst{\ref{aff11},\ref{aff21},\ref{aff22}}
    \and G.~Seidel\orcid{0000-0003-2907-353X}\inst{\ref{aff62}}
    \and S.~Serrano\orcid{0000-0002-0211-2861}\inst{\ref{aff38},\ref{aff91},\ref{aff16}}
    \and C.~Sirignano\orcid{0000-0002-0995-7146}\inst{\ref{aff8},\ref{aff51}}
    \and G.~Sirri\orcid{0000-0003-2626-2853}\inst{\ref{aff25}}
    \and L.~Stanco\orcid{0000-0002-9706-5104}\inst{\ref{aff51}}
    \and J.~Steinwagner\inst{\ref{aff26}}
    \and C.~Surace\orcid{0000-0003-2592-0113}\inst{\ref{aff72}}
    \and P.~Tallada-Cresp\'{i}\orcid{0000-0002-1336-8328}\inst{\ref{aff35},\ref{aff36}}
    \and D.~Tavagnacco\orcid{0000-0001-7475-9894}\inst{\ref{aff11}}
    \and A.~N.~Taylor\inst{\ref{aff42}}
    \and H.~I.~Teplitz\orcid{0000-0002-7064-5424}\inst{\ref{aff92}}
    \and I.~Tereno\inst{\ref{aff48},\ref{aff93}}
    \and R.~Toledo-Moreo\orcid{0000-0002-2997-4859}\inst{\ref{aff94}}
    \and F.~Torradeflot\orcid{0000-0003-1160-1517}\inst{\ref{aff36},\ref{aff35}}
    \and I.~Tutusaus\orcid{0000-0002-3199-0399}\inst{\ref{aff87}}
    \and L.~Valenziano\orcid{0000-0002-1170-0104}\inst{\ref{aff2},\ref{aff95}}
    \and T.~Vassallo\orcid{0000-0001-6512-6358}\inst{\ref{aff27},\ref{aff11}}
    \and G.~Verdoes~Kleijn\orcid{0000-0001-5803-2580}\inst{\ref{aff96}}
    \and A.~Veropalumbo\orcid{0000-0003-2387-1194}\inst{\ref{aff20},\ref{aff30},\ref{aff97}}
    \and Y.~Wang\orcid{0000-0002-4749-2984}\inst{\ref{aff92}}
    \and J.~Weller\orcid{0000-0002-8282-2010}\inst{\ref{aff27},\ref{aff26}}
    \and E.~Zucca\orcid{0000-0002-5845-8132}\inst{\ref{aff2}}
    \and A.~Biviano\orcid{0000-0002-0857-0732}\inst{\ref{aff11},\ref{aff21}}
    \and A.~Boucaud\orcid{0000-0001-7387-2633}\inst{\ref{aff75}}
    \and C.~Burigana\orcid{0000-0002-3005-5796}\inst{\ref{aff7},\ref{aff95}}
    \and M.~Calabrese\orcid{0000-0002-2637-2422}\inst{\ref{aff98},\ref{aff34}}
    \and J.~A.~Escartin~Vigo\inst{\ref{aff26}}
    \and J.~Gracia-Carpio\inst{\ref{aff26}}
    \and N.~Mauri\orcid{0000-0001-8196-1548}\inst{\ref{aff40},\ref{aff25}}
    \and A.~Pezzotta\orcid{0000-0003-0726-2268}\inst{\ref{aff26}}
    \and M.~P\"ontinen\orcid{0000-0001-5442-2530}\inst{\ref{aff67}}
    \and C.~Porciani\orcid{0000-0002-7797-2508}\inst{\ref{aff70}}
    \and V.~Scottez\inst{\ref{aff77},\ref{aff99}}
    \and M.~Tenti\orcid{0000-0002-4254-5901}\inst{\ref{aff25}}
    \and M.~Viel\orcid{0000-0002-2642-5707}\inst{\ref{aff21},\ref{aff11},\ref{aff23},\ref{aff22},\ref{aff100}}
    \and M.~Wiesmann\orcid{0009-0000-8199-5860}\inst{\ref{aff56}}
    \and Y.~Akrami\orcid{0000-0002-2407-7956}\inst{\ref{aff101},\ref{aff102}}
    \and V.~Allevato\orcid{0000-0001-7232-5152}\inst{\ref{aff9}}
    \and S.~Anselmi\orcid{0000-0002-3579-9583}\inst{\ref{aff51},\ref{aff8},\ref{aff103}}
    \and M.~Ballardini\orcid{0000-0003-4481-3559}\inst{\ref{aff104},\ref{aff2},\ref{aff105}}
    \and P.~Bergamini\orcid{0000-0003-1383-9414}\inst{\ref{aff106},\ref{aff2}}
    \and M.~Bethermin\orcid{0000-0002-3915-2015}\inst{\ref{aff107},\ref{aff72}}
    \and A.~Blanchard\orcid{0000-0001-8555-9003}\inst{\ref{aff87}}
    \and L.~Blot\orcid{0000-0002-9622-7167}\inst{\ref{aff108},\ref{aff103}}
    \and S.~Borgani\orcid{0000-0001-6151-6439}\inst{\ref{aff109},\ref{aff21},\ref{aff11},\ref{aff22}}
    \and S.~Bruton\orcid{0000-0002-6503-5218}\inst{\ref{aff110}}
    \and R.~Cabanac\orcid{0000-0001-6679-2600}\inst{\ref{aff87}}
    \and A.~Calabro\orcid{0000-0003-2536-1614}\inst{\ref{aff39}}
    \and G.~Canas-Herrera\orcid{0000-0003-2796-2149}\inst{\ref{aff79},\ref{aff111}}
    \and A.~Cappi\inst{\ref{aff2},\ref{aff112}}
    \and C.~S.~Carvalho\inst{\ref{aff93}}
    \and T.~Castro\orcid{0000-0002-6292-3228}\inst{\ref{aff11},\ref{aff22},\ref{aff21},\ref{aff100}}
    \and K.~C.~Chambers\orcid{0000-0001-6965-7789}\inst{\ref{aff113}}
    \and S.~Contarini\orcid{0000-0002-9843-723X}\inst{\ref{aff26}}
    \and T.~Contini\orcid{0000-0003-0275-938X}\inst{\ref{aff87}}
    \and A.~R.~Cooray\orcid{0000-0002-3892-0190}\inst{\ref{aff114}}
    \and O.~Cucciati\orcid{0000-0002-9336-7551}\inst{\ref{aff2}}
    \and S.~Davini\orcid{0000-0003-3269-1718}\inst{\ref{aff30}}
    \and B.~De~Caro\inst{\ref{aff34}}
    \and G.~Desprez\inst{\ref{aff115}}
    \and A.~D\'iaz-S\'anchez\orcid{0000-0003-0748-4768}\inst{\ref{aff116}}
    \and S.~Di~Domizio\orcid{0000-0003-2863-5895}\inst{\ref{aff29},\ref{aff30}}
    \and H.~Dole\orcid{0000-0002-9767-3839}\inst{\ref{aff17}}
    \and S.~Escoffier\orcid{0000-0002-2847-7498}\inst{\ref{aff65}}
    \and A.~G.~Ferrari\orcid{0009-0005-5266-4110}\inst{\ref{aff40},\ref{aff25}}
    \and P.~G.~Ferreira\orcid{0000-0002-3021-2851}\inst{\ref{aff117}}
    \and I.~Ferrero\orcid{0000-0002-1295-1132}\inst{\ref{aff56}}
    \and A.~Finoguenov\orcid{0000-0002-4606-5403}\inst{\ref{aff67}}
    \and F.~Fornari\orcid{0000-0003-2979-6738}\inst{\ref{aff95}}
    \and L.~Gabarra\orcid{0000-0002-8486-8856}\inst{\ref{aff117}}
    \and K.~Ganga\orcid{0000-0001-8159-8208}\inst{\ref{aff75}}
    \and J.~Garc\'ia-Bellido\orcid{0000-0002-9370-8360}\inst{\ref{aff101}}
    \and V.~Gautard\inst{\ref{aff118}}
    \and E.~Gaztanaga\orcid{0000-0001-9632-0815}\inst{\ref{aff16},\ref{aff38},\ref{aff119}}
    \and F.~Giacomini\orcid{0000-0002-3129-2814}\inst{\ref{aff25}}
    \and F.~Gianotti\orcid{0000-0003-4666-119X}\inst{\ref{aff2}}
    \and G.~Gozaliasl\orcid{0000-0002-0236-919X}\inst{\ref{aff120},\ref{aff67}}
    \and A.~Hall\orcid{0000-0002-3139-8651}\inst{\ref{aff42}}
    \and S.~Hemmati\orcid{0000-0003-2226-5395}\inst{\ref{aff121}}
    \and H.~Hildebrandt\orcid{0000-0002-9814-3338}\inst{\ref{aff122}}
    \and J.~Hjorth\orcid{0000-0002-4571-2306}\inst{\ref{aff123}}
    \and A.~Jimenez~Mu\~noz\orcid{0009-0004-5252-185X}\inst{\ref{aff124}}
    \and S.~Joudaki\orcid{0000-0001-8820-673X}\inst{\ref{aff119}}
    \and J.~J.~E.~Kajava\orcid{0000-0002-3010-8333}\inst{\ref{aff125},\ref{aff126}}
    \and V.~Kansal\orcid{0000-0002-4008-6078}\inst{\ref{aff127},\ref{aff128}}
    \and D.~Karagiannis\orcid{0000-0002-4927-0816}\inst{\ref{aff129},\ref{aff130}}
    \and C.~C.~Kirkpatrick\inst{\ref{aff64}}
    \and J.~Le~Graet\orcid{0000-0001-6523-7971}\inst{\ref{aff65}}
    \and L.~Legrand\orcid{0000-0003-0610-5252}\inst{\ref{aff131}}
    \and A.~Loureiro\orcid{0000-0002-4371-0876}\inst{\ref{aff132},\ref{aff133}}
    \and J.~Macias-Perez\orcid{0000-0002-5385-2763}\inst{\ref{aff124}}
    \and G.~Maggio\orcid{0000-0003-4020-4836}\inst{\ref{aff11}}
    \and M.~Magliocchetti\orcid{0000-0001-9158-4838}\inst{\ref{aff50}}
    \and C.~Mancini\orcid{0000-0002-4297-0561}\inst{\ref{aff34}}
    \and F.~Mannucci\orcid{0000-0002-4803-2381}\inst{\ref{aff13}}
    \and R.~Maoli\orcid{0000-0002-6065-3025}\inst{\ref{aff134},\ref{aff39}}
    \and C.~J.~A.~P.~Martins\orcid{0000-0002-4886-9261}\inst{\ref{aff135},\ref{aff3}}
    \and S.~Matthew\orcid{0000-0001-8448-1697}\inst{\ref{aff42}}
    \and L.~Maurin\orcid{0000-0002-8406-0857}\inst{\ref{aff17}}
    \and R.~B.~Metcalf\orcid{0000-0003-3167-2574}\inst{\ref{aff1},\ref{aff2}}
    \and P.~Monaco\orcid{0000-0003-2083-7564}\inst{\ref{aff109},\ref{aff11},\ref{aff22},\ref{aff21}}
    \and C.~Moretti\orcid{0000-0003-3314-8936}\inst{\ref{aff23},\ref{aff100},\ref{aff11},\ref{aff21},\ref{aff22}}
    \and G.~Morgante\inst{\ref{aff2}}
    \and Nicholas~A.~Walton\orcid{0000-0003-3983-8778}\inst{\ref{aff136}}
    \and L.~Patrizii\inst{\ref{aff25}}
    \and V.~Popa\inst{\ref{aff84}}
    \and D.~Potter\orcid{0000-0002-0757-5195}\inst{\ref{aff137}}
    \and I.~Risso\orcid{0000-0003-2525-7761}\inst{\ref{aff97}}
    \and P.-F.~Rocci\inst{\ref{aff17}}
    \and M.~Sahl\'en\orcid{0000-0003-0973-4804}\inst{\ref{aff138}}
    \and A.~Schneider\orcid{0000-0001-7055-8104}\inst{\ref{aff137}}
    \and M.~Schultheis\inst{\ref{aff112}}
    \and M.~Sereno\orcid{0000-0003-0302-0325}\inst{\ref{aff2},\ref{aff25}}
    \and P.~Simon\inst{\ref{aff70}}
    \and A.~Spurio~Mancini\orcid{0000-0001-5698-0990}\inst{\ref{aff139},\ref{aff140}}
    \and S.~A.~Stanford\orcid{0000-0003-0122-0841}\inst{\ref{aff141}}
    \and K.~Tanidis\inst{\ref{aff117}}
    \and C.~Tao\orcid{0000-0001-7961-8177}\inst{\ref{aff65}}
    \and G.~Testera\inst{\ref{aff30}}
    \and R.~Teyssier\orcid{0000-0001-7689-0933}\inst{\ref{aff142}}
    \and S.~Toft\orcid{0000-0003-3631-7176}\inst{\ref{aff61},\ref{aff143},\ref{aff144}}
    \and S.~Tosi\orcid{0000-0002-7275-9193}\inst{\ref{aff29},\ref{aff30}}
    \and A.~Troja\orcid{0000-0003-0239-4595}\inst{\ref{aff8},\ref{aff51}}
    \and M.~Tucci\inst{\ref{aff6}}
    \and C.~Valieri\inst{\ref{aff25}}
    \and J.~Valiviita\orcid{0000-0001-6225-3693}\inst{\ref{aff67},\ref{aff68}}
    \and D.~Vergani\orcid{0000-0003-0898-2216}\inst{\ref{aff2}}
    \and G.~Verza\orcid{0000-0002-1886-8348}\inst{\ref{aff145},\ref{aff146}}
    \and I.~A.~Zinchenko\inst{\ref{aff27}}
    \and G.~Rodighiero\orcid{0000-0002-9415-2296}\inst{\ref{aff8},\ref{aff14}}
    \and M.~Talia\orcid{0000-0003-4352-2063}\inst{\ref{aff1},\ref{aff2}}
    }
    										   
    \institute{Dipartimento di Fisica e Astronomia "Augusto Righi" - Alma Mater Studiorum Universit\`a di Bologna, via Piero Gobetti 93/2, 40129 Bologna, Italy\label{aff1}
    \and
    INAF-Osservatorio di Astrofisica e Scienza dello Spazio di Bologna, Via Piero Gobetti 93/3, 40129 Bologna, Italy\label{aff2}
    \and
    Instituto de Astrof\'isica e Ci\^encias do Espa\c{c}o, Universidade do Porto, CAUP, Rua das Estrelas, PT4150-762 Porto, Portugal\label{aff3}
    \and
    DTx -- Digital Transformation CoLAB, Building 1, Azur\'em Campus, University of Minho, 4800-058 Guimar\~aes, Portugal\label{aff4}
    \and
    Faculdade de Ci\^encias da Universidade do Porto, Rua do Campo de Alegre, 4150-007 Porto, Portugal\label{aff5}
    \and
    Department of Astronomy, University of Geneva, ch. d'Ecogia 16, 1290 Versoix, Switzerland\label{aff6}
    \and
    INAF, Istituto di Radioastronomia, Via Piero Gobetti 101, 40129 Bologna, Italy\label{aff7}
    \and
    Dipartimento di Fisica e Astronomia "G. Galilei", Universit\`a di Padova, Via Marzolo 8, 35131 Padova, Italy\label{aff8}
    \and
    INAF-Osservatorio Astronomico di Capodimonte, Via Moiariello 16, 80131 Napoli, Italy\label{aff9}
    \and
    INFN section of Naples, Via Cinthia 6, 80126, Napoli, Italy\label{aff10}
    \and
    INAF-Osservatorio Astronomico di Trieste, Via G. B. Tiepolo 11, 34143 Trieste, Italy\label{aff11}
    \and
    Dipartimento di Fisica e Astronomia, Universit\`{a} di Firenze, via G. Sansone 1, 50019 Sesto Fiorentino, Firenze, Italy\label{aff12}
    \and
    INAF-Osservatorio Astrofisico di Arcetri, Largo E. Fermi 5, 50125, Firenze, Italy\label{aff13}
    \and
    INAF-Osservatorio Astronomico di Padova, Via dell'Osservatorio 5, 35122 Padova, Italy\label{aff14}
    \and
    Instituto de Astrof\'isica de Canarias (IAC); Departamento de Astrof\'isica, Universidad de La Laguna (ULL), 38200, La Laguna, Tenerife, Spain\label{aff15}
    \and
    Institute of Space Sciences (ICE, CSIC), Campus UAB, Carrer de Can Magrans, s/n, 08193 Barcelona, Spain\label{aff16}
    \and
    Universit\'e Paris-Saclay, CNRS, Institut d'astrophysique spatiale, 91405, Orsay, France\label{aff17}
    \and
    ESAC/ESA, Camino Bajo del Castillo, s/n., Urb. Villafranca del Castillo, 28692 Villanueva de la Ca\~nada, Madrid, Spain\label{aff18}
    \and
    School of Mathematics and Physics, University of Surrey, Guildford, Surrey, GU2 7XH, UK\label{aff19}
    \and
    INAF-Osservatorio Astronomico di Brera, Via Brera 28, 20122 Milano, Italy\label{aff20}
    \and
    IFPU, Institute for Fundamental Physics of the Universe, via Beirut 2, 34151 Trieste, Italy\label{aff21}
    \and
    INFN, Sezione di Trieste, Via Valerio 2, 34127 Trieste TS, Italy\label{aff22}
    \and
    SISSA, International School for Advanced Studies, Via Bonomea 265, 34136 Trieste TS, Italy\label{aff23}
    \and
    Dipartimento di Fisica e Astronomia, Universit\`a di Bologna, Via Gobetti 93/2, 40129 Bologna, Italy\label{aff24}
    \and
    INFN-Sezione di Bologna, Viale Berti Pichat 6/2, 40127 Bologna, Italy\label{aff25}
    \and
    Max Planck Institute for Extraterrestrial Physics, Giessenbachstr. 1, 85748 Garching, Germany\label{aff26}
    \and
    Universit\"ats-Sternwarte M\"unchen, Fakult\"at f\"ur Physik, Ludwig-Maximilians-Universit\"at M\"unchen, Scheinerstrasse 1, 81679 M\"unchen, Germany\label{aff27}
    \and
    INAF-Osservatorio Astrofisico di Torino, Via Osservatorio 20, 10025 Pino Torinese (TO), Italy\label{aff28}
    \and
    Dipartimento di Fisica, Universit\`a di Genova, Via Dodecaneso 33, 16146, Genova, Italy\label{aff29}
    \and
    INFN-Sezione di Genova, Via Dodecaneso 33, 16146, Genova, Italy\label{aff30}
    \and
    Department of Physics "E. Pancini", University Federico II, Via Cinthia 6, 80126, Napoli, Italy\label{aff31}
    \and
    Dipartimento di Fisica, Universit\`a degli Studi di Torino, Via P. Giuria 1, 10125 Torino, Italy\label{aff32}
    \and
    INFN-Sezione di Torino, Via P. Giuria 1, 10125 Torino, Italy\label{aff33}
    \and
    INAF-IASF Milano, Via Alfonso Corti 12, 20133 Milano, Italy\label{aff34}
    \and
    Centro de Investigaciones Energ\'eticas, Medioambientales y Tecnol\'ogicas (CIEMAT), Avenida Complutense 40, 28040 Madrid, Spain\label{aff35}
    \and
    Port d'Informaci\'{o} Cient\'{i}fica, Campus UAB, C. Albareda s/n, 08193 Bellaterra (Barcelona), Spain\label{aff36}
    \and
    Institute for Theoretical Particle Physics and Cosmology (TTK), RWTH Aachen University, 52056 Aachen, Germany\label{aff37}
    \and
    Institut d'Estudis Espacials de Catalunya (IEEC),  Edifici RDIT, Campus UPC, 08860 Castelldefels, Barcelona, Spain\label{aff38}
    \and
    INAF-Osservatorio Astronomico di Roma, Via Frascati 33, 00078 Monteporzio Catone, Italy\label{aff39}
    \and
    Dipartimento di Fisica e Astronomia "Augusto Righi" - Alma Mater Studiorum Universit\`a di Bologna, Viale Berti Pichat 6/2, 40127 Bologna, Italy\label{aff40}
    \and
    Instituto de Astrof\'isica de Canarias, Calle V\'ia L\'actea s/n, 38204, San Crist\'obal de La Laguna, Tenerife, Spain\label{aff41}
    \and
    Institute for Astronomy, University of Edinburgh, Royal Observatory, Blackford Hill, Edinburgh EH9 3HJ, UK\label{aff42}
    \and
    Jodrell Bank Centre for Astrophysics, Department of Physics and Astronomy, University of Manchester, Oxford Road, Manchester M13 9PL, UK\label{aff43}
    \and
    European Space Agency/ESRIN, Largo Galileo Galilei 1, 00044 Frascati, Roma, Italy\label{aff44}
    \and
    Universit\'e Claude Bernard Lyon 1, CNRS/IN2P3, IP2I Lyon, UMR 5822, Villeurbanne, F-69100, France\label{aff45}
    \and
    Institute of Physics, Laboratory of Astrophysics, Ecole Polytechnique F\'ed\'erale de Lausanne (EPFL), Observatoire de Sauverny, 1290 Versoix, Switzerland\label{aff46}
    \and
    UCB Lyon 1, CNRS/IN2P3, IUF, IP2I Lyon, 4 rue Enrico Fermi, 69622 Villeurbanne, France\label{aff47}
    \and
    Departamento de F\'isica, Faculdade de Ci\^encias, Universidade de Lisboa, Edif\'icio C8, Campo Grande, PT1749-016 Lisboa, Portugal\label{aff48}
    \and
    Instituto de Astrof\'isica e Ci\^encias do Espa\c{c}o, Faculdade de Ci\^encias, Universidade de Lisboa, Campo Grande, 1749-016 Lisboa, Portugal\label{aff49}
    \and
    INAF-Istituto di Astrofisica e Planetologia Spaziali, via del Fosso del Cavaliere, 100, 00100 Roma, Italy\label{aff50}
    \and
    INFN-Padova, Via Marzolo 8, 35131 Padova, Italy\label{aff51}
    \and
    Universit\'e Paris-Saclay, Universit\'e Paris Cit\'e, CEA, CNRS, AIM, 91191, Gif-sur-Yvette, France\label{aff52}
    \and
    Institut de Ciencies de l'Espai (IEEC-CSIC), Campus UAB, Carrer de Can Magrans, s/n Cerdanyola del Vall\'es, 08193 Barcelona, Spain\label{aff53}
    \and
    School of Physics, HH Wills Physics Laboratory, University of Bristol, Tyndall Avenue, Bristol, BS8 1TL, UK\label{aff54}
    \and
    Istituto Nazionale di Fisica Nucleare, Sezione di Bologna, Via Irnerio 46, 40126 Bologna, Italy\label{aff55}
    \and
    Institute of Theoretical Astrophysics, University of Oslo, P.O. Box 1029 Blindern, 0315 Oslo, Norway\label{aff56}
    \and
    Jet Propulsion Laboratory, California Institute of Technology, 4800 Oak Grove Drive, Pasadena, CA, 91109, USA\label{aff57}
    \and
    Department of Physics, Lancaster University, Lancaster, LA1 4YB, UK\label{aff58}
    \and
    Felix Hormuth Engineering, Goethestr. 17, 69181 Leimen, Germany\label{aff59}
    \and
    Technical University of Denmark, Elektrovej 327, 2800 Kgs. Lyngby, Denmark\label{aff60}
    \and
    Cosmic Dawn Center (DAWN), Denmark\label{aff61}
    \and
    Max-Planck-Institut f\"ur Astronomie, K\"onigstuhl 17, 69117 Heidelberg, Germany\label{aff62}
    \and
    Department of Physics and Astronomy, University College London, Gower Street, London WC1E 6BT, UK\label{aff63}
    \and
    Department of Physics and Helsinki Institute of Physics, Gustaf H\"allstr\"omin katu 2, 00014 University of Helsinki, Finland\label{aff64}
    \and
    Aix-Marseille Universit\'e, CNRS/IN2P3, CPPM, Marseille, France\label{aff65}
    \and
    Universit\'e de Gen\`eve, D\'epartement de Physique Th\'eorique and Centre for Astroparticle Physics, 24 quai Ernest-Ansermet, CH-1211 Gen\`eve 4, Switzerland\label{aff66}
    \and
    Department of Physics, P.O. Box 64, 00014 University of Helsinki, Finland\label{aff67}
    \and
    Helsinki Institute of Physics, Gustaf H{\"a}llstr{\"o}min katu 2, University of Helsinki, Helsinki, Finland\label{aff68}
    \and
    NOVA optical infrared instrumentation group at ASTRON, Oude Hoogeveensedijk 4, 7991PD, Dwingeloo, The Netherlands\label{aff69}
    \and
    Universit\"at Bonn, Argelander-Institut f\"ur Astronomie, Auf dem H\"ugel 71, 53121 Bonn, Germany\label{aff70}
    \and
    INFN-Sezione di Roma, Piazzale Aldo Moro, 2 - c/o Dipartimento di Fisica, Edificio G. Marconi, 00185 Roma, Italy\label{aff71}
    \and
    Aix-Marseille Universit\'e, CNRS, CNES, LAM, Marseille, France\label{aff72}
    \and
    Department of Physics, Institute for Computational Cosmology, Durham University, South Road, DH1 3LE, UK\label{aff73}
    \and
    Institut d'Astrophysique de Paris, UMR 7095, CNRS, and Sorbonne Universit\'e, 98 bis boulevard Arago, 75014 Paris, France\label{aff74}
    \and
    Universit\'e Paris Cit\'e, CNRS, Astroparticule et Cosmologie, 75013 Paris, France\label{aff75}
    \and
    University of Applied Sciences and Arts of Northwestern Switzerland, School of Engineering, 5210 Windisch, Switzerland\label{aff76}
    \and
    Institut d'Astrophysique de Paris, 98bis Boulevard Arago, 75014, Paris, France\label{aff77}
    \and
    Institut de F\'{i}sica d'Altes Energies (IFAE), The Barcelona Institute of Science and Technology, Campus UAB, 08193 Bellaterra (Barcelona), Spain\label{aff78}
    \and
    European Space Agency/ESTEC, Keplerlaan 1, 2201 AZ Noordwijk, The Netherlands\label{aff79}
    \and
    School of Mathematics, Statistics and Physics, Newcastle University, Herschel Building, Newcastle-upon-Tyne, NE1 7RU, UK\label{aff80}
    \and
    Department of Physics and Astronomy, University of Aarhus, Ny Munkegade 120, DK-8000 Aarhus C, Denmark\label{aff81}
    \and
    Space Science Data Center, Italian Space Agency, via del Politecnico snc, 00133 Roma, Italy\label{aff82}
    \and
    Centre National d'Etudes Spatiales -- Centre spatial de Toulouse, 18 avenue Edouard Belin, 31401 Toulouse Cedex 9, France\label{aff83}
    \and
    Institute of Space Science, Str. Atomistilor, nr. 409 M\u{a}gurele, Ilfov, 077125, Romania\label{aff84}
    \and
    Departamento de Astrof\'isica, Universidad de La Laguna, 38206, La Laguna, Tenerife, Spain\label{aff85}
    \and
    Institut f\"ur Theoretische Physik, University of Heidelberg, Philosophenweg 16, 69120 Heidelberg, Germany\label{aff86}
    \and
    Institut de Recherche en Astrophysique et Plan\'etologie (IRAP), Universit\'e de Toulouse, CNRS, UPS, CNES, 14 Av. Edouard Belin, 31400 Toulouse, France\label{aff87}
    \and
    Universit\'e St Joseph; Faculty of Sciences, Beirut, Lebanon\label{aff88}
    \and
    Departamento de F\'isica, FCFM, Universidad de Chile, Blanco Encalada 2008, Santiago, Chile\label{aff89}
    \and
    Universit\"at Innsbruck, Institut f\"ur Astro- und Teilchenphysik, Technikerstr. 25/8, 6020 Innsbruck, Austria\label{aff90}
    \and
    Satlantis, University Science Park, Sede Bld 48940, Leioa-Bilbao, Spain\label{aff91}
    \and
    Infrared Processing and Analysis Center, California Institute of Technology, Pasadena, CA 91125, USA\label{aff92}
    \and
    Instituto de Astrof\'isica e Ci\^encias do Espa\c{c}o, Faculdade de Ci\^encias, Universidade de Lisboa, Tapada da Ajuda, 1349-018 Lisboa, Portugal\label{aff93}
    \and
    Universidad Polit\'ecnica de Cartagena, Departamento de Electr\'onica y Tecnolog\'ia de Computadoras,  Plaza del Hospital 1, 30202 Cartagena, Spain\label{aff94}
    \and
    INFN-Bologna, Via Irnerio 46, 40126 Bologna, Italy\label{aff95}
    \and
    Kapteyn Astronomical Institute, University of Groningen, PO Box 800, 9700 AV Groningen, The Netherlands\label{aff96}
    \and
    Dipartimento di Fisica, Universit\`a degli studi di Genova, and INFN-Sezione di Genova, via Dodecaneso 33, 16146, Genova, Italy\label{aff97}
    \and
    Astronomical Observatory of the Autonomous Region of the Aosta Valley (OAVdA), Loc. Lignan 39, I-11020, Nus (Aosta Valley), Italy\label{aff98}
    \and
    Junia, EPA department, 41 Bd Vauban, 59800 Lille, France\label{aff99}
    \and
    ICSC - Centro Nazionale di Ricerca in High Performance Computing, Big Data e Quantum Computing, Via Magnanelli 2, Bologna, Italy\label{aff100}
    \and
    Instituto de F\'isica Te\'orica UAM-CSIC, Campus de Cantoblanco, 28049 Madrid, Spain\label{aff101}
    \and
    CERCA/ISO, Department of Physics, Case Western Reserve University, 10900 Euclid Avenue, Cleveland, OH 44106, USA\label{aff102}
    \and
    Laboratoire Univers et Th\'eorie, Observatoire de Paris, Universit\'e PSL, Universit\'e Paris Cit\'e, CNRS, 92190 Meudon, France\label{aff103}
    \and
    Dipartimento di Fisica e Scienze della Terra, Universit\`a degli Studi di Ferrara, Via Giuseppe Saragat 1, 44122 Ferrara, Italy\label{aff104}
    \and
    Istituto Nazionale di Fisica Nucleare, Sezione di Ferrara, Via Giuseppe Saragat 1, 44122 Ferrara, Italy\label{aff105}
    \and
    Dipartimento di Fisica "Aldo Pontremoli", Universit\`a degli Studi di Milano, Via Celoria 16, 20133 Milano, Italy\label{aff106}
    \and
    Universit\'e de Strasbourg, CNRS, Observatoire astronomique de Strasbourg, UMR 7550, 67000 Strasbourg, France\label{aff107}
    \and
    Kavli Institute for the Physics and Mathematics of the Universe (WPI), University of Tokyo, Kashiwa, Chiba 277-8583, Japan\label{aff108}
    \and
    Dipartimento di Fisica - Sezione di Astronomia, Universit\`a di Trieste, Via Tiepolo 11, 34131 Trieste, Italy\label{aff109}
    \and
    Minnesota Institute for Astrophysics, University of Minnesota, 116 Church St SE, Minneapolis, MN 55455, USA\label{aff110}
    \and
    Institute Lorentz, Leiden University, Niels Bohrweg 2, 2333 CA Leiden, The Netherlands\label{aff111}
    \and
    Universit\'e C\^{o}te d'Azur, Observatoire de la C\^{o}te d'Azur, CNRS, Laboratoire Lagrange, Bd de l'Observatoire, CS 34229, 06304 Nice cedex 4, France\label{aff112}
    \and
    Institute for Astronomy, University of Hawaii, 2680 Woodlawn Drive, Honolulu, HI 96822, USA\label{aff113}
    \and
    Department of Physics \& Astronomy, University of California Irvine, Irvine CA 92697, USA\label{aff114}
    \and
    Department of Astronomy \& Physics and Institute for Computational Astrophysics, Saint Mary's University, 923 Robie Street, Halifax, Nova Scotia, B3H 3C3, Canada\label{aff115}
    \and
    Departamento F\'isica Aplicada, Universidad Polit\'ecnica de Cartagena, Campus Muralla del Mar, 30202 Cartagena, Murcia, Spain\label{aff116}
    \and
    Department of Physics, Oxford University, Keble Road, Oxford OX1 3RH, UK\label{aff117}
    \and
    CEA Saclay, DFR/IRFU, Service d'Astrophysique, Bat. 709, 91191 Gif-sur-Yvette, France\label{aff118}
    \and
    Institute of Cosmology and Gravitation, University of Portsmouth, Portsmouth PO1 3FX, UK\label{aff119}
    \and
    Department of Computer Science, Aalto University, PO Box 15400, Espoo, FI-00 076, Finland\label{aff120}
    \and
    Caltech/IPAC, 1200 E. California Blvd., Pasadena, CA 91125, USA\label{aff121}
    \and
    Ruhr University Bochum, Faculty of Physics and Astronomy, Astronomical Institute (AIRUB), German Centre for Cosmological Lensing (GCCL), 44780 Bochum, Germany\label{aff122}
    \and
    DARK, Niels Bohr Institute, University of Copenhagen, Jagtvej 155, 2200 Copenhagen, Denmark\label{aff123}
    \and
    Univ. Grenoble Alpes, CNRS, Grenoble INP, LPSC-IN2P3, 53, Avenue des Martyrs, 38000, Grenoble, France\label{aff124}
    \and
    Department of Physics and Astronomy, Vesilinnantie 5, 20014 University of Turku, Finland\label{aff125}
    \and
    Serco for European Space Agency (ESA), Camino bajo del Castillo, s/n, Urbanizacion Villafranca del Castillo, Villanueva de la Ca\~nada, 28692 Madrid, Spain\label{aff126}
    \and
    ARC Centre of Excellence for Dark Matter Particle Physics, Melbourne, Australia\label{aff127}
    \and
    Centre for Astrophysics \& Supercomputing, Swinburne University of Technology,  Hawthorn, Victoria 3122, Australia\label{aff128}
    \and
    Department of Physics and Astronomy, University of the Western Cape, Bellville, Cape Town, 7535, South Africa\label{aff129}
    \and
    School of Physics and Astronomy, Queen Mary University of London, Mile End Road, London E1 4NS, UK\label{aff130}
    \and
    ICTP South American Institute for Fundamental Research, Instituto de F\'{\i}sica Te\'orica, Universidade Estadual Paulista, S\~ao Paulo, Brazil\label{aff131}
    \and
    Oskar Klein Centre for Cosmoparticle Physics, Department of Physics, Stockholm University, Stockholm, SE-106 91, Sweden\label{aff132}
    \and
    Astrophysics Group, Blackett Laboratory, Imperial College London, London SW7 2AZ, UK\label{aff133}
    \and
    Dipartimento di Fisica, Sapienza Universit\`a di Roma, Piazzale Aldo Moro 2, 00185 Roma, Italy\label{aff134}
    \and
    Centro de Astrof\'{\i}sica da Universidade do Porto, Rua das Estrelas, 4150-762 Porto, Portugal\label{aff135}
    \and
    Institute of Astronomy, University of Cambridge, Madingley Road, Cambridge CB3 0HA, UK\label{aff136}
    \and
    Department of Astrophysics, University of Zurich, Winterthurerstrasse 190, 8057 Zurich, Switzerland\label{aff137}
    \and
    Theoretical astrophysics, Department of Physics and Astronomy, Uppsala University, Box 515, 751 20 Uppsala, Sweden\label{aff138}
    \and
    Department of Physics, Royal Holloway, University of London, TW20 0EX, UK\label{aff139}
    \and
    Mullard Space Science Laboratory, University College London, Holmbury St Mary, Dorking, Surrey RH5 6NT, UK\label{aff140}
    \and
    Department of Physics and Astronomy, University of California, Davis, CA 95616, USA\label{aff141}
    \and
    Department of Astrophysical Sciences, Peyton Hall, Princeton University, Princeton, NJ 08544, USA\label{aff142}
    \and
    Cosmic Dawn Center (DAWN)\label{aff143}
    \and
    Niels Bohr Institute, University of Copenhagen, Jagtvej 128, 2200 Copenhagen, Denmark\label{aff144}
    \and
    Center for Cosmology and Particle Physics, Department of Physics, New York University, New York, NY 10003, USA\label{aff145}
    \and
    Center for Computational Astrophysics, Flatiron Institute, 162 5th Avenue, 10010, New York, NY, USA\label{aff146}
    }    

    \date{Received July 9, 2024; accepted September 12, 2024}

 
   \abstract
   {\Euclid will collect an enormous amount of data during the mission's lifetime, observing billions of galaxies in the extragalactic sky. Along with traditional template-fitting methods, numerous machine learning (ML) algorithms have been presented for computing their photometric redshifts and physical parameters (PPs), requiring significantly less computing effort while producing equivalent performance measures. However, their performance is limited by the quality and amount of input information entering the model (the features), to a level where the recovery of some well-established physical relationships between parameters might not be guaranteed -- for example, the star-forming main sequence (SFMS).
   
   To forecast the reliability of \Euclid photo-$z$s and PPs calculations, we produced two mock catalogs simulating the photometry with the UNIONS $ugriz$ and \Euclid filters. We simulated the Euclid Wide Survey (EWS) and Euclid Deep Fields (EDF), alongside two auxiliary fields. We tested the performance of a template-fitting algorithm (\phosphoros) and four ML methods in recovering photo-$z$s, PPs (stellar masses and star formation rates), and the SFMS on the simulated \Euclid fields. To mimic the \Euclid processing as closely as possible, the models were trained with \phosphoros-recovered labels and tested on the simulated ground truth. For the EWS, we found that the best results are achieved with a mixed labels approach, training the models with wide survey features and labels from the \phosphoros\ results on deeper photometry, that is, with the best possible set of labels for a given photometry. This imposes a prior to the input features, helping the models to better discern cases in degenerate regions of feature space, that is, when galaxies have similar magnitudes and colors but different redshifts and PPs, with performance metrics even better than those found with \phosphoros. We found no more than 3\% performance degradation using a COSMOS-like reference sample or removing $u$ band data, which will not be available until after data release DR1. The best results are obtained for the EDF, with appropriate recovery of photo-$z$, PPs, and the SFMS.}


   \titlerunning{\Euclid\/: Forecasting physical parameters and relations with ML}
   \authorrunning{Euclid Collaboration: A. Enia et al.}
   \maketitle
%
\section{\label{sc:Intro}Introduction}
\Euclid\footnote{\url{https://sci.esa.int/euclid/}} is an European Space Agency mission whose primary objective is to reveal the geometry of the Universe by measuring precise distances and shapes of $\sim 10^9$ galaxies up to $z \sim 3$, while it is also predicted to observe millions of galaxies at $3 < z < 6$ \citep{EuclidSkyOverview}. \Euclid will observe the extragalactic sky in four optical and \gls{NIR} filters: $\IE$, corresponding to $r$, $i$, and $z$ filters \citep{EuclidSkyVIS}; and $\YE$, $\JE$, and $\HE$ on the Near Infrared Spectrometer and Photometer \citep[NISP:][]{EuclidSkyNISP}. Such a wealth of data will dramatically improve our knowledge of the evolution of galaxies throughout cosmic time.

The \gls{EWS} will cover $\num{13345}$ deg$^2$ of the sky up to a $5 \sigma$ point-like source depth of $26.2$ mag in $\IE$ and $24.5$ mag in $\YE$, $\JE$, and $\HE$ \citep{Scaramella-EP1, Schirmer-EP18}. The \gls{EDF} will probe a smaller ($\sim 53$ deg$^2$) area to a targeted $5 \sigma$ point-like source depth of $28.2$ in $\IE$ and $26.5$ in $\YE$, $\JE$, and $\HE$. In total, \Euclid is expected to detect approximately ten billion sources and determine roughly $30$ million spectroscopic redshifts \citep[e.g.,][]{Laureijs11}. The \Euclid observations will be complemented with ground-based data from the Ultraviolet Near-Infrared Optical Northern Survey \citep[UNIONS, e.g.,][]{2017ApJ...848..128I}, the Legacy Survey of Space and Time \citep[LSST,][]{2008SerAJ.176....1I, 2009arXiv0912.0201L}, and the Dark Energy Survey \citep[DES,][]{2015AJ....150..150F, 2016MNRAS.460.1270D}, in order to have a complete wavelength coverage between $0.3\,\micron$ and $1.8\,\micron$.

Such a vast amount of data are out of computational reach for traditional template-fitting algorithms, which aim to model the observed \gls{sed} with a set of synthetic templates searching for the best fit parameters (i.e., photometric redshifts, stellar masses, and star formation rates) with computational times scaling linearly with the number of objects involved. For this reason, a wide set of \gls{ml} techniques have been proposed, developed, tested, and used to extract the maximum scientific information from such a huge amount of data, especially for the photo-$z$s \citep[][requiring a precision of $\sigma_z < 0.05$ and $< 10\%$ outlier fraction]{Desprez-EP10}, with the intention of speeding up the computational efforts while yielding comparable (or even better) performance in recovering the quantities of interest.

The past decade has seen an incredible surge in the use of \gls{ml} methods for astrophysical data analysis in virtually every possible subfield, from identification and modeling of strong lensing systems \citep{2017Natur.548..555H, 2022MNRAS.510..500G, 2023MNRAS.522.5442G, Leuzzi-TBD}, to classification tasks aiming to automatically identifying objects in images and catalogs, or to measure morphologies \citep{2015ApJS..221....8H, 2015MNRAS.450.1441D, 2018MNRAS.475..894T, 2021MNRAS.501.4579B, 2021MNRAS.506.2471G, 2022A&A...666A..87C, 2022A&A...666A..85L, EP-Aussel, Signor24}, to regression tasks, for example in finding the relationship between the photometric redshifts and/or physical properties from the observed photometry \citep{2003LNCS.2859..226T, 2004PASP..116..345C, 2013ApJ...772..140B, 2017MNRAS.466.2039C, 2018A&A...609A.111D, 2018MNRAS.477.1484U, 2019A&A...622A.137B, 2019MNRAS.486.1377D, 2019A&A...621A..26P, 2020MNRAS.493.4808S, 2021MNRAS.502.2770M, 2021MNRAS.507.5034R, 2021ApJ...908...47S, 2022A&A...665A..34D, 2022MNRAS.509.2289L, 2023A&A...679A.101C, Bisigello-EP23, 2023ApJS..264...29A, 2023ApJS..264...23L, 2024arXiv240200935A,2024arXiv240619437T}. Astrophysics has entered the big data era, and the potential of \gls{ml} methods has been revealed to the whole community.

However, as powerful as they can be, \gls{ml} techniques are not flawless. The goodness of the predicted quantities is inevitably limited by the quality (and size) of the input information used to train the model. Noisy features hamper a plain association between them and the desired outputs, degrading the final performance to a level where the optimal recovery of the most important quantities to place an observational constraint on galaxy evolution models might not be guaranteed at all. Some kind of agnostic analysis on the performance of \gls{ml} methods is necessary, as it is determining how those benchmark against classical methods (i.e., template-fitting).

Therefore, it is crucial to evaluate the \Euclid (and complementary data) capability to recover photometric redshifts, \gls{pp}, and the relationships between those, such as the star forming main sequence \citep[SFMS,][]{2007ApJ...670..156D, 2014MNRAS.443...19R}, and doing so in the most realistic way possible. This will help put the forthcoming \gls{EWS} and \gls{EDF} results into a more stable context and could act as a benchmark for those that will be obtained by the forthcoming large-area surveys of the next decade, LSST with the {\it Vera C. Rubin} Observatory \citep{2008SerAJ.176....1I}, and the {\it Nancy Roman} Space Telescope \citep{2019arXiv190205569A}.

\Euclid was successfully launched on July 1, 2023, reaching its observing orbit around the second Lagrange point (L2) the following month. The first public Data Release (DR1), covering $\sim 2500\,\mathrm{deg}^2$, is expected to be in June 2026. In the meantime, in order to estimate the performance of the survey's retrieved physical parameters (and relations), we make use of mock catalogs built from simulations, for which the ground truth (i.e., the real value of the physical parameters) is known.

This paper is outlined as follows: In Sect.\,\ref{sec:Mocks}, we describe the simulations from which we built \Euclid and ground-based photometry as inputs to the \gls{ml} models and test their performance. In Sect.\,\ref{sec:Methodology}, we describe the template-fitting and \gls{ml} methods used. In Sect.\,\ref{sec:Results}, we report the results, focusing in particular on the \gls{EWS} and \gls{EDF} and what can be done to improve the recovery of photo-$z$s and physical parameters with the \Euclid data products. In Sect.\,\ref{sec:Summary}, we present our conclusions and perspectives on other upcoming wide-area surveys.

In this work, we adopt a flat Lambda cold dark matter ($\Lambda$CDM) cosmology with $H_0 = 70\,\kmsMpc$, $\Omega_{\rm m} = 0.3$, and $\Omega_{\Lambda} = 0.7$, and assume a \citet{2003PASP..115..763C} initial mass function (IMF). The magnitudes are given in the AB photometric system \citep{1983ApJ...266..713O}.

\section{Building the mock catalogs}\label{sec:Mocks}
\begin{figure}
    \centering  
    \includegraphics[width=\hsize]{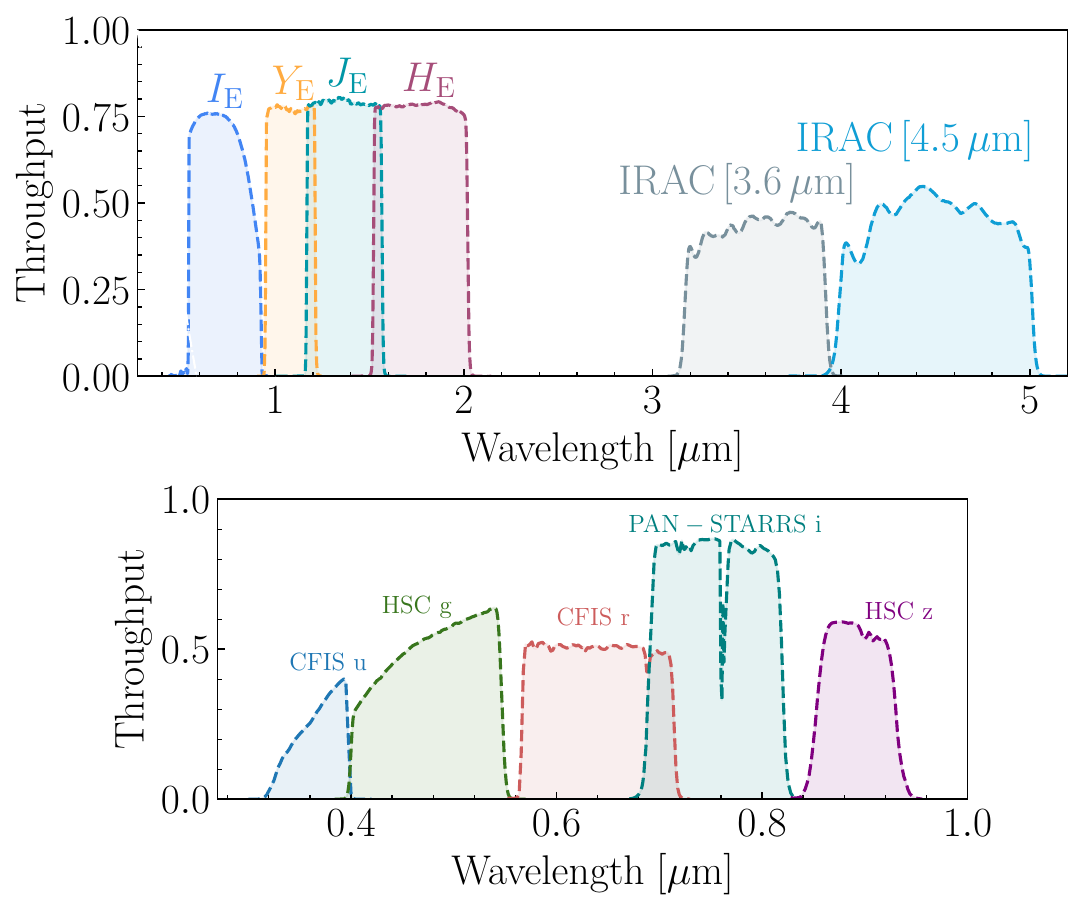}
    \caption{Throughput of the filters used through this work. On the top panel we show the four \Euclid filters: $\IE$, $\YE$, $\JE$, and $\HE$, along with two IRAC filters at $3.6\,\micron$ and $4.5\,\micron$. In the bottom panel, we include the four UNIONS filters that will complement the \Euclid data in the northern sky: CFIS $u$ and $r$, HSC $g$ and $z$, and PAN-STARRS $i$.} 
    \label{fig:filters}
\end{figure}
Assessing how good the \Euclid observations will yield to photometric redshifts and physical parameters necessarily passes through the use of simulated data, for which the ground truth is known. We want these simulations to be as close as possible to the real \Euclid data, which will not be available until DR1.

\subsection{The {\sc mambo} workflow}\label{sec:mambo}
In this work, we use the {\it Mocks with Abundance Matching in BOlogna} ({\sc mambo}) workflow \citep[see][for a thorough description]{Girelliphd}. {\sc mambo} starts from an N-body dark matter simulation to build an empirical mock catalog of galaxies, reproducing their observed physical properties and observables with high accuracy. The cosmological simulation used here is the Millennium dark matter N-body Simulation \citep{2005Natur.435..629S}, matched to the Planck cosmology following \cite{2010MNRAS.405..143A}, with a lightcone taken from \cite{2015MNRAS.451.2663H}, covering $3.14\,\mathrm{deg}^2$ with sub-halo masses $M_{200} > 1.7 \times 10^{10}\,h^{-1}\,M_\odot$ up to $z = 6$. Considering a typical stellar-to-halo mass relation \citep[SHMR,][]{2020A&A...634A.135G}, the corresponding stellar mass at low redshift is on the order of $\Mstarwun = 7.5$. In COSMOS2020 \citep{2022ApJS..258...11W}, galaxies with such a small stellar mass at low redshift are characterized by a $H$ band magnitude of $m_H \sim 25.2$. This is therefore the limit to be considered for the completeness of the {\sc mambo} simulation at very low redshifts $z < 0.2$; however, given that the volume of the simulation is very small at such redshifts, the incompleteness in the case of the simulated \gls{EDF} is negligible, and the simulation can be considered complete in all the explored regimes. The simulation extends to higher redshifts, in principle, but we cut it at $z = 6$, as it is the default limit of the main \Euclid pipeline for photometric redshifts.

Starting from the lightcone, the main parameters that we use are the position of each halo in RA and Dec, its redshift $z$, and the DM sub-halo mass. {\sc mambo} assigns to each galaxy its properties following empirical prescriptions with a scatter that randomizes the properties. In this way, not only do we ensure a better representation of the observed universe, but we also avoid the possible replication of galaxies that would be caused by a deterministic approach. As for the stellar masses $M_{\star}$, those come from a SHMR developed using a sub-halo abundance matching technique based on observed stellar mass functions (SMFs) on the SDSS, COSMOS, and CANDELS fields \citep{2020A&A...634A.135G}. The SMFs are:
\begin{itemize}
    \item \cite{2010ApJ...721..193P}, measured in the SDSS survey and divided into passive and star-forming using the rest-frame $(U-B)$ color at $z \sim 0$;
    \item \cite{2013A&A...556A..55I}, measured in COSMOS and classified into red or blue using the rest frame color selection $(NUV -r)$ vs $(r-J)$ at  $0.2 < z < 4$;
    \item \cite{2015A&A...575A..96G}, measured in CANDELS at $z \geq 4$.
\end{itemize} 

Every galaxy is randomly assigned a star-forming or passive and quiescent label based on the ratio of the stellar mass functions (SMFs) for the blue and red populations. Due to the high observational uncertainties of the fraction of SF/Q galaxies at $z > 4$ \citep{2018MNRAS.473.2098M, 2019A&A...632A..80G}, the star-forming fraction $f_{\rm SF}$ was extrapolated from the results at lower redshifts with a limit of $f_{\rm SF} = 99\%$ up to $z = 6$. 

All the other properties, for example, SFR, metallicity, rest-frame, and observed photometry from UV to submillimeter in the desired bands, are extracted with the Empirical Galaxy Generator \citep[\texttt{EGG},][]{2017A&A...602A..96S}, a C++ code that creates a mock catalog of galaxies from a simulated lightcone, whose empirical nature assures that the retrieved physical properties are realistic -- as long as the \texttt{EGG} models are. In the configuration of \texttt{EGG} used for {\sc mambo}, each galaxy SED is assigned from a pre-built library of templates from the \citet{2003MNRAS.344.1000B} models covering the $UVJ$-plane \citep{2009ApJ...691.1879W}. Models in the library are derived with a Salpeter IMF \citep{1955ApJ...121..161S}, but we subsequently converted stellar masses and SFRs to a Chabrier IMF \citep{2003PASP..115..763C}. The physical properties (and type, i.e., star-forming or quiescent) are randomly extracted using empirical relations starting from the stellar mass previously assigned, once again covering the full $UVJ$-plane.
\begin{figure*}
    \centering
    \includegraphics[width=\hsize]{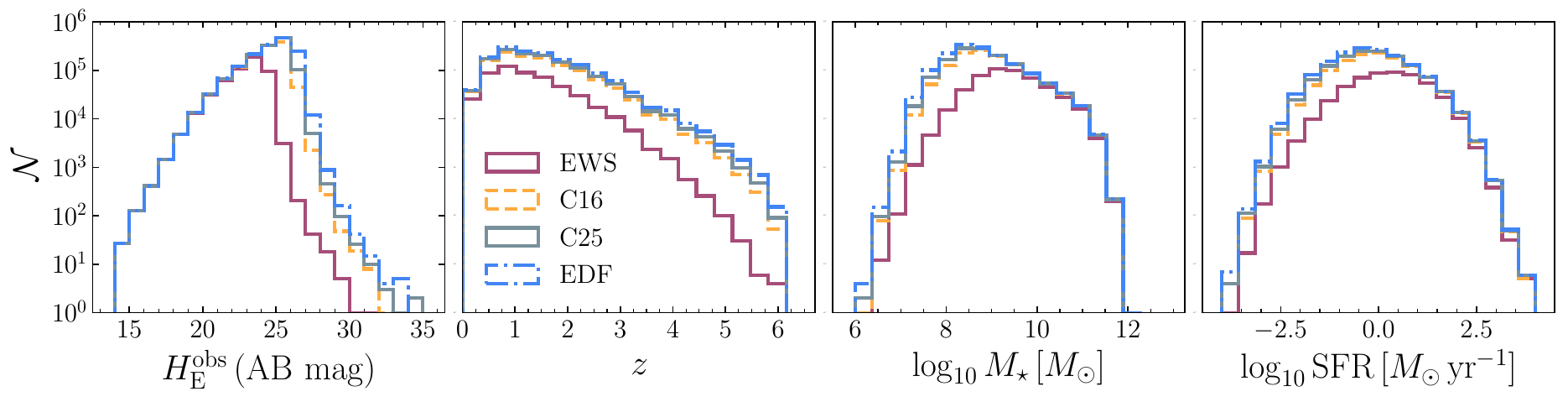}
    \caption{Four simulated \Euclid catalogs used in this work (solid purple line, \gls{EWS}; dashed orange line, C16; solid gray line, C25; dashed-dotted blue line, \gls{EDF}) shown as the number of sources as a function of the \Euclid $\HE$ band magnitude (leftmost panel), redshift (center left), stellar mass (center right), and star formation rate (rightmost panel). We notice that the magnitude cut upon which the fields are built is an OR condition on the ${\rm S/N}$ in $\HE$, $\IE$ filters; as such, simulated galaxies are found below the nominal limiting magnitude cut for $\HE$ band, as those are detected at ${\rm S/N} > 10$ in $\IE$.} 
    \label{fig:catalog_hist}
\end{figure*}
\begin{table}[]
    \centering
    \caption{Set of filters used in this work.}\label{tab:bands}
    \begin{tabular}{lccc}
        \hline
        \noalign{\vskip 1pt}
        Band &  $\lambda_{\rm eff}$ [\micron] & \gls{EWS} \\
        \hline
        \noalign{\vskip 1pt}
        CFHT/MegaCam $u$ & $0.372$ & $23.6$ \\
        HSC $g$          & $0.480$ & $24.5$ \\
        CFHT/MegaCam $r$ & $0.640$ & $24.1$ \\
        PAN-STARRS $i$   & $0.755$ & $23.2$ \\
        HSC $z$          & $0.891$ & $23.4$ \\
        VIS/\IE          & $0.715$ & $25.0$ \\
        NISP/\YE         & $1.085$ & $23.5$ \\
        NISP/\JE         & $1.375$ & $23.5$ \\
        NISP/\HE         & $1.773$ & $23.5$ \\
        \hline
        \hline
        \noalign{\vskip 1pt}
        IRAC/[$3.6\,\micron$] & $3.550$ & $24.8$ \\
        IRAC/[$4.5\,\micron$] & $4.493$ & $24.7$ \\
        \hline
    \end{tabular}
    \tablefoot{Reported magnitudes are the $10\sigma$ expected observational depths for an extended source in a $2 \arcsec$ diameter aperture. IRAC observations will be available only for the \gls{EDF}; as such, their reported magnitudes are the ones measured in the EDF-N and \gls{EDF}-F \citep[see][]{Moneti-EP17, EP-McPartland}.}
\end{table}

With {\sc mambo}, we generate a mock catalog of roughly five million galaxies between redshifts zero and six, with the same photometric filters as the ones expected for DR1 in the \gls{EWS} in the northern hemisphere, where a network of multiple collaborations will obtain data in different bands as part of the Ultraviolet Near-Infrared Optical Northern Survey (UNIONS), whose throughput is shown in Fig.\,\ref{fig:filters}. These are the Canada-France Imaging Survey \citep[CFIS;][on the Canada-France-Hawaii Telescope CFHT]{2017ApJ...848..128I} for bands $u$ and $r$; Subaru Hyper Suprime-Cam \citep[HSC:][]{2018PASJ...70S...1M} observations for $z$ and $g$ bands as part of the Wide Imaging with Subaru HSC of the \Euclid Sky (WISHES) and the Waterloo Hawaii IfA G band Survey (WHIGS); PAN-STARRS in band $i$ \citep{2016arXiv161205560C}, and the \Euclid $\IE$, $\YE$, $\JE$ and $\HE$ filters \citep{2016SPIE.9904E..0QC, 2016SPIE.9904E..0TM}.

The \gls{EDF} has already been observed with the {\it Spitzer} Space Telescope’s Infrared Array Camera \citep[IRAC,][]{2004ApJS..154....1W, 2004ApJS..154...10F} at $3.6\,\micron$ and $4.5\,\micron$. These observations are described in detail in \citet{Moneti-EP17} and \citet{EP-McPartland}. When dealing with the \gls{EDF}, we also include these two photometric filters, assuming the same observation depth reported in \citet{EP-McPartland}.

For convenience, the full set of filters is also listed in Table\,\ref{tab:bands}, with the corresponding expected $10 \sigma$ observation depths for a generic extended source (in a $2\arcsec$ aperture, i.e., a typical \Euclid extended source) per band in the \gls{EWS} -- with attached IRAC observed depths in the same aperture for the \gls{EDF}.

\subsection{The \Euclid simulated fields}\label{sec:simfields}
\begin{table}[]
    \centering
    \caption{Four simulated \Euclid catalogs.}\label{tab:catalogs}
    \begin{tabular}{lcccc}
        \hline
        \noalign{\vskip 1pt}
        {}       & ROS & Sources & $\IE$ lim & $\HE$ lim \\
        \hline
        \noalign{\vskip 1pt}
        \gls{EWS}    & $\phantom{0}1$ & $\phantom{0}\num{512527}$  & $25.00$  & $23.50$  \\
        C16 & $16$ & $\num{1209598}$ & $26.50$  & $25.00$  \\
        C25 & $25$ & $\num{1361041}$ & $26.75$  & $25.25$  \\
        \gls{EDF}    & $40$ & $\num{1534023}$ & $27.00$  & $25.50$  \\
        \hline
    \end{tabular}
    \tablefoot{The limits are the $10\sigma$ ($\IE$) and $5\sigma$ ($\HE$) expected observational depths for an extended source in a $2 \arcsec$ diameter aperture.}
\end{table}
We simulate different versions of \Euclid observations by adding realistic photometric noise to each band depending on the number of reference observation sequences that are going to be observed \citep[ROS, see Fig. 8 of][]{Scaramella-EP1} and the expected limiting magnitudes of the survey. A galaxy is kept in the catalog if it is detected either in $\HE$ at ${\rm S/N} > 5$ or in $\IE$ at ${\rm S/N} > 10$, given the expected limiting magnitude. Those limits were used because they enable a posteriori selections for other \Euclid analyses, such as cluster detection and weak lensing analysis.

The four simulated catalogs (see Table\,\ref{tab:catalogs} and Fig.\,\ref{fig:catalog_hist}) are:
\begin{itemize}
    \item Wide, a single ROS at limiting magnitudes of $H_{\sfont{E},{\rm lim}} = 23.5$ and $I_{\sfont{E},{\rm lim}} = 25.0$, simulating what is expected from the \gls{EWS} \citep{Scaramella-EP1}.
    \item C16, 16 ROS at limiting magnitudes of $H_{\sfont{E},{\rm lim}} = 25.0$ and $I_{\sfont{E},{\rm lim}} = 26.5$, corresponding to a limit $1.5$ mags deeper than the \gls{EWS}. This simulates the so-called \Euclid auxiliary fields \citep{Scaramella-EP1}, six well-known regions with vast ancillary information, observed for photometric and color calibration; 16 ROS are expected to be observed by the time of DR1.
    \item C25, 25 ROS at limiting magnitudes of $H_{\sfont{E},{\rm lim}} = 25.25$ and $I_{\sfont{E},{\rm lim}} = 26.75$, corresponding to a limit of $1.75$ mags deeper than \gls{EWS}. This simulates the expected final average number of ROS to the \Euclid auxiliary fields.
    \item Deep, 40 ROS reaching limiting magnitudes of $H_{\sfont{E},{\rm lim}} = 25.5$ and $I_{\sfont{E},{\rm lim}} = 27.0$, corresponding to an expected limiting magnitude of $2$ mags deeper than \gls{EWS}, simulating the minimum number of ROS of the different fields composing the \gls{EDF} (north, south, and Fornax).
\end{itemize}

We notice that the magnitude limits reported here are different from the ones in \cite{Scaramella-EP1}, which refer to point-sources at $5\sigma$. Here instead, we convert those to $10\sigma$ limits for an extended source with a $2\arcsec$ aperture (as a proxy for a typical \Euclid extended source).
 
We are building the calibration fields by improving the \gls{EWS} photometry on the \Euclid and UNIONS filters; however, the real auxiliary fields, such as the Cosmic Evolution Survey field \citep[COSMOS;][]{2007ApJS..172....1S} will benefit from a wealth of multiwavelength ancillary data \citep[i.e., the COSMOS2020 catalog,][]{2022ApJS..258...11W} that will yield better photometric redshifts and physical parameter estimation with respect to what we report in this work.

There are a few caveats about the simulated catalogs. While the photometric noise for all the considered mock catalogs is simulated in the most realistic way possible, we are still dealing with an idealized situation where the photometric procedures are bypassed. Moreover, within the catalogs, we are considering galaxies only, without accounting for any source of contamination that the real \Euclid data will have to deal with: contaminants such as stars, photometric masks (e.g., from stars) and defects (snowballs, cosmic rays, persistence from solar flares), AGN and QSOs \citep{2024arXiv240906700L}, under-deblended and over-deblended objects, Local Universe extended objects, and low surface brightness galaxies. While all of those are expected to be reduced to the minimum possible \citep[i.e., by exploiting \gls{ml} to automatically classify stars and galaxies, see e.g.][whose reported F1-scores are $\sim 98\%$]{2022A&A...666A..87C}, some degree of performance degradation will be unavoidable.

Finally, as the absolute best-case scenario, yielding the best possible value for each quality metric, we report the results coming from the unperturbed version of the survey, that is, the {\sc mambo} generated catalog without any photometric noise added, run on the ground-truth magnitudes. Regardless of the flaws that may be inherent in the simulations, whatever uncertainty is generated from this set of photometric values and parameters depends only on the technique used to derive the second from the first, such as badly interpolated holes in the feature space for \gls{ml} algorithms or a lack of SED models and degeneracies between colors and physical properties for template-fitting algorithms.

\section{Methodology}\label{sec:Methodology}
In this section, we describe the algorithms and metrics used to assess the model's performance in recovering the ground truth. In particular, we focus on the recovery of photometric redshifts $z$ and two physical parameters: stellar masses $\Mstarwun$ and star formation rates $\sfrwun$, and the relation between them, the \gls{ms}.

\subsection{Feature, labels, and samples}\label{sec:labels}
In line with the standard \gls{ml} terminology, we now designate the catalogs' photometry and subproducts (i.e., broad-band photometry and colors) as features, and the model output (i.e., redshifts and physical properties) as labels. In this work, we address two versions of the latter:
\begin{itemize}
    \item the true labels, which are the ground-truth $z$ and physical properties extracted from {\sc mambo};
    \item the recovered labels, whose values have been obtained by running a traditional template-fitting code (\phosphoros, see Sect.\,\ref{sec:phosph}) on the (simulated) \Euclid observed features.
\end{itemize}

Thus, we can check what the best possible performance for a particular run is (i.e., when the redshifts and physical parameters are perfectly known) to compare with the more realistic ones that will be obtained with \Euclid data, when the ground truth will be inevitably unknown and recovered labels will be built from the observed features as the model input samples. This information is useful, especially in cases where the \gls{EWS} performance are evaluated with a reference sample built from the calibration fields (see in the next paragraphs). As reported in Sect.\,\ref{sec:simfields}, the simulated ones have labels recovered from the same set of filters as the \gls{EWS}, but the real ones will benefit from lots more multiwavelength data, with better recovered photo-$z$ and \gls{pp}. The expected real performance of such cases should therefore be in between the recovered and true labels performance.

Every supervised \gls{ml} application is composed of a training (or reference\footnote{For some \gls{ml} methods (i.e., the nearest-neighbors algorithms), there is no training phase (or it could be considered instantaneous training); as such, the sample from which the predictions are inferred is not referred to as the training sample but as the reference sample.}) sample from which the relations between features and labels are inferred, and a target (or test\footnote{Similarly, both terms are applied to data samples that will not be utilized for the model's training or as a reference sample. The subtle distinction lies in the utilization of a test sample, specifically employed for evaluating the model's performance.}) sample on which the models are applied. As described in Sect.\,\ref{sec:Mocks}, we have four different simulated versions of \Euclid observations, mimicking the expected outcome of the \gls{EWS}, \gls{EDF}, and two calibration (auxiliary) fields. As common practice in \gls{ml} applications, we split those catalogs, using part of the sample for training (or as references) and the rest equally split for cross-validation and testing. In this work, we used $90\%$ of the samples for training (translating into training sets between $500$\,k and one million galaxies), ensuring at least $\sim 50$\,k galaxies in the test sample, which is more than enough to evaluate the model's performance.

To understand the actual performance expected from the observed \Euclid data, we explore the predictive capabilities of models trained on deeper photometry when applied to a shallower one (in this case, the \gls{EWS}). For instance, this is achieved by training a model using the \gls{EDF} catalog and subsequently evaluating it on the \gls{EWS} catalog. In those cases, at test time, we share the same set of train and test sources between the catalogs to be as consistent as possible. The same is done for every \gls{ml} method used in this work, and we share the same train, reference, and test samples for the same catalog permutation between different models (e.g., when testing the performance of a model trained on the \gls{EDF} catalog and tested on the \gls{EWS} one, the training and test source IDs are the same for all the methods considered).

When dealing with recovered labels, in order to simulate a typical application where the ground truth is unknown as \Euclid will observe photometry from which the photo-$z$s and physical parameters will be derived, we train the models on those and test on the true labels.

\subsection{Features engineering}\label{sec:feateng}
As reported in Sect.\,\ref{sec:Mocks}, in standard \gls{ml} terminology, the catalogs observed photometric values are the features of the models. At the base level, each entry in the features space is a single galaxy's simulated photometry, in magnitudes; that is, the nine \Euclid + UNIONS bands for the \gls{EWS}, with the addition of two IRAC bands for the \gls{EDF}. In order to improve the quality of the models, thus the model inferences, we also include derived features as the colors (pairwise differences of the magnitudes, excluding permutations), increasing the number of total features to $45$ (\gls{EWS}) and $66$ (\gls{EDF}). This is the number of features for each of the previously described methods, with the notable exception of the \gls{CCR}, where the inferred labels are added on top of those as new training features at each iteration, as described in Sect.\,\ref{sec:ChRegr}.

All of the methods presented in this section are not sensitive to the dynamic range and scales of the input features, except for the \gls{dlnn}. In that case, we scale the features to a similar dynamic range with a standard Z-score normalization.

\subsection{\phosphoros}\label{sec:phosph}
\phosphoros\footnote{\url{https://phosphoros.readthedocs.io}} (Paltani et al., in prep) is a Bayesian template-fitting tool for galaxies \gls{sed} developed within the \Euclid collaboration. In the \Euclid photo-$z$ data challenge, which evaluated the performance metrics of different template-fitting and \gls{ml} codes in retrieving the photometric redshift of a mock catalog, \phosphoros\ yielded the best performance along with LePhare \citep{Desprez-EP10}.

\phosphoros\ can be used to evaluate at the same time the photometric redshift and the physical properties of galaxies that have to be provided as tags for the templates. In the present work, we have used $1254$ templates from \citet[][in the 2016 version\footnote{\url{http://www.bruzual.org/bc03/Updated_version_2016/}}]{2003MNRAS.344.1000B} with Chabrier IMF \citep{2003PASP..115..763C}, considering exponentially declining (e-folding timescale $\tau = 0.1, 0.3, 1, 2, 3, 5, 10, 15, 30$\,Gyr) and delayed (characteristic timescale $\tau = 1, 3$\,Gyr) star formation histories, 2 metallicities ($Z=0.008, 0.02$) and $57$ ages between $0.01$ and $13.5$\,Gyr. The internal dust attenuation has been modeled with Calzetti's law \citep{Calzetti2000} with $E(B-V)$ values in the range $[0.0,0.5]$. We tested whether the IMF choice for the templates might bias the performance by running \phosphoros\ with templates built with Salpeter IMF \citep{1955ApJ...121..161S}, finding identical results in terms of performance metrics (see Sec.\,\ref{sec:metrics}), though almost monolithically shifted by a factor $0.23$ dex in logarithm with respect to the Chabrier results, shown throughout the paper.

As a first step, a grid with model photometry is derived for all the templates in the redshift range $z\in[0.0,6.0]$ with steps of $\diff z=0.01$. When comparing the model to the observed photometry, the only factor considered prior is the “volume-prior”, proportional to the redshift-dependent differential comoving volume. Upper limits are treated in a statistical sense, as models with fluxes over the limit in the undetected bands are still considered when looking for the best-fit model; in those cases, the $\chi^2$ evaluation follows the indications in \citet{2012PASP..124.1208S}. In the version that will be used for \Euclid data, there will also be the possibility of consistently dereddening the photometry for the Galactic extinction \citep{2017A&A...598A..20G} and considering the variability of the filter transmission functions across the field of view \citep{EP-Paltani}. A recipe to add emission lines is also implemented in \phosphoros, but not used in this work.

The final result of the computation is the characterization of the multidimensional posterior with a density sampling of $100$ values for each galaxy, as well as the values of the physical properties and redshift from the best posterior model as the mean, the median (used in this work), or the mode of the distribution.

In this work, we use \phosphoros\ results to benchmark \gls{ml} methods against a standard template-fitting algorithm and as a necessary step to build the reference sample to use as input for \nnpz.

\subsection{\nnpz}\label{sec:nnpz}
The Nearest-Neighbors Photometric Redshift \citep[\nnpz, see][for a first application on the HSC-SSP survey]{2018PASJ...70S...9T} is a supervised-learning technique mapping a given set of features to known labels with an upgraded version of the $k$-nearest neighbors algorithm ($k$-NN). In its most simple form, a $k$-NN algorithm combines an integer number of $k$ neighbors in a reference sample closest to the target in feature space with respect to some distance metric (e.g., Euclidean) and predicts a label based on some user-defined combination of the metrics of the $k$-NN reference sample labels (e.g., a mean weighted by the distance in feature space). The same conceptual approach can be employed to provide a posterior distribution function (PDF) for the desired label by combining some a priori known PDFs for the reference sample under the assumption that similar observations with similar uncertainties would naturally produce similar results. Predictions and confidence intervals will naturally follow from the output PDF.

This is the concept behind \nnpz\ in a nutshell. The reference sample is built starting from \phosphoros\ as a set of objects whose full parameters' PDF has been sampled with $100$ randomly extracted points according to the PDF density distribution. The samples of the $k$-NNs are then combined to produce the target PDF, from which a punctual prediction is obtained from the mean, the median, or the mode of the distribution.


In this work, we used the 1.2.2 version of \nnpz\ available on the \Euclid\ {\it LOcal DEvelopment ENvironment} (LODEEN) version 3.1.0, a virtual machine containing all of the \Euclid software and pipelines. As for the code hyperparameters, after a first skim in a batch of at least $1000$ nearest neighbors in the target space obtained from a space partition with KDTree (necessary to speed up the whole process instead of simply brute-forcing the search), we fix the final $k$ from which the target labels are evaluated to $30$ nearest neighbors. To generate a prediction, each neighbor is weighted with its $\chi^2$ likelihood, which is the $\chi^2$ distance between the reference neighbor and the target point in the feature space. \nnpz\ combines the posterior coming from all the nearest neighbors and produces a PDF for the predicted target galaxy, from which we extract the point prediction as the median value of the distribution. We perform the same tests presented in this work with the mode of the distribution (i.e., the maximum-likelihood estimator) as the point prediction without noticing a significant change in the results.

In fact, returning a source's multivariate PDF samples as output instead of a single-point prediction is one of the great advantages of \nnpz. This information is in principle recoverable with other \gls{ml} algorithms, such as \catboost\ \citep[see][for an application to a simple random forest]{2021MNRAS.502.2770M} if considering all the training samples in a particular leaf as PDF samples, though this is computationally and memory-wise less feasible than the $\sim 100$ samples per galaxy of \nnpz.

\subsection{Other machine learning techniques}
Apart from \nnpz, we performed similar tests using previously tested \gls{ml} techniques that have been shown to be extremely efficient for redshift and galaxy property estimation: \gls{GBDT} and \gls{dlnn}.

\subsubsection{\catboost\ single-model regressor}\label{sec:Catboost}
A \gls{GBDT} is rooted in decision trees, a building block of widely used and successful techniques for regression and classification tasks. In a decision tree, the data is recursively split into smaller subsets based on the features that best separate the data according to some information gain (for classification) or variance minimization (for regression) criteria, until a stopping criterion is met. The result is a tree-like structure, with each internal node representing a feature, each branch representing a potential value for that feature, and each final node (leaf) representing a class (for classification) or predicted value (for regression). This scheme has been improved by what is called gradient boosting, which decreases the randomness improvement in training by starting with a set of imprecise decision trees ({\it “weak learners”}) and iteratively improving them, focusing on what these are predicting wrong rather than generating a new random subset of the data.

\catboost\ \citep{NEURIPS2018_14491b75}\footnote{\url{https://catboost.ai/}} is a cutting-edge \gls{ml} algorithm specifically designed for gradient boosting on decision trees. There are some specific features that help reduce some typical issues in gradient boosting algorithm implementations, such as the ordered boosting to reduce overfitting and the oblivious trees to regularize while increasing speed.

In this work, we use \catboost\ in two different ways. With the \gls{CSMR}, we train a single model to solve a multiregression problem. Each set of features is associated with a pool of labels ($z_{\rm phot}$, $M_\star$, ${\rm SFR}$) and not just a single label per time (as in Sect.\,\ref{sec:ChRegr}), finding the best model with a Multivariate Root Mean Square Error ({\it MultiRMSE}) loss. In each case presented, the final model is trained with 1000 estimators, allowing for a maximum depth of 11.

\subsubsection{\catboost\ chained regressors}\label{sec:ChRegr}
With the \gls{CCR} we train on a set of features one label per time, and iteratively append the predicted labels to the features up until convergence, allowing the model to naturally learn the correlation between parameters through an iterative approach. A thorough and more detailed description can be found in Humphrey et al., (in prep). Here, we summarize it in the following paragraph.

We start with a training set ($X_{\rm train}$, $y_{\rm train}$) and a test set ($X_{\rm test}$, $y_{\rm test}$). The first iteration goes as follows:
\begin{enumerate}
    \item the model is trained on $X_{\rm train}$ whose features are the full set of colors and magnitudes (with permutations, see in Sect.\,\ref{sec:feateng}), with only $z_{\rm phot}$ as the lone label in $y_{\rm train}$. From this model, we can predict some $z_{\rm phot}$ and evaluate their performance metrics on the test sample.
    \item Now, the model is trained on a new $X_{\rm train}$ that is composed of the previous ones (magnitudes and colors) and the $z_{\rm phot}$ predicted in 1. From this model, we predict $\Mstarwun$ and evaluate the $\Mstarwun$ performance metrics on the test sample; of course, the $X_{\rm test}$ has been extended to incorporate the new feature from the predicted $z_{\rm phot}$ on the test sample.
    \item Then, the model is trained on another $X_{\rm train}$ composed of the previous features plus the predicted $\Mstarwun$ in the previous step. With this model, we predict $\sfrwun$ and evaluate the $\sfrwun$ performance metrics on the (once again, extended) test sample.
\end{enumerate}
Now the second iteration starts.
\begin{enumerate}[resume]
    \item The model is trained on an $X_{\rm train}$ composed of the previous features -- including the $z_{\rm phot}$ and $\Mstarwun$ predicted in steps (1) and (2) -- plus the predicted SFRs in step (3), and with this model, we re-predict $z_{\rm phot}$ and evaluate their performance metrics.
    \item Again, another model is retrained with the previous features plus the new $z_{\rm phot}$ predicted in the previous step. $\Mstarwun$ is re-predicted, and the model performance on the label is evaluated.
\end{enumerate}
The whole procedure goes on for four iterations, when we observe a convergence of the evaluated metrics in agreement with Humphrey et al., (in prep).

As such, the model features are (in square brackets, the step in which they have been evaluated):
\begin{enumerate}
    \item magnitudes and colors;
    \item magnitudes, colors, and $z_{\rm phot}$ [1];
    \item magnitudes, colors, $z_{\rm phot}$ [1], and $\Mstarwun$ [2];
    \item magnitudes, colors, $z_{\rm phot}$ [1], $\Mstarwun$ [2], and $\sfrwun$ [3];
    \item magnitudes, colors, $z_{\rm phot}$ [1], $\Mstarwun$ [2], $\sfrwun$ [3], and $z_{\rm phot}$ [4].
\end{enumerate}
And so on, for four iterations.

Finally, we store the final set of label predictions for the test set on which we evaluate the performance metrics (see Sect.\,\ref{sec:metrics}). In running \catboost\, we use the same set of hyperparameters as in \gls{CSMR} and in Humphrey et al., (in prep).

\subsubsection{\label{sc:DLNN}Deep learning neural network}
\begin{table}[]
    \centering
    \caption{Architecture of the \gls{dlnn} used in this work.}\label{tab:DLNNarch}
    \begin{tabular}{lcc}
        \hline
        Layer & $N_{\rm in}$ & $N_{\rm out}$ \\
        \hline
        input & $N_{\rm feat}$ & 4096 \\
        dense & 4096 & 2084 \\
        dense & 2048 & 1204 \\
        dense & 1024 & \phantom{0}512 \\
        dense & \phantom{0}512 & \phantom{00}64 \\
        output & \phantom{00}64 & \phantom{000}3 \\
        \hline
    \end{tabular}
    \tablefoot{All layers are dense, fully connected with a ReLU activation function. $N_{\rm feat}$ values are reported in Sect.\,\ref{sec:feateng}.}
\end{table}

As we only deal with structured (i.e., tabular) data, we also test the performance of a simple, multilayered \gls{dlnn}. Here we adopt a typical architecture that has been widely used in the literature in searching for photometric redshifts and physical parameters \citep[e.g.][]{2003MNRAS.339.1195F, 2004PASP..116..345C, Bisigello-EP23}.

The \gls{dlnn} inputs are the training features (magnitudes and colors, with permutations, see Sect.\,\ref{sec:feateng}), and the output is a set of three labels ($z_{\rm phot}$, $M_\star$, $\mathrm{SFR}$). The \gls{dlnn} architecture (described in Table\,\ref{tab:DLNNarch}) consists of five fully connected layers with a decreasing power of two hidden units for each layer. The adopted activation function for each layer is a Rectified Linear Unit \citep[ReLU,][]{nair2010rectified}, Mean Squared Error (MSE) for the loss function with $L_2$ regularization to avoid overfitting, and the model is optimized with the ADaptive Moment estimator \citep[Adam,][]{2014arXiv1412.6980K}.

The \gls{dlnn} for each model are trained and tested on the same train and test samples as for all the other methods. We run the training on mini-batches of size $512$.

\subsection{Metrics and quality assessment}\label{sec:metrics}
We use standard metrics to quantify the model's performance. Those are defined differently when referring to redshifts or \gls{pp}.

The first is the normalized median absolute deviation, defined as:
\begin{equation}
{\rm NMAD} = 1.48\,\times\,{\rm median}
\begin{cases}
 \,\dfrac{|z_{\rm pred} - z_{\rm test}|}{1+z_{\rm test}} - b & \text{for redshifts;} \\
 \,|y_{\rm pred} - y_{\rm test}| - b & \text{for \gls{pp},}
\end{cases}
\end{equation}
with $b$ being the model bias (see below).

The outlier fraction $f_{\rm out}$ is defined as the fraction of catastrophic outliers \citep{2010A&A...523A..31H} over a certain threshold (in log space for physical parameters, linear for redshifts):
\begin{equation}
f_{\rm out}\,:
    \begin{cases}
    \,\dfrac{|z_{\rm pred} - z_{\rm test}|}{1+z_{\rm test}} > 0.15 & \text{for redshifts;} \\
    \,|y_{\rm pred} - y_{\rm test}| > t_{\rm out} & \text{for \gls{pp}.} 
    \end{cases}
\end{equation}
These thresholds have been evaluated looking at the standard deviation of the \gls{pp} distribution, considering only sources with a good photo-$z$ recovery -- that is, below the $0.15 (1+z)$ threshold -- for all the methods considered, trained with true labels (see Sec.\,\ref{sec:labels}). As a consequence, the \gls{pp} thresholds are not the same for stellar masses or SFRs. The chosen thresholds are two times the mean standard deviation between the prediction and the true values found for all the considered methods, rounded to the nearest decimal. The values are $t_{\rm out} = 0.4$ dex for stellar masses and $t_{\rm out} = 0.8$ dex for SFRs.

Defining the catastrophic outliers in this way is different than assuming a plain 0.3 dex difference between the prediction and the true values of the physical parameters (corresponding to a factor of two) that is found in recent literature \citep[e.g.][]{Bisigello-EP23}. Fixing the same value in dex space for every PP is a penalizing choice, especially for SFRs, where a 0.3 dex difference might even be the order of (or even below) $1\sigma$ of the distribution, which is too little to define a galaxy as a \textit{catastrophic} outlier. In this way, we adopt a more robust definition from a statistical sense, which actually returns an informative quantitative description of what a catastrophic outlier is for a stellar mass estimate or a star formation rate.

Finally, a model's overall bias $b$ is:
\begin{equation}
b\,={\rm median}
    \begin{cases}
    \,\left( \dfrac{z_{\rm pred} - z_{\rm test}}{1+z_{\rm test}} \right) & \text{for redshifts;} \\
    \,(y_{\rm pred} - y_{\rm test}) & \text{for \gls{pp}.}
    \end{cases}
\end{equation}

In all three cases, the closer to zero, the better the predicted values resemble the test ones. Of all three, only the bias can take either positive or negative values.

We notice that those metrics are different from the \Euclid requirements, that depend on the redshift probability distribution functions \citep[PDZ, see their definition in Sect.\,4.2 of][]{Desprez-EP10}. We use photo-$z$ and \gls{pp} point estimates instead.

As for the \gls{ms}, we evaluate the performance of the recovered relation by evaluating three parameters:
\begin{itemize}
    \item the relation slope $m$, measured with an \gls{odr};
    \item the fraction of passive galaxies $f_{\rm p}$, defined as the fraction of objects with specific-SFR $\logten ({\rm sSFR}/\si{G\year}^{-1}) < -1$. This limit has been determined looking at the divide between passive and nonpassive galaxies in {\sc mambo} and is in accordance with values found in the literature \citep[e.g.][]{2010A&A...523A..13P, 2013A&A...556A..55I};
    \item the relation scatter $\sigma$, measured only on nonpassive sources.
\end{itemize}

\section{Results}\label{sec:Results}
In this section, we present the results of the methods presented in Sect.\,\ref{sec:Methodology} on the simulated \Euclid fields described in Sect.\,\ref{sec:Mocks}. Our primary objective is to evaluate the performance of the methods described in Sect.\,\ref{sec:Methodology} and find the optimal strategies to extract the maximum amount of information available in the \Euclid survey, in particular the \gls{EWS} and \gls{EDF}. We ought to do so by dealing with realistic information, that is, what the survey will actually deliver as a scientific product. As such, we present results for the \gls{EWS} and for the \gls{EDF}, obtained by combining the \phosphoros\ results (photometric redshifts, physical parameters) in the fields with the available photometry. As described in Sect.\,\ref{sec:Mocks}, we deal with two kinds of training labels: the true ones (i.e., the ground-truth simulated parameters), which of course are unknown in a real-life application, and as such, we only use them to assess what the best-case scenario for a particular field could be, and the recovered labels, whose values are the \phosphoros\ outputs resulting from the observed photometry, which is what an actual application to real \Euclid data will have to deal with. However, it is worth noticing that spectroscopic redshifts will be available for a smaller sample (still around the order of millions of sources), which is the closest thing to true labels that the \gls{EWS} and \gls{EDF} will yield. A similar argument (with lots of attached caveats) could apply to those sources with H$\alpha$-derived SFRs, though the numbers in this case are sensibly reduced with respect to the spectroscopic redshifts. For each trained model, we carefully checked that the performance metrics evaluated on the training set do not differ significantly from the ones obtained by applying the model to the test set, thus excluding any kind of overfitting to the training set.

\subsection{Computational performance}\label{sec:comp_perf}
Most of the runs presented here -- specifically, the \gls{CSMR}, \gls{CCR} and \gls{dlnn} results -- are performed on Galileo100, a high-performance computing (HPC) system located at Cineca, within the {\it Italian SuperComputing Resource Allocation} (ISCRA\footnote{\url{https://iscra.cineca.it/}}) Class C program, as part of the PPRESCIA-HP10CBAZOH program (PI, Enia). Galileo100\footnote{\url{https://www.hpc.cineca.it/systems/hardware/galileo100/}} is a DUal-Socket Dell PowerEdge cluster, hosting 636 computing nodes each with two x86 Intel(R) Xeon(R) Platinum 8276-8276L, with 24 cores each. In fact, the main advantage of \gls{ml} methods over classic template-fitting templates is the dramatic speed-up to the inference problem, at least when dealing with point value prediction for the parameters, coupled with the improved computational performance in training these models in HPC systems.
\begin{figure*}
    \centering
    \includegraphics[width=\hsize]{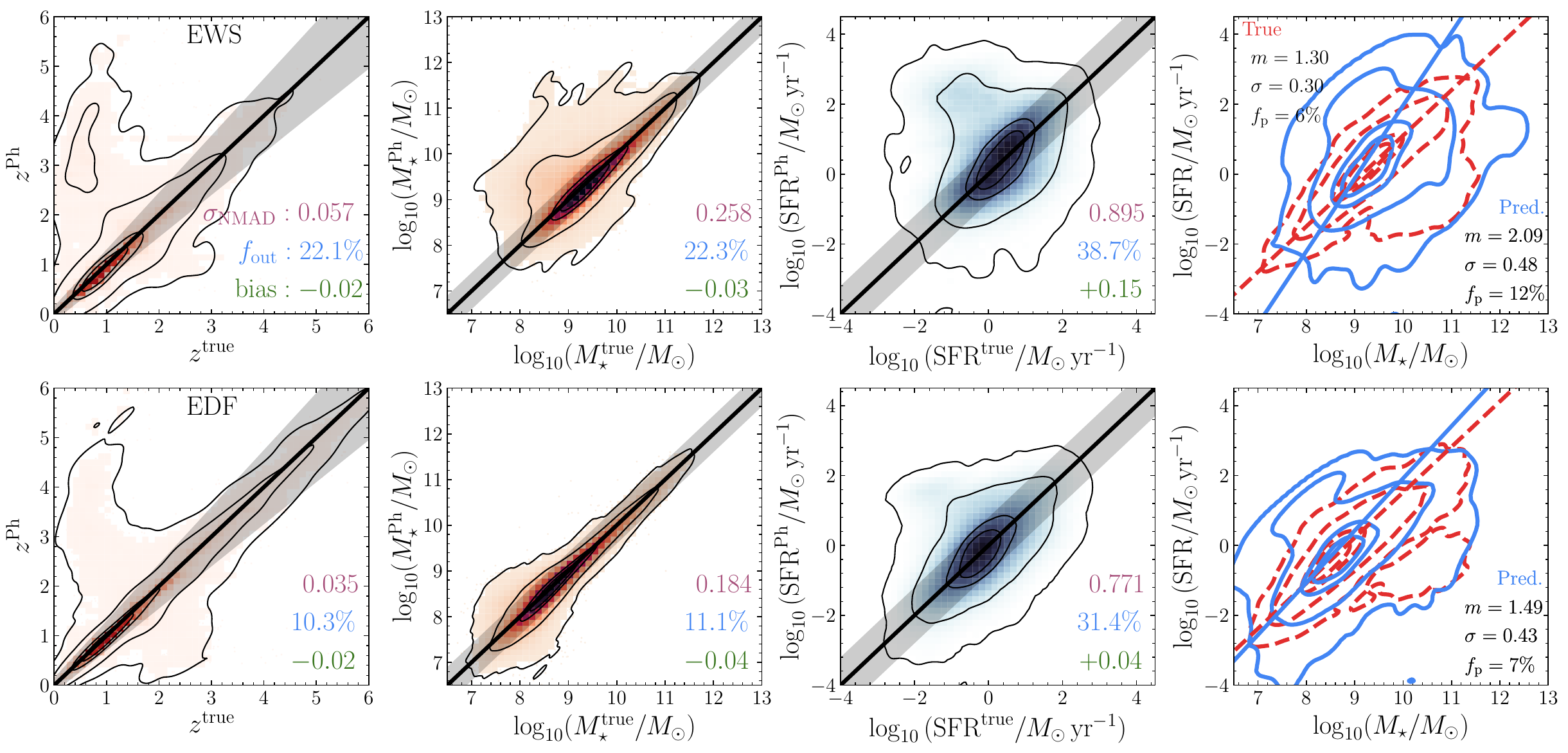}
    \caption{\phosphoros\ results on two simulated \Euclid catalogs, \gls{EWS} (top panels) and \gls{EDF} (bottom panels), with true values plotted against the \phosphoros\ recovered ones. The black line is the 1:1 relation; the shaded area is the region beyond which a prediction is an outlier. In every plot, the four contours are the area containing $98\%$, $86\%$, $39\%$ (corresponding to the $3\sigma$, $2\sigma$ and $1\sigma$ levels for a 2D histogram) and $20\%$ of the sample. For \gls{ms} the true distribution is reported in red (dashed), the predicted one in blue (solid). The lines are the \gls{odr} best-fit to the (passive-removed) distribution. The reported metrics are NMAD (purple), the outlier fraction $f_{\rm out}$ (blue) and the bias (green) for the photometric redshifts and physical parameters, and the slope $m$, scatter $\sigma$ and fraction of passive galaxies $f_{\rm p}$ for the \gls{ms}, all defined in Sect.\,\ref{sec:metrics}.}
    \label{fig:results_Phosphoros}
\end{figure*}

\phosphoros\ and \nnpz\ are run on a PowerEdge T640 machine with an Intel(R) Xeon(R) Silver 4116 CPU @ 2.10GHz processor, and 24 available cores. A typical run of \phosphoros\ requires $\sim 0.8$ seconds per galaxy; for a number of galaxies around a million (as in our cases), it translates into uninterrupted runs of a couple of weeks on the 24 available cores of our workstation.

With \nnpz, which technically does not need training as the whole computational load is on the shoulders of the neighbor search and PDF combination, a typical run on a target sample of $\sim 50$\,k galaxies requires $\sim 8$ minutes of time, or $0.006$ seconds per galaxy, a speed-up of a factor $100$ with respect to \phosphoros.

\catboost-based runs and \gls{dlnn} require training instead, after which the inference is almost instantaneous. How long those methods will run depends on the size of the training set, how complex the model is allowed to be and the number of training epochs for \gls{dlnn}; for a typical training set size of a million galaxies, it translates into training runs of $\sim 15$ minutes for \gls{CSMR} on 16 cores, $\sim 0.002$ seconds of training time per galaxy. For \gls{CCR} the training time per galaxy is similar, though the final run is of course longer since a model is trained at every iteration for each label. We run the \gls{CCR} on Galileo100, asking for a single node of 48 cores, whose overall run lasted for $\sim 1$ hour time.

Finally, on the same HPC system, we trained the \gls{dlnn} for 300 epochs, translating into $\sim 7$ hours of training time ($\sim 80$ seconds per epoch) for a $\sim$ million galaxies in the training sample, a training time of $\sim 0.003$ seconds per galaxy.

\subsection{\phosphoros\ results}\label{sec:res_phosp}
The first results we present are the template-fitting runs with \phosphoros\ on all the galaxies present in the training (or reference) samples. We refer the reader to Sect.\,\ref{sec:phosph} for further details on how \phosphoros\ has been run. The results are shown in Fig.\,\ref{fig:results_Phosphoros}, for the simulated \gls{EWS} and \gls{EDF}. In Appendix \ref{app:phospcalib} we also show the results for the two auxiliary fields at 16 and 25 ROS. In each plot, the true values are plotted against the recovered ones, and the performance metrics are reported in the bottom right of each plot.
\begin{table*}
    \caption{Metrics for the unperturbed simulation.}\label{tab:Unp}
    \centering
    \begin{tabular}{lccc|ccc|ccc|ccc}
    \hline
                  & \multicolumn{3}{c}{\gls{CSMR}} & \multicolumn{3}{c}{\gls{CCR}} & \multicolumn{3}{c}{DLNN} & \multicolumn{3}{c}{\nnpz} \\
    {}            &  NMAD  & $f_{\rm out}$ & bias  &  NMAD  & $f_{\rm out}$ & bias &  NMAD  & $f_{\rm out}$ & bias  &  NMAD  & $f_{\rm out}$ & bias \\
    \hline
    \noalign{\vskip 1pt}
    $z$      & $0.005$ & $\phantom{0}0.1\%$ & $\phantom{0} < 10^{-3}$ & $0.002$ & $0.0\%$ &  $< 10^{-3}$ & $0.017$ & $1.4\%$ & $0.002$      & $0.001$ & $0.1\%$ & $ <10^{-3}$ \\
    $M_\star$ & $0.032$ & $\phantom{0}0.1\%$ & $\phantom{0} < 10^{-3}$ & $0.022$ & $0.1\%$ &  $< 10^{-3}$ & $0.083$ & $1.5\%$ & $< 10^{-3}$ & $0.019$ & $0.1\%$ & $<10^{-3}$ \\
    ${\rm SFR}$ & $0.178$ & $\phantom{0}0.2\%$ & $\phantom{-}0.005$ & $0.168$ & $0.2\%$ &  $< 10^{-3}$ & $0.275$ & $1.5\%$ & $0.004$ & $0.132$ & $0.2\%$ & $<10^{-3}$     \\
    \hline
    \end{tabular}
    \tablefoot{Here, the methods try to find the mapping between the true labels and the noise-free features. As such, these should be intended as the absolute lower limits for each parameter. $M_\star$ refers to $\Mstarwun$, SFR to $\sfrwun$.}
\end{table*}

These results are sort of the blueprint for all the others found in this work. The first thing that jumps out is the difficulty in recovering the correct SFR, as both the \gls{EWS} and \gls{EDF} simulations display high NMADs ($0.90$--$0.77$, respectively) and fractions of outliers ($>30\%$). The recovered SFRs for the \gls{EWS} are also biased toward higher values by a factor $\sim 1.4$ (a bias of $0.13$--$0.15$ in the logarithm).

Optimal recovery is obtained for photometric redshifts instead, with NMADs that improve from $0.057$ to $0.035$ passing from Wide to Deep photometry -- and the addition of the two IRAC bands --  and $f_{\rm out}$ reducing from $22\%$ to $10\%$, with half of this reduction the consequence of an improvement in correctly distinguishing faint low-$z$, low-mass objects from high-$z$, high-mass ones. For the \gls{EWS}, worse results are obtained for the stellar masses, with higher NMADs ($0.258$) and fractions of outliers ($22\%$). The combined effect of deeper photometry plus the two IRAC bands sensibly improve the recovered stellar masses in the \gls{EDF}, with NMADS decreasing to $0.18$ and $f_{\rm out}$ to $11\%$. Both the recovered photometric redshifts and stellar masses show low biases (absolute values smaller than $0.04$) with respect to the ones found in the SFRs.

These are not unexpected findings, given the specific set of filters used as input. As reported in Sect.\,\ref{sec:Mocks} (see also Table\,\ref{tab:bands} and Fig.\,\ref{fig:filters}), for the \gls{EWS} we use 9 filters with rest-frame $\lambda_{\rm eff}$ between $0.37\,\micron$ and $1.77\,\micron$. As the photometric redshifts are more sensitive to colors in the \gls{UV}-to-\gls{NIR} part of the spectrum, these are well recovered with the given wavelength range and the number of filters. Moreover, dropouts in different filters are an excellent proxy for high-$z$ galaxies. Stellar masses correlate well with rest-frame \gls{NIR} photometry, in particular the $\HE$ band, and most of our simulated sample ($> 60\%$) reside between $0 < z < 1.5$ where NIR is still sampled by \Euclid filters. The addition of the first two IRAC channels helps significantly in improving the stellar masses recovery. Things are harder for SFRs, as they correlate the most with mid-IR to far-IR photometry \citep{2012ARA&A..50..531K}, tracing obscured star formation, and secondly with UV rest-frame monochromatic fluxes at $1550\,\AA$ \citep[FUV,][]{2001ApJ...548..681B} and $2800\,\AA$ \citep[NUV,][]{2005ApJ...625...23B}, tracing unobscured star formation. The former, stronger proxy is inaccessible with the chosen set of filters, while the latter is a weaker one. This makes the recovery of SFRs difficult even in an ideal, pristine situation (see Sect.\,\ref{sec:res_unp} and Table\,\ref{tab:Unp}) and extremely complicated when more sources of uncertainty are added. These could be improved by imposing some SFR-related priors to the template-fitting algorithm, something that will be carefully considered when dealing with real \Euclid data.

The main fraction of photo-$z$ catastrophic outliers (around $10\%$ for the \gls{EWS}, $5\%$ for the \gls{EDF}) is composed of faint low-redshift ($z^{\rm true} < 1$), low-mass [$\logten(M_\star^{\rm true}/M_\odot) < 9$] and low-SFR galaxies [$\logten (\mathrm{SFR}^{\rm true} / M_\odot\,{\rm yr}^{-1}) < 0$] that are instead misplaced at higher redshifts ($z > 2$) with at least one order of magnitude higher masses [$\Mstarwun > 10$] and SFRs [$\sfrwun > 1$]. This is reflected in the \gls{ms}. In the \gls{EWS} case, the higher SFR overestimation with respect to stellar masses yields a fitted relation with a sensibly higher slope ($m = 2.1$) with respect to the true one ($m = 1.3$). The uncertainties on the recovered parameters translate also into a higher scatter of $\sigma = 0.48$ (ground truth of $0.30$) and a fraction of passive galaxies higher ($f_{\rm p} = 12 \%$ instead of $6 \%$). Things get better for the \gls{EDF}, with metrics still distant from the true ones though.

\subsection{The unperturbed simulation}\label{sec:res_unp}

One might wonder what the absolute best-case scenario is in terms of performance when applying the methods described in Sect.\,\ref{sec:Methodology} to a pristine, unperturbed set of features mapping to the true labels. This is the same as asking what order of magnitude the irremovable inherent uncertainty of those methods is, which will always affect the measured metrics, even in a more realistic scenario where the noise affecting the features (and labels) will dominate.

To answer this question, we run the methods defined in Sect.\,\ref{sec:Methodology} on an unperturbed, noise-free features version of the {\sc mambo} catalog with true labels (see Sect.\,\ref{sec:Mocks} for definitions). Any uncertainty depends only on the specifics of the technique used to map features to labels and, from a broader perspective, on how well those specific features (magnitudes and colors) are able to recover those particular labels (photometric redshifts, stellar masses, and star formation rates).
\begin{figure*}
    \centering
    \includegraphics[width=\hsize]{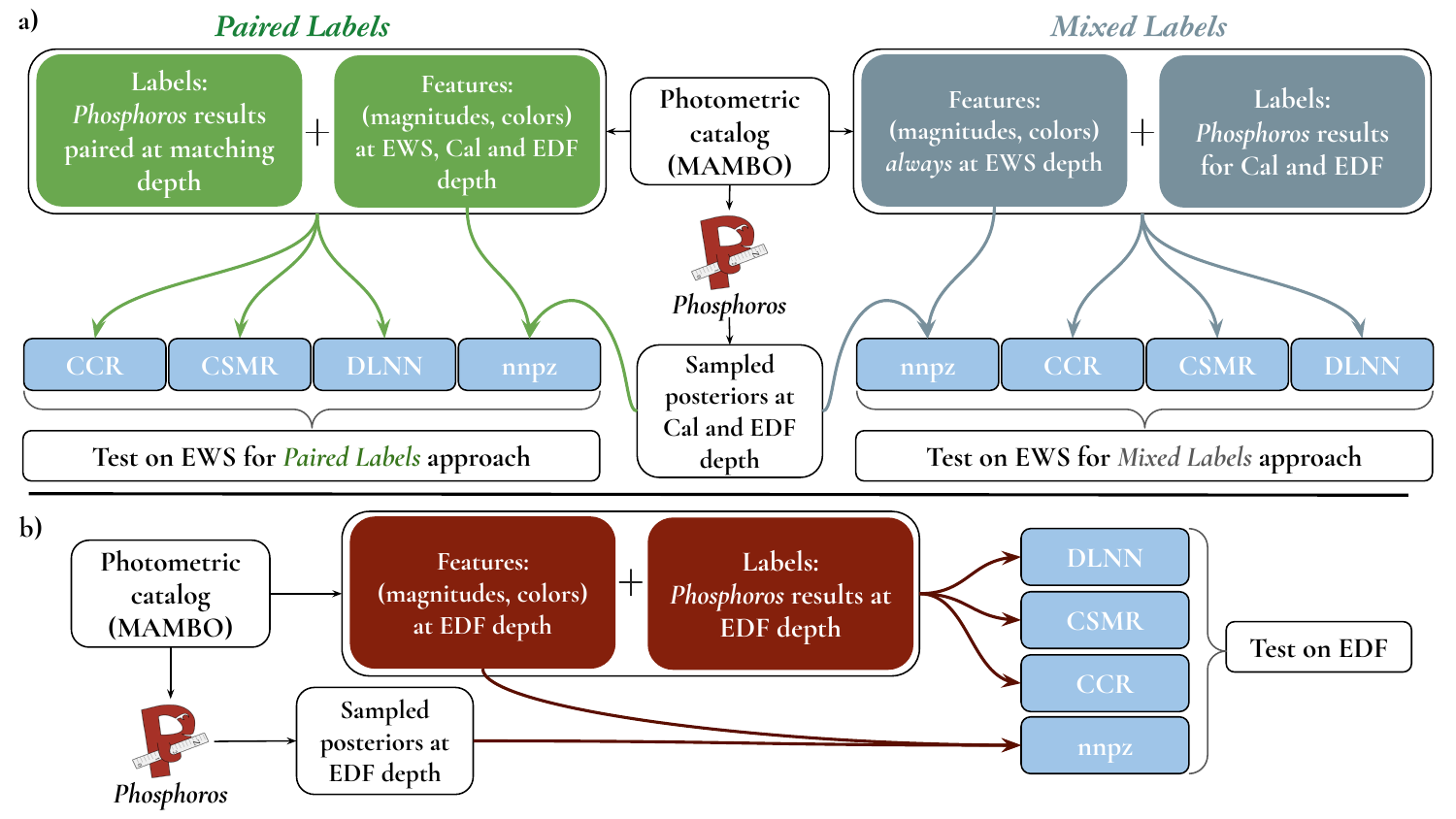}
    \caption{Flowchart followed for the reported results on the \gls{EWS} and \gls{EDF}. Panel a) summarizes what has been done for the \gls{EWS}. In this case, we employed two different approaches: pairing features to labels coming from \phosphoros\ results at the corresponding depth (paired labels), or with features always from the Wide simulated catalog and labels coming from \phosphoros\ results at the Calibration and Deep fields depth for the corresponding sources (mixed labels). These pairs of (features, labels) -- or (features, posteriors for \nnpz\ (see Sect.\,\ref{sec:nnpz}) -- are thus given as input for the \gls{ml} models described in Sect.\,\ref{sec:Methodology}. Panel b) illustrates the straighter flowchart for the \gls{EDF}, where the pairs (features, labels) or (features, posteriors) always come from the simulated Deep field.}
    \label{fig:flowchart}
\end{figure*}

The results are reported in Table\,\ref{tab:Unp}. What stands out is the perfect recovery of the photometric redshifts, up to a $0.1\%$ fraction of outliers. Those nine filters and the associated colors are able to correctly put a galaxy in its right place in the cosmic picture (see Appendix\,\ref{app:featimp} for a quantification on the feature importance). This is similar for stellar masses, though with metrics degraded by an order of magnitude, NMADs of $\sim 10^{-2}$ vs. $\sim 10^{-3}$ for photo-$z$s -- and comparable outlier fractions -- as expected once considering that the rest-frame $H$ band is a well-known tracer for correctly identifying the galaxy mass content, which is still true as the majority of the simulated sources are at $z < 1.5$.

Star formation rates are harder to recover, though. Even in an ideal, perfect scenario, it is impossible to go below NMADs of $\sim 0.13$ for that particular set of features. Of course, there is room for improvement if adding other features more sensitive to the star formation processes when available, for instance, mid-IR or far-IR photometry, or spectral features, such as the H$\alpha$ emission line. A more detailed dissertation is beyond the scope of this work, which focuses mainly on the \gls{EWS} and \gls{EDF}, without considering other ancillary (or new) filters. However, the complete exploitation of the full spectrophotometric (and morphologic) information in \Euclid will be explored in a forthcoming work.


\subsection{Results for the Euclid Wide Survey}\label{sec:res_wide}
The unperturbed case gives back an extremely optimistic best-case scenario. In reality, all the observed photometry in \Euclid will be affected by some degree of uncertainty, whose effect is to make the feature space noisier, mixing together sources with different labels. At times, even with extremely different ones, in degenerate regions of the feature space (i.e., fainter and less massive or brighter and more massive), making it hard -- or even impossible -- to correctly understand which label is associated with that particular set of features. This unavoidably degrades the quality of the model and the performance metrics when applied to a sizable sample of data.
\begin{figure*}
    \centering
    \includegraphics[width=\hsize]{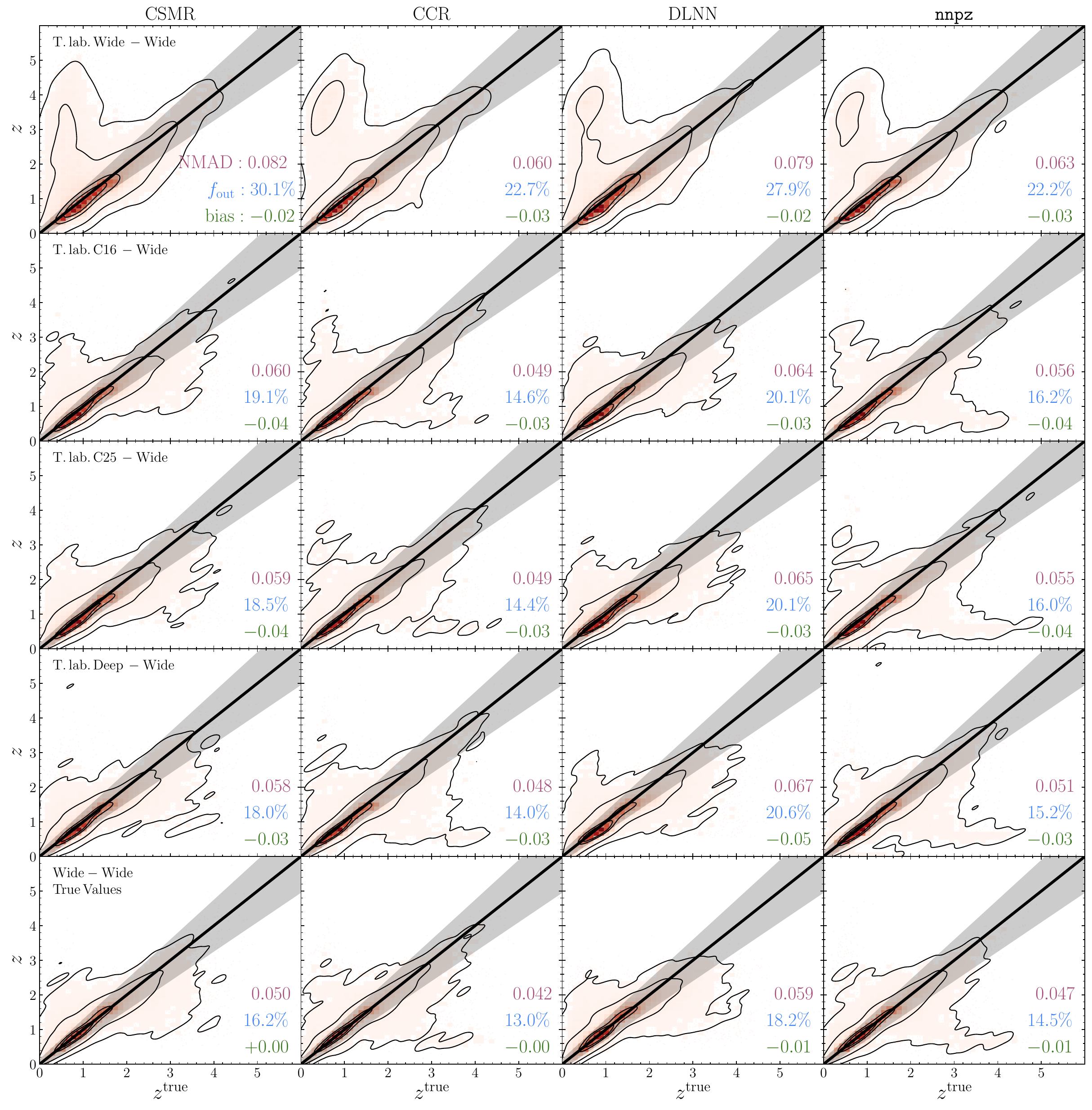}
    \caption{Results for the \gls{EWS} with the mixed labels approach. The true values on the $x$-axis are plotted against the predicted values on $y$. The black line is the 1:1 relation; the shaded area is the region beyond which a prediction is an outlier. Contours are the area containing $98\%$, $86\%$, $39\%$ (corresponding to the $3\sigma$, $2\sigma$ and $1\sigma$ levels for a 2D histogram) and $20\%$ of the sample. Each column represents the results for the methods described in Sect.\,\ref{sec:Methodology}. In the first four rows, the training labels are the recovered ones, coming from \phosphoros\ results to the mock photometry at the same depth of the field reported in the leftmost plot legend and tested on the \gls{EWS} (see Sect.\,\ref{sec:Mocks} for further details). The T.lab Wide-Wide case is exactly the same as the Wide-Wide case in Table\,\ref{tab:WideRec}. In the fifth row, we show the results of the \gls{EWS} training the models with their true labels as the best-case scenario for that particular field. The reported metrics are NMAD (purple), the outlier fraction $f_{\rm out}$ (blue) and the bias (green) for the photometric redshifts and physical parameters, as well as the slope $m$, scatter $\sigma$ and fraction of passive galaxies $f_{\rm p}$ for the \gls{ms}, all defined in Sect.\,\ref{sec:metrics}).}
    \label{fig:WIDE_zphot}
\end{figure*}
\begin{figure*}[h]
    \centering
    \includegraphics[width=\hsize]{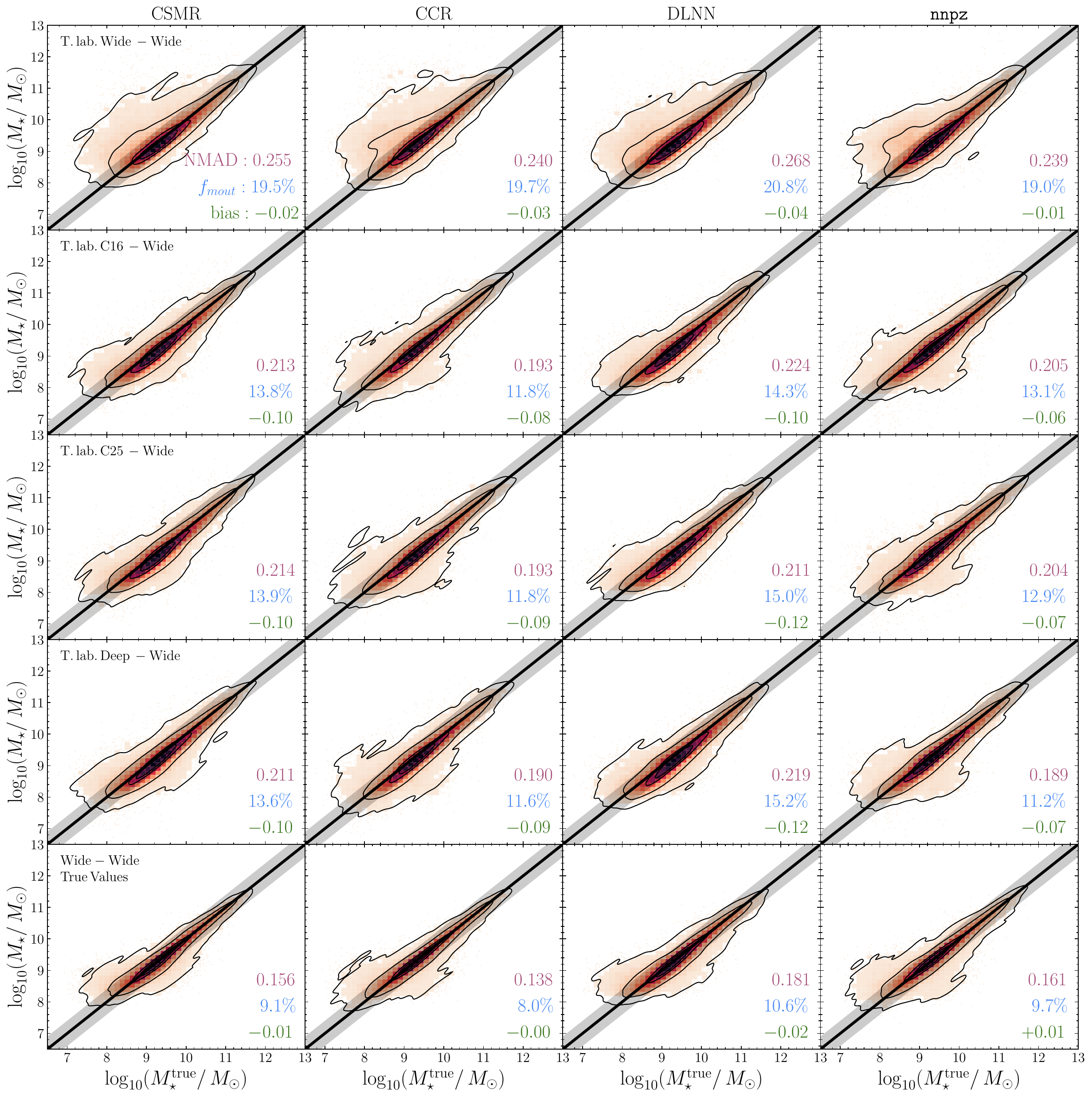}
    \caption{Same as in Fig.\,\ref{fig:WIDE_zphot}, but for stellar masses.}
    \label{fig:WIDE_M}
\end{figure*}
\begin{figure*}[h]
    \centering
    \includegraphics[width=\hsize]{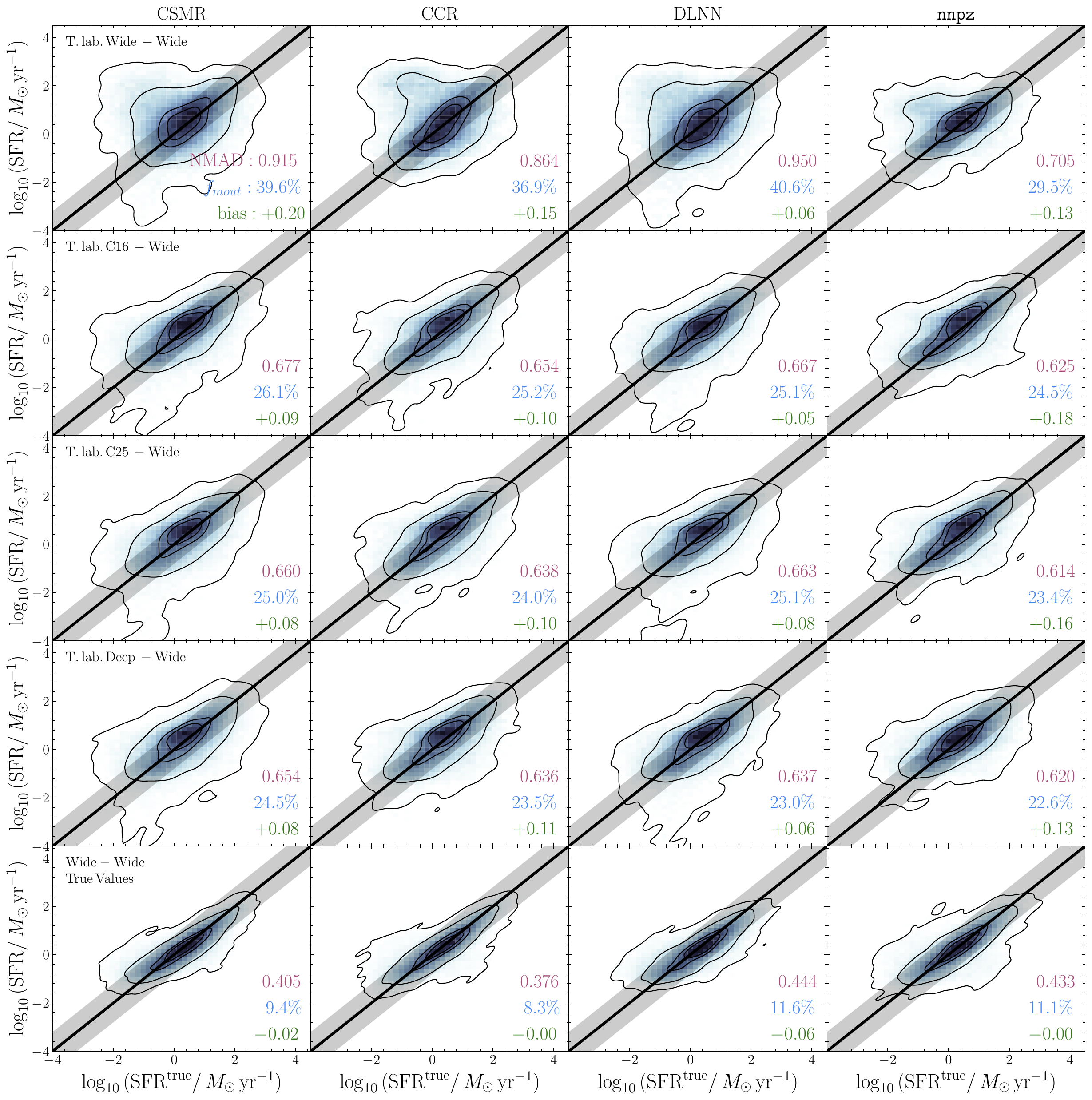}
    \caption{Same as in Fig.\,\ref{fig:WIDE_zphot}, but for star formation rates.}
    \label{fig:WIDE_SFR}
\end{figure*}
\begin{figure*}[h]
    \centering
    \includegraphics[width=\hsize]{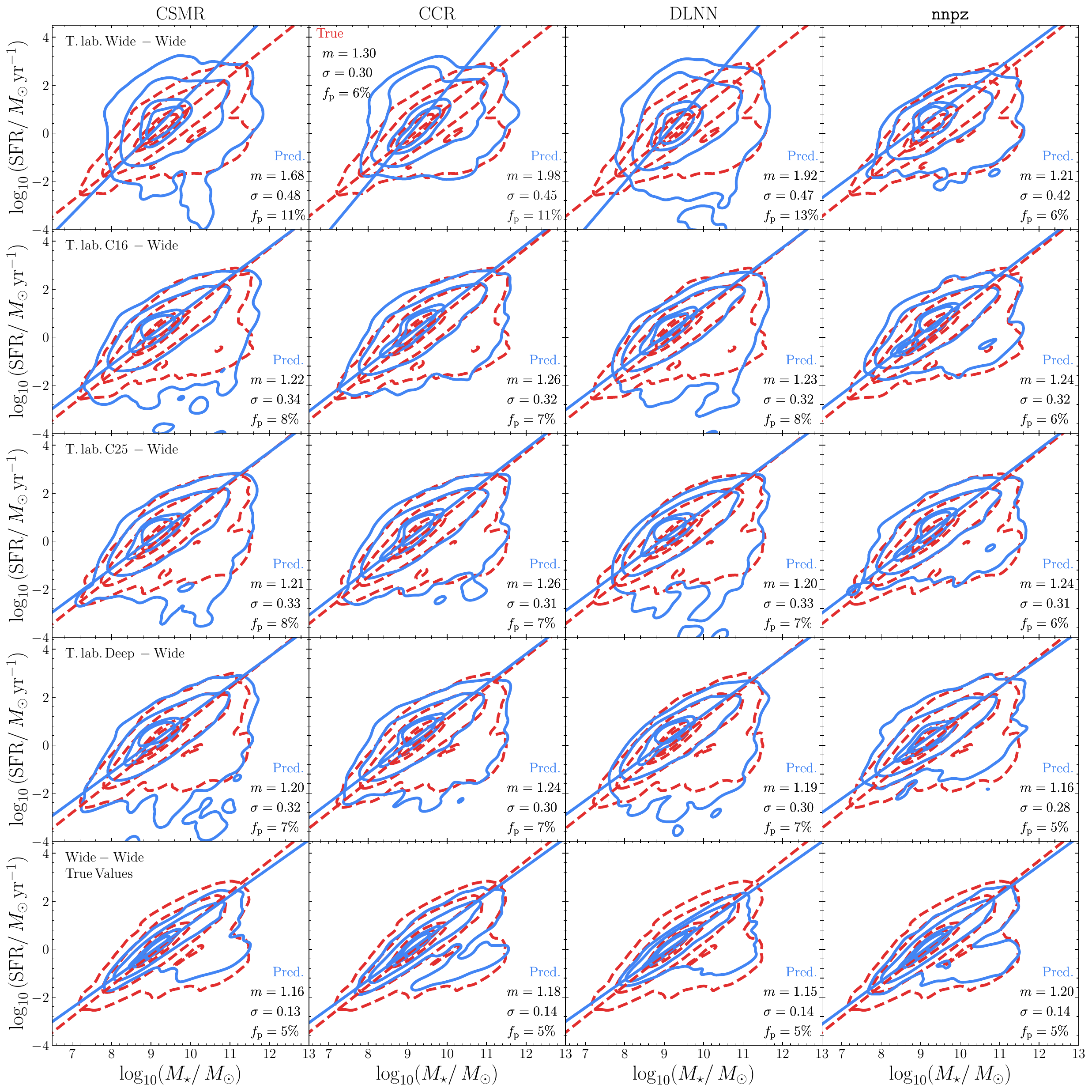}
    \caption{Same as in Fig.\,\ref{fig:WIDE_zphot}, but for the \gls{ms}. Dashed red contours are the test \gls{ms} (i.e., the true values), solid blue is the predicted one. Contour levels are the same as reported in Fig.\,\ref{fig:WIDE_zphot}. The lines are the \gls{odr} best-fit to the passive-removed distribution (dashed for test \gls{ms}, solid for predicted). The reported metrics are the \gls{ms} slope, scatter, and fraction of passive galaxies, defined in Sect.\,\ref{sec:metrics}.}
    \label{fig:WIDE_MS}
\end{figure*}

As reported in Sect.\,\ref{sec:Mocks}, we simulate four different versions of \Euclid observed catalogs: the \gls{EWS} and \gls{EDF}, and two calibration fields with 16 and 25 ROS, respectively, mimicking the \Euclid auxiliary fields for photometric and color gradient calibration \citep{Scaramella-EP1}. In this section, we focus on the \gls{EWS}, and the performance observed when training the models on deeper samples.

We present two possible approaches for this task. In the flowchart shown in Fig.\,\ref{fig:flowchart}, we summarize what has been done in obtaining the reported results for the \gls{EWS} (top panel) and the \gls{EDF} (bottom panel). The flowchart describes the different approaches employed when dealing with simulations at different depths of the same field.

\subsubsection{Paired labels approach}
The first one is the paired labels approach. Here, we train each model (or build a reference sample) with features and labels coming both from a particular field (\gls{EWS}, \gls{EDF}, or the two calibration fields), and test on the \gls{EWS}. The labels are the recovered ones (see Sect.\,\ref{sec:labels}), that is, the \phosphoros\ results for photo-$z$ and physical parameters on the field-correspondent photometry. The results are summarized in Table\,\ref{tab:WideRec}, where for each pair of training/reference - test field we report the performance metrics for all the considered labels, and Table\,\ref{tab:WideRecMS}, where we report the same for the \gls{ms} results.

The photometric redshifts performance is good, in line with the template-fitting results in Sect.\,\ref{sec:res_phosp} (see top panels of Fig.\,\ref{fig:WIDE_zphot}). There is a slight improvement in training the model with photometry and labels coming from deeper fields, with NMADs reducing by $\sim 0.01$ and outliers by $\sim 5\%$ at best. \nnpz\ has the best results overall (${\rm NMAD} \sim 0.06$, $f_{\rm out} \sim 18\%$), for every possible case of training field involved. The vast majority of outliers -- raising the NMAD too -- are $z < 1.5$ galaxies mistakenly assumed to be higher redshift ones at $z > 2$ (more on that in the next paragraphs). When looking at their magnitudes, these objects are revealed to be faint galaxies, with a distribution peaking close to the magnitude limits for each band.

This wrong distance attribution is carried over to the stellar mass prediction. A part of the degradation in the NMAD and most of the one in $f_{\rm out}$ is a consequence of those lower-$z$, lower-mass galaxies mistakenly assumed to be as high-$z$, high-mass ones. At best, with the given features and true labels in the training sample, no less than ${\rm NMAD} \sim 0.14$ and $f_{\rm out} \sim 13\%$ is expected (with the \gls{CCR}, see bottom panel of Fig.\,\ref{fig:WIDE_M}). For stellar masses, no improvement is observed when using deeper calibration fields for training but rather a degradation (see Table\,\ref{tab:WideRec}, with the exception of \gls{EDF} field, with the two IRAC channels). This is not unexpected, as it is common in \gls{ml} applications to see cases where training and testing on noisier data altogether yields better results than training with better features and testing on the noisy ones.

The most worrisome metrics are the ones associated with SFRs. The outlier fraction, defined as points with predicted SFR above or below a certain threshold to the true value ($0.8$ difference in $\log$ space, Sect.\,\ref{sec:metrics}), is over $30\%$ for every method with the notable exception of \nnpz, where it stays between $26\%$ and $30\%$. We already showed in Sect.\,\ref{sec:res_unp} how recovering SFR with the given set of features is harder than photo-$z$s or stellar masses even in the ideal, unperturbed case. In a more realistic scenario, the results of the \gls{EWS} are far from ideal, even when the true values for SFR are used in the training process (no less than an NMAD of $0.38$ and $10\%$ of outliers, see bottom panel of Fig.\,\ref{fig:WIDE_SFR}). The template-fitting algorithm finds it hard to recover SFR indeed, as reported in Sect.\,\ref{sec:res_phosp} ($\sim 39\%$). Differently than stellar masses, this is not just a matter of the wrong photo-$z$s attribution affecting the SFRs (i.e., closer and less star-forming vs. farther and more star-forming; more on that in the following paragraphs), but an inherent degeneracy due to the filters and colors used in the inference process.
\begin{table*}[]
    \caption{Metrics for the \gls{EWS}, with the mixed labels approach.}\label{tab:WideRec_ml}
    \centering
    \begin{tabular}{llccc|ccc|ccc|ccc}
    \hline
                                  &     & \multicolumn{3}{c}{\gls{CSMR}}   & \multicolumn{3}{c}{\gls{CCR}}     & \multicolumn{3}{c}{DLNN} & \multicolumn{3}{c}{\nnpz} \\
                                  &     &  NMAD &  $f_{\rm out}$ &  bias &  NMAD &  $f_{\rm out}$ &  bias &  NMAD &  $f_{\rm out}$ &  bias &  NMAD &  $f_{\rm out}$ &  bias\\
    \hline
    \noalign{\vskip 1pt}
    \multirow{3}{*}{\gls{EWS}}    & $z$         & $0.08$ & $30\%$ & $-0.02$ & $0.06$ & $23\%$ & $-0.03$ & $0.08$ & $28\%$ & $-0.02$ & $0.06$ & $22\%$ & $-0.03$ \\
                                  & $M_\star$   & $0.25$ & $20\%$ & $-0.02$ & $0.24$ & $20\%$ & $-0.03$ & $0.27$ & $20\%$ & $-0.04$ & $0.24$ & $19\%$ & $-0.01$ \\
                                  & ${\rm SFR}$ & $0.92$ & $40\%$ & $\phantom{+}0.20$ & $0.86$ & $37\%$ & $\phantom{+}0.15$ & $0.95$ & $41\%$ & $\phantom{+}0.06$ & $0.71$ & $30\%$ & $\phantom{+}0.13$ \\
                                  \hline
    \multirow{3}{*}{C16} & $z$      & $0.06$ & $19\%$ & $-0.04$ & $0.05$ & $15\%$ & $-0.03$ & $0.06$ & $20\%$ & $-0.03$ & $0.06$ & $16\%$ & $-0.04$ \\
                                  & $M_\star$    & $0.21$ & $14\%$ & $-0.10$ & $0.19$ & $12\%$ & $-0.08$ & $0.22$ & $14\%$ & $-0.10$ & $0.21$ & $13\%$ & $-0.06$ \\
                                  & ${\rm SFR}$  & $0.68$ & $26\%$ & $\phantom{+}0.09$ & $0.65$ & $25\%$ & $\phantom{+}0.10$ & $0.67$ & $25\%$ & $\phantom{+}0.05$ & $0.62$ & $25\%$ & $\phantom{+}0.18$ \\
                                  \hline
    \multirow{3}{*}{C25} & $z$      & $0.06$ & $19\%$ & $-0.04$ & $0.05$ & $14\%$ & $-0.03$ & $0.07$ & $20\%$ & $-0.03$ & $0.06$ & $16\%$ & $-0.04$ \\
                                  & $M_\star$    & $0.21$ & $14\%$ & $-0.10$ & $0.19$ & $12\%$ & $-0.09$ & $0.21$ & $15\%$ & $-0.12$ & $0.20$ & $13\%$ & $-0.07$ \\
                                  & ${\rm SFR}$  & $0.66$ & $25\%$ & $\phantom{+}0.08$ & $0.64$ & $24\%$ & $\phantom{+}0.11$ & $0.66$ & $25\%$ & $\phantom{+}0.08$ & $0.61$ & $23\%$ & $\phantom{+}0.16$ \\
                                  \hline
    \multirow{3}{*}{\gls{EDF}}    & $z$      & $0.06$ & $18\%$ & $-0.03$ & $0.05$ & $14\%$ & $-0.03$ & $0.07$ & $21\%$ & $-0.05$ & $0.05$ & $15\%$ & $-0.03$ \\
                                  & $M_\star$    & $0.21$ & $14\%$ & $-0.10$ & $0.19$ & $12\%$ & $-0.09$ & $0.22$ & $15\%$ & $-0.12$ & $0.19$ & $11\%$ & $-0.07$ \\
                                  & ${\rm SFR}$  & $0.65$ & $25\%$ & $\phantom{+}0.08$ & $0.64$ & $24\%$ & $\phantom{+}0.11$ & $0.64$ & $23\%$ & $\phantom{+}0.06$ & $0.62$ & $23\%$ & $\phantom{+}0.13$ \\
    \hline
    \multirow{3}{*}{True}    & $z$      & $0.05$ & $16\%$ & $\phantom{+}0.00$ & $0.04$ & $13\%$ & $-0.00$ & $0.06$ & $18\%$ & $-0.01$ & $0.05$ & $15\%$ & $-0.01$ \\
                                  & $M_\star$    & $0.15$ & $\phantom{0}9\%$ & $-0.01$ & $0.14$ & $\phantom{0}8\%$ & $-0.00$ & $0.18$ & $11\%$ & $-0.02$ & $0.16$ & $10\%$ & $\phantom{+}0.01$ \\
                                  & ${\rm SFR}$  & $0.41$ & $\phantom{0}9\%$ & $-0.02$ & $0.38$ & $\phantom{0}8\%$ & $-0.00$ & $0.44$ & $12\%$ & $-0.06$ & $0.43$ & $11\%$ & $-0.01$ \\
    \hline
    \end{tabular}
    \tablefoot{Leftmost column refers to the training sample, i.e., \gls{EDF} means a model trained with \gls{EWS} features and labels from \phosphoros\ results to the \gls{EDF} photometry, True means a model trained with \gls{EWS} features and the ground truth labels. All of those models are then tested on galaxies with features from the \gls{EWS} survey and ground truth as labels. The reported metrics are the ones presented in Sect.\,\ref{sec:metrics}. $M_\star$ refers to $\Mstarwun$, SFR to $\sfrwun$.}
\end{table*}
\begin{table*}[]
    \caption{Metrics for the recovered \gls{ms} in the \gls{EWS}, with the mixed labels approach.}\label{tab:WideRecMS_ml}
    \centering
    \begin{tabular}{llccc|ccc|ccc|ccc}
    \hline
                                  &     & \multicolumn{3}{c}{\gls{CSMR}}   & \multicolumn{3}{c}{\gls{CCR}}     & \multicolumn{3}{c}{DLNN} & \multicolumn{3}{c}{\nnpz} \\
                                  &     &  $m$ &  $\sigma$ &  $f_{\rm p}$ &  $m$ &  $\sigma$ &  $f_{\rm p}$ & $m$ &  $\sigma$ &  $f_{\rm p}$ & $m$ &  $\sigma$ &  $f_{\rm p}$ \\
    \hline
    \noalign{\vskip 1pt}
    \gls{EWS}    &  \multirow{5}{*}{\gls{ms}}  & $1.68$ & $0.48$ & $0.11$ &  $1.98$ & $0.45$ & $0.11$ & $1.92$ & $0.47$ & $0.13$ & $1.22$ & $0.41$ & $0.08$ \\
    C16 &   & $1.22$ & $0.34$ & $0.08$ &  $1.26$ & $0.32$ & $0.08$ & $1.23$ & $0.32$ & $0.08$ & $1.24$ & $0.32$ & $0.06$ \\
    C25 &   & $1.21$ & $0.33$ & $0.08$ &  $1.26$ & $0.31$ & $0.07$ & $1.20$ & $0.33$ & $0.07$ & $1.24$ & $0.31$ & $0.06$ \\
    \gls{EDF}    &   & $1.20$ & $0.32$ & $0.07$ &  $1.24$ & $0.30$ & $0.07$ & $1.19$ & $0.30$ & $0.07$ & $1.16$ & $0.28$ & $0.05$ \\
    True    &   & $1.16$ & $0.13$ & $0.05$ &  $1.18$ & $0.14$ & $0.05$ & $1.15$ & $0.14$ & $0.05$ & $1.20$ & $0.14$ & $0.05$ \\
    \hline
    \end{tabular}
    \tablefoot{Leftmost column refers to the training sample, i.e., \gls{EDF} means a model trained with \gls{EWS} features and labels from \phosphoros\ results to the \gls{EDF} photometry, True to models trained with \gls{EWS} features and the ground truth labels. The reported metrics are the ones presented in Sect.\,\ref{sec:metrics}. The \gls{ms} ground truth values, injected in the simulation, are $m = 1.30$, $\sigma = 0.30$, $f_p = 0.06$.}
\end{table*}

The occurrence of simultaneous wrong predictions for stellar masses and SFRs (both overestimated or underestimated) mitigates the impact on the recovered \gls{ms}, at least regarding the relation slope $m$, when training with deeper photometry (Table\,\ref{tab:WideRecMS}). However, with the notable exception of \nnpz, which yields the best performance in terms of SFRs, the recovered fraction of passive galaxies [$\logten ({\rm sSFR}/\si{G\year}^{-1}) < -1$] is usually well overestimated by a factor of two, the true one being $6\%$. No method at whatever training depth is able to recover the correct relation scatter ($\sigma = 0.24$).

\subsubsection{Mixed labels approach}
Trying to mitigate the effect of the aforementioned cloud of catastrophic outliers, we tried another approach, rooted in the belief that better performance should arise in training the models with the best possible set of labels for a given set of features. We refer to this one as the mixed labels approach, whose results are reported in Figs.\,\ref{fig:WIDE_zphot}--\ref{fig:WIDE_MS} and Tables\,\ref{tab:WideRec_ml}--\ref{tab:WideRecMS_ml}. 

Differently from the previous approach, here we train the models with features (magnitudes and colors) always coming from the \gls{EWS} catalog. However, for the deeper fields, the training labels are the \phosphoros\ results obtained with the corresponding photometry. This is specified in the plot with the text Training Label (T.lab.) followed by the name of the field. The model is then tested on features and (true) labels of the \gls{EWS}.

To give an example, when referring to T.lab Deep - Wide we mean:
\begin{itemize}
    \item training features from the \gls{EWS};
    \item training labels from the \phosphoros\ results obtained with the \Euclid Deep photometry;
    \item test with features from the \gls{EWS} and as labels the true values for $z_{\rm phot}$, $\Mstarwun$ and $\sfrwun$.
\end{itemize}

The mixed labels T.lab. Wide-Wide case is exactly the same as the paired labels Wide-Wide case, so the first rows of Table\,\ref{tab:WideRec} report the same values as the first rows of Table\,\ref{tab:WideRec_ml} and shown in Figs.\,\ref{fig:WIDE_zphot} -- \ref{fig:WIDE_SFR}. Notice also that with this approach we reduce the number of galaxies in the reference/training samples, as only those detected in the \gls{EWS} will have \phosphoros\ recovered labels, thus the number of training galaxies passes from the $\sim$ one million for the Deep and calibration fields to $\sim 500$\,k (in Sect.\,\ref{sec:cosmosref} we reduce it to $\sim 230$\,k to simulate a more realistic COSMOS-alike reference sample).

However, despite the reduced training set, this approach improves the overall performance when applying the models to the test sample.\footnote{In typical \gls{ml} applications, the relation between the size of the training sample and the quality metrics scales in logarithm scale and saturates after a while; as such, adding (or removing) a factor of two from the training sample could not significantly impact the final metrics.} In fact, attaching labels obtained with the best possible available photometry acts similar to a prior, which is able to guide the model in better distinguishing the cases in which there are degeneracies in the feature space where two close sets of features yield drastically different solutions (and catastrophic outliers, e.g., two faint galaxies with similar features can be either low-$z$, low-mass or high-$z$, high-mass objects), an improvement that totally compensates the loss in sheer number of training examples. This behavior in feature space translates in the photo-$z$ predictions as a vertical strip at $z_{\rm phot} \sim 1$, which are $z < 1.5$ galaxies mistakenly assumed as being farther away (see upper left panel of Fig.\,\ref{fig:results_Phosphoros} or the top row of Fig.\,\ref{fig:WIDE_zphot}), and generating a cloud of higher mass, higher SFR galaxies in their respective plots. The wrong photo-$z$ attribution is dragged onto the stellar masses (top rows of Fig.\,\ref{fig:WIDE_M}), where the outliers cloud is less prominent as in the photometric redshift case but still present as a stripe of higher mass galaxies than expected. This also applies to SFRs, which are also heavily affected by the reduced predictive power of the chosen features, as previously described.
\begin{figure}[h]
    \centering
    \includegraphics[width=\hsize]{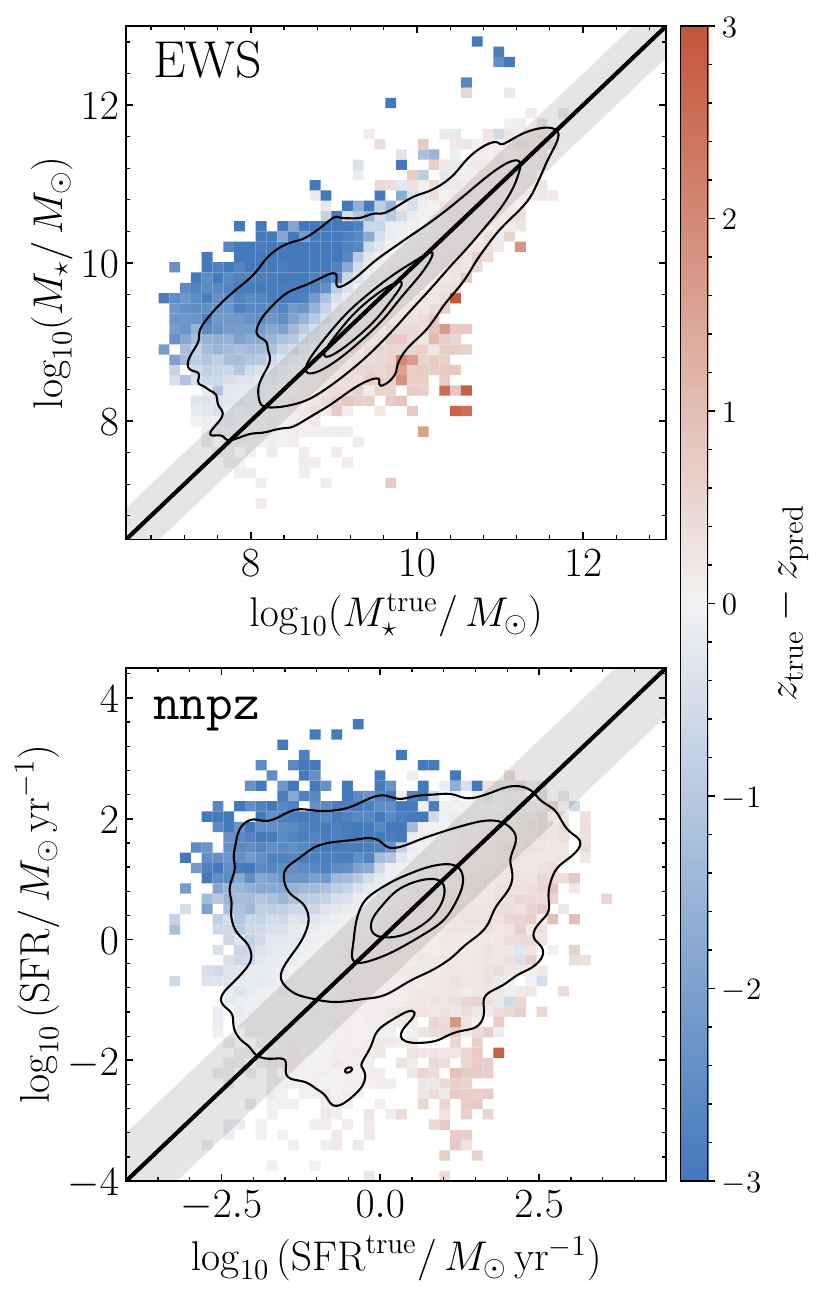}
    \caption{Stellar masses (top panel) and SFRs (bottom panel), color coded with the difference between the true and predicted redshifts, for the \nnpz\ results in the \gls{EWS} trained with wide features (the upper-right plots in Figs.\,\ref{fig:WIDE_M}--\ref{fig:WIDE_SFR}). Similar results are obtained for all the other methods considered. The black line is the 1:1 relation. Shaded area is the region beyond which a prediction is an outlier (0.4 dex for stellar masses, 0.8 dex for SFRs, see Sect.\,\ref{sec:metrics}). Contours are the area containing $98\%$, $86\%$, $39\%$ (corresponding to the $3\sigma$, $2\sigma$ and $1\sigma$ levels for a 2D histogram) and $20\%$ of the sample. The $z_{\rm pred} - z_{\rm true}$ are measured as the median of the values falling in each true vs. predicted bin. It is clearly visible how the wrong photo-$z$ attribution generates the main bulk of catastrophic outliers. However, the distribution scatter along the 1:1 relation -- i.e., the NMAD -- is mainly due to the inherent difficulties of the methods in assigning the correct \gls{pp} given the set of input features independently from the inferred photo-$z$. This is particularly true for SFRs. This plot also illustrates how penalizing a 0.3 dex definition for catastrophic outliers would be with respect with the ones chosen for this work, which actually follows better the distribution of true and predicted \gls{pp}.}
    \label{fig:WIDE_Deltaz}
\end{figure}

To better illustrate what was described in the previous paragraphs, in Fig.\,\ref{fig:WIDE_Deltaz} we show the distribution of true vs. predicted \gls{pp} as a function of the difference between $z_{\rm pred}$ and $z_{\rm true}$. In the true PP -- predicted PP plane, $z_{\rm pred} - z_{\rm true}$ is measured as the median of the values falling within each bin. We report the results obtained with \nnpz on \gls{EWS} (i.e., the T.lab. Wide-Wide case), but similar things are observed for all the other methods and cases considered.

Both for stellar masses (top panel) and SFRs (bottom panel), the cloud of photo-$z$ catastrophic outliers is visible as a region of blue squares -- meaning low-$z$ objects mistakenly placed at high-$z$ -- while a minor impact is due to the opposite, in red. These regions are correctly placed outside the defined thresholds for \gls{pp} outliers (shaded area), described in Sect.\,\ref{sec:metrics}, thus showing how they better relate with the distribution statistics with respect to a fixed 0.3 dex threshold. Another thing to notice is how the presence of those photo-$z$s outliers has a limited effect on the \gls{pp} distributions scatter (i.e., the NMADs). The majority of the distributions ($>80\%$ of points) have $|z_{\rm pred} - z_{\rm true}| < 0.1$. Even removing all the catastrophic photo-$z$ outliers, the NMADs would still be $0.19$ (for stellar masses) and $0.58$ (for SFRs). This is a consequence of the inherent difficulties of the methods (template-fitting and \gls{ml}) in recovering \gls{pp} with the given set of features, that is, filters chosen to sample the galaxies emissions, even in cases where photo-$z$ is correctly measured.

This fraction ($\sim 9$--$10\%$) of low-$z$ galaxies mistakenly assumed to be high-$z$ skew the respective luminosity functions to overestimation. This is also observed in the \gls{ms}, where the overall effect is somehow compensated by the mass-SFR scaling in the same (wrong) direction. The fit to the relation is usually way steeper than the true one when the training sample is at the same depth as the \gls{EWS}, with the exception of \nnpz.

With the mixed labels approach, the models will preferentially place fainter objects at lower-$z$ with lower masses instead of the opposite. The other side of the coin is that now a fraction of truly high-$z$ galaxies will be placed at lower redshifts. This is particularly visible with \nnpz\ (right column of Fig.\,\ref{fig:WIDE_zphot}), and a little less with the other models. However, the net result is an improvement, as only $\sim 1$--$2\%$ of these kinds of outliers are present in our results, while the number of low-$z$ objects previously mistaken for high-$z$ ones reduces by a half (from $\sim 10\%$ to $\sim 5\%$).

We find that \nnpz\ and \gls{CCR} return the best performance of all the \gls{ml} methods described in Sect.\,\ref{sec:Methodology}, followed by \gls{CSMR} and the \gls{dlnn}. In particular, \nnpz\ returns the best results for the stellar masses and SFRs for the \gls{EWS}, reducing the outlier fraction from the $\sim 28\%$ of the paired labels approach to the (still high) $\sim 20$--$30\%$ in the mixed labels one, and NMADs down to $0.61$ from $0.67$.

\nnpz\ is also the best method for recovering the \gls{ms}, as shown in Fig.\,\ref{fig:WIDE_MS}. In the T.lab Deep - Wide case, all the methods get close to the true values, with \nnpz\ being the closest. In fact, even in the worst case where Wide features and labels are employed in training, \nnpz\ results are better than the ones obtained with the other methods, with a close to correct recovery of the slope and fraction of passive galaxies, despite a higher scatter ($\sigma = 0.42$ instead of $\sigma = 0.30$) and a very small parallel displacement of the relation due to an overall overestimation of the SFRs.

The whole \gls{EDF} will not be finalized until the end of the mission. The \gls{EWS} results will be largely inferred from the auxiliary fields. These results show that this will not significantly affect the \gls{EWS} scientific outcomes, as the performance of the auxiliary fields at $16$ or $25$ ROS is only slightly (a few percentage points) worse than the ones with the full \gls{EDF} photometry available.

We find photo-$z$ metrics slightly outside the mission requirements ($\sigma_z < 0.05$, $f_{\rm out}< 10\%$) As reported in Sec\,\ref{sec:metrics}, the metrics reported here are the ones measured on point predictions, while the requirements in \cite{Desprez-EP10} depend on the PDZ. Moreover, the calibration fields used here have recovered labels from applying \phosphoros\ to the 9 bands described in Sec.\,\ref{sec:Mocks}, while the real calibration fields will benefit from more multiwavelength observations. The net effect will be more reliable labels for training and, thus, improved metrics. While our findings lie just outside the mission requirements for photo-$z$, we can likely assume that real ones will meet them.

\subsubsection{A COSMOS-like reference sample}\label{sec:cosmosref}

The photometric and color gradient calibration in \Euclid will be performed observing six of the most observed fields in the sky \citep[the so-called \Euclid auxiliary fields; see Sect. 6.2 in][for further details]{Scaramella-EP1}. The COSMOS field \citep{2007ApJS..172....1S} will be one of the first to be observed, and the widest, covering $\sim 2\ {\rm deg}^2$.

When applying the same expected depth cuts of the \Euclid auxiliary fields to the COSMOS2020 catalog \citep{2022ApJS..258...11W}, we are left with $\sim 230$\,k galaxies. As such, we try to quantify if and how much the performance degrades with such a reduced number of training galaxies. Therefore, we run all the previously reported tests with this smaller sample of $\sim 230$\,k galaxies for all the reference fields with the mixed labels approach and test on the \gls{EWS}. This corresponds to a $\sim 50\%$ cut to the training samples, though, as previously reported (Sect.\,\ref{sec:res_wide}), this does not automatically translate into a catastrophic reduction in performance metrics. We observe a reduction in NMAD and $f_{\rm out}$ between less than $1\%$ and $2\%$--$3\%$ indeed, a sign that a $\sim 230$\,k COSMOS-like reference sample is enough to reach close to the saturation limit for performance metrics given the specifics of the surveys.

\subsubsection{Removing the $u$ band in the target sample}

The final design (and timing) for the \gls{EWS} are still under redefinition with respect to what was reported in \cite{Scaramella-EP1}. However, we already know that for DR1 we will have different observations in the northern and southern sky, with the latter lacking a $u$ band filter in the complementary ground-based observations.

Therefore, we perform another test to quantify the performance degradation once we remove the $u$ filter from the target sample. As reported in Appendix\,\ref{app:featimp}, except for the $u-g$ color, which is the second most important one, the $u$ band is typically absent from the first $\sim 30$ features in terms of importance and usually appears with less than $0.5\%$ importance. As such, we observe a small reduction in the metrics performance, of the order of $\sim 3\%$, for all the methods and fields considered.

\subsection{Results for the Euclid Deep Fields}\label{sec:res_deep}
For the \gls{EDF}, \Euclid will observe $53\,\mathrm{deg}^2$ with at least $40$ ROS, pushing the expected magnitude limits two magnitudes deeper in all the bands. Moreover, the \gls{EDF} will benefit from two additional bands at $3.6\,\micron$ and $4.5\,\micron$. This will not mean just deeper data but also more robust photometry with smaller uncertainties, translating into a more reliable estimation of photometric redshifts and physical parameters, especially stellar masses.
\begin{figure*}
    \centering
    \includegraphics[width=\hsize]{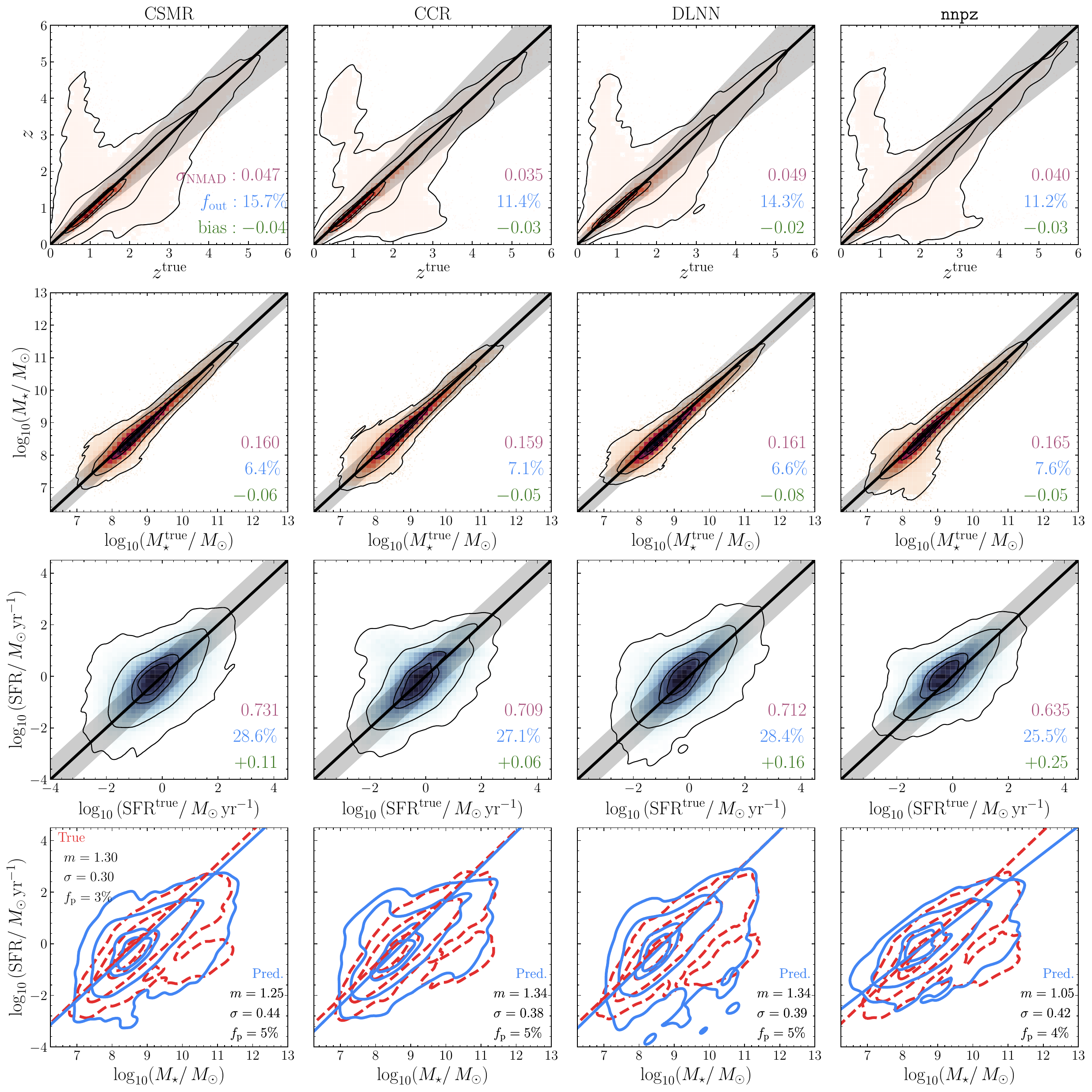}
    \caption{Results for the \gls{EDF}. Each column represents the results for the methods described in Sect.\,\ref{sec:Methodology}. The first three rows are the labels, with the true value plotted against the predicted one. The fourth column is the \gls{ms}, with true values in red (dashed) and predicted ones in blue (solid). The reported metrics are the NMAD (purple), outlier fraction $f_{\rm out}$ (blue), the bias (green), and for the \gls{ms} the slope ($\alpha$), the scatter ($\sigma$) and fraction of passive galaxies ($f_{\rm p}$), all defined in Sect.\,\ref{sec:metrics}.}
    \label{fig:DEEP_full}
\end{figure*}
\begin{table*}
    \caption{Metrics for the \gls{EDF}, with the same \gls{EWS} photometric cuts to the test galaxies.}\label{tab:Deep_Widecut}
    \centering
    \begin{tabular}{lccc|ccc|ccc|ccc}
    \hline
                  & \multicolumn{3}{c}{\gls{CSMR}} & \multicolumn{3}{c}{\gls{CCR}} & \multicolumn{3}{c}{DLNN} & \multicolumn{3}{c}{\nnpz}\\
    {}            &  NMAD  & $f_{\rm out}$ & bias &  NMAD  & $f_{\rm out}$ & bias &  NMAD  & $f_{\rm out}$ & bias &  NMAD  & $f_{\rm out}$ & bias \\
    \hline
    $z$      & $0.02$ & $\phantom{0}1.8\%$ & $-0.03$ & $0.02$ & $\phantom{0}0.9\%$ & $-0.03$ & $0.03$ & $3.2\%$  & $\phantom{+}0.00$ & $0.02$ & $\phantom{0}0.9\%$ & $-0.02$ \\  
    $M_\star$ & $0.12$ & $\phantom{0}2.5\%$ & $-0.10$ & $0.12$ & $2.5\%$ & $-0.09$ & $0.12$ & $4.7\%$ & $\phantom{-}0.01$ & $0.13$ & $2.2\%$ & $-0.06$ \\
    ${\rm SFR}$ & $0.60$ & $23.4\%$ & $\phantom{+}0.13$ & $0.61$ & $22.1\%$ & $\phantom{+}0.13$ & $0.63$ & $25.4\%$ & $\phantom{+}0.22$ & $0.63$ & $22.5\%$ & $\phantom{+}0.13$ \\
    \hline
    \end{tabular}
    \tablefoot{$M_\star$ refers to $\Mstarwun$, SFR to $\sfrwun$.}
\end{table*}

\begin{figure*}
    \centering
    \includegraphics[width=\hsize]{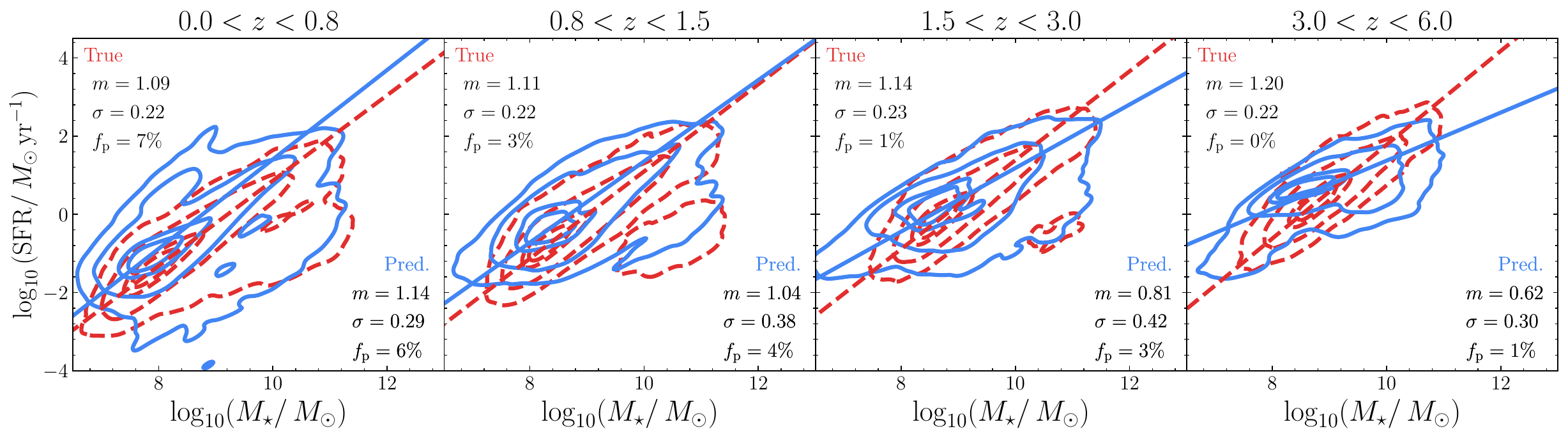}
    \caption{Results for the recovered \gls{ms} in the \gls{EDF}, in four different redshift bins. Test values are in red (dashed) and predicted ones in blue (solid). The reported metrics are defined in Sect.\ref{sec:metrics}.}
    \label{fig:DEEP_MSbin}
\end{figure*}

To quantify how more reliable the \gls{EDF} will be with respect to the \gls{EWS}, we perform the same tests in Sect.\,\ref{sec:res_wide} on a set of training (reference) and test (target) samples coming both from the {\sc mambo} simulated \gls{EDF}, with 40 ROS, the minimum expected for the Deep fields (see Sect.\,\ref{sec:Mocks} for further details). It is useful to point out that these deeper data push significantly toward higher-$z$, lower stellar masses, and lower SFRs. As shown in Fig.\,\ref{fig:catalog_hist}, the number of $z > 4$ galaxies increases from a few $10^3$ to $10^4$ (in the simulated 3.14 $\deg^2$ of the lightcone), one order of magnitude higher. A similar increase is observed for galaxies with $\Mstarwun < 8$, $\sfrwun < 0$. Correctly predicting those values will become increasingly difficult as they become more distant and less massive or star-forming, as a consequence of their inherently lower S/N and the shifting of informative wavelengths out of the region sampled by \Euclid. As such, those will hamper the naturally expected metrics improvement for Wide-alike sources observed at Deep uncertainty level. For the sake of comparison, in Table\,\ref{tab:Deep_Widecut} we also report the corresponding metrics obtained by applying the same photometric cuts from the \gls{EWS} to the {\gls{EDF}} test galaxies (see Sect.\,\ref{sec:Mocks} for further details).

The results are reported in Fig.\,\ref{fig:DEEP_full}. It is immediately visible how the EDF goes farther in distance (contours extending significantly at $z > 4$) and at lower masses and SFRs. The photo-$z$s NMAD values are comparable, though lower ($0.04$--$0.05$) than the best ones reached for the \gls{EWS} ($0.05$--$0.06$). The same is true for outliers, with a reduction of $1\%$--$3\%$ despite the presence of a cloud of low-$z$ galaxies spread over $1.5 < z_{\rm phot} < 5$. Those are fainter, low-mass galaxies with $\Mstarwun < 8$ which were marginally detected in the \gls{EWS} but are one order of magnitude more present in the {\gls{EDF}}. If we apply the Wide cuts to the Deep test galaxies, we observe a dramatic improvement in the metrics, as the NMAD for photo-$z$ falls to $\sim 0.02$--$0.04$ with only $\sim 1\%$ of outliers for both \nnpz\ and \gls{CCR}.

A significant improvement is observed also for the stellar masses, even without parceling out the fainter galaxies from the Wide-alike, as for the photo-$z$. This is principally a consequence of the addition of the two IRAC filters and, secondly, of the improved photometry. All the codes show a net improvement for NMADs, outlier fractions, and biases, even for the full set of test galaxies (down to NMADs of $\sim 0.16$ and $f_{\rm out} \sim 13 \%$. If accounting only for the Wide cut, the net improvement gets important, falling to ${\rm NMAD} \sim 0.12$ and $f_{\rm out} \sim 2.5\%$.

SFRs show a different behavior for the full set of test data. This is due to a particular set of outliers, low star-forming galaxies [$\sfrwun < -1$] that are mistakenly predicted as higher $\sfrwun > 0$ due to a wrong photo-$z$ attribution that impacts the SFRs, visible as the strip of $z^{\rm true} \sim 1$ galaxies in the top panels of Fig.\,\ref{fig:DEEP_full}. When applying the \gls{EWS} photometric cuts, the NMADs are lower than those found in the mixed labels Deep - Wide case. The same applies for $f_{\rm out}$ and the bias.

There is a notable exception, as \nnpz\ is able to reduce the impact of these outliers, to the point where even the full set of test data gives back comparable results to the mixed labels Deep - Wide case and better for the \gls{EWS}-cut ones. This is both a consequence of a better photo-$z$ estimation and an overall better ability of \nnpz\ in recovering the SFRs given the input set of features, as also observed in Sect.\,\ref{sec:res_wide}.


Overall, the \gls{ms} recovery is optimal in the \gls{EDF}. Of all the considered methods, \gls{CCR} and \nnpz\ are the ones returning the best overall results, as the former optimally recovers photo-zs and stellar masses, while the latter recovers the SFRs the best. If we perform a binning in redshift, we observe how the best results are obtained when the stellar masses and SFRs are optimally recovered (as the true redshift is) -- thus at  $0.8 < z < 1.5$ -- and worse ones at higher-$z$, where lots of low-$z$ objects are mistakenly placed at high-$z$ with greatly enhanced stellar masses (and to a lesser extent the SFRs), thus bending the whole relation by a significant amount (see rightmost plot in Fig.\,\ref{fig:DEEP_MSbin}). At lower-$z$, we notice a symmetric issue, with lots of low-mass and $\sfrwun \sim 0$ galaxies removed from the binning as the models place those at $z > 2$, and thus the predicted relation differs significantly from the true one.

There are reasons for optimism. First of all, we stress that these results should be considered lower-limit performance. Forty ROS are the minimum number expected for the \gls{EDF}, with the highest going up to 53. We expect that increasing the number of ROS will produce better performance, at least slightly (the order of a few percentage points), even though it is not straightforward to assess as mentioned before. In fact, the improvement in metrics seems to saturate after a certain number of ROS. We can roughly quantify, by extrapolating from the different realizations of the metrics for the four different number of ROS used in this work, a reduction of a few percentage points in NMAD and $f_{\rm out}$ with 53 ROS, both with template-fitting and \gls{ml} methods.

All those results are limited to the UNIONS+\Euclid+two IRAC filters, which do not extend over the $4.5\,\micron$ observed-frame, with a gap between the $\HE$ band and IRAC. The more the galaxies move to higher redshift, the more these particular sets of features will probe the source UV-rest frame, which is less sensible to stellar masses and more to SFRs. However, the \gls{EDF}, given their extension and importance in galaxy formation and evolution, will benefit from a wealth of ancillary and upcoming multiwavelength data from UV to radio. For example, the \gls{EDF}-South has been observed with the MeerKAT program MKT-23041 (P.I. Prandoni); the \gls{EDF}-N is currently proposed to be observed with the LOFAR2.0 Ultra-Deep Observation (LUDO), whose $2\,\mu{\rm Jy \,beam^{-1}}$ at $150$ MHz would translate into one of the deepest radio observations ever, and it has already been observed at $144\,{\rm MHz}$ with a central rms of $32\,\mu{\rm Jy \,beam^{-1}}$ \citep{2024A&A...683A.179B}; as illustrated, all the \gls{EDF} has been covered with {\it Spitzer} at 3.6 and 4.5 $\micron$ \citep{Moneti-EP17}; deep observations in U band with CFHT \citep{EP-Zalesky} are currently ongoing; the same for deep optical observations with Hyper Suprime Cam \citep{EP-McPartland}; K band observations with VISTA of the EDF-S have been taken as part of the EDFS-Ks program, down to a limiting magnitude of 23.5 (PI Nonino), covering the gap between the $\HE$ band and IRAC. Extending the feature space with mid-IR to submillimeter and radio fluxes will undoubtedly produce better, more reliable physical parameters than those reported in this work.

\section{Summary}\label{sec:Summary}

\Euclid and the forthcoming large-scale surveys -- {\it Rubin}/LSST, {\it Roman}) -- will benefit from unprecedentedly sampled areas of the sky, with an estimated number of observed sources up to the order of billions. At those scales, automated, accurate, and rapid methods to assess photometric redshifts and physical properties from the observables must be developed and tested. 

In this study, we evaluated the performance of \phosphoros, a template-fitting algorithm, with four \gls{ml} methods for the recovery of photo-$z$s, stellar masses, star formation rates, and the \gls{ms}: two \catboost-based methods rooted in \gls{GBDT}, a single-pass regressor (CSMR) and a chained regressor ensemble (CCR); a simple and plain \gls{dlnn}; and \nnpz, an enhanced nearest-neighbors algorithm capable of handling the full parameter posteriors. As it is typical in \gls{ml} applications, the quality of the recovered labels is inevitably limited by the number and quality of input information entering the model. Noisy features hamper a plain association between those and the labels, degrading the final performance.

In order to realistically quantify how reliable \Euclid photo-$z$s and physical parameters will be, we simulated observations of both the \gls{EWS} and the \gls{EDF} with ground-based $ugriz$ and \Euclid filters (plus two IRAC channels for the latter). The simulations are obtained within the {\sc mambo} workflow, an empirical method to extract galaxies' physical information from simulated lightcones. We also simulated an intermediate number of ROS mimicking what is expected from the \Euclid auxiliary fields, observations of well-known fields in the sky where a wealth of ancillary multiwavelength data is available for optimal photometric and color calibration. Finally, we run all the methods on an unperturbed version of the mocks, that is, without any photometric noise added and trained on the labels true values, serving as an unrealistic best-case scenario for the performance given that particular set of features (magnitudes and pairwise differences, the colors).

We found how, in the unperturbed catalog, that set of 45 features is more than enough for the models to almost perfectly recover photometric redshifts and stellar masses ($\mathrm{NMAD} < 0.03$, $f_{\rm out} < 0.3\%$). Things are more complicated for star formation rates, with NMADs never falling below $0.16$. This is expected, as the SFR correlates weakly with the input labels that sample the $0.3$--$1.8\,\micron$ observed-frame wavelength range but correlates strongly with the $8$--$1000\,\micron$ integrated luminosity (and monochromatic flux in the UV at $2800\,\AA$ rest-frame).

When feeding the mock photometry to \phosphoros\ either for the \gls{EWS} or the \gls{EDF}, we observe a typical pattern where the vast majority of outliers are generated by low-$z$ galaxies ($z < 1$) misplaced at $1.5< z < 5$, with predicted higher values of $\Mstarwun > 10$ instead of $\sim 8.5$, and $\sfrwun > 0.5$ instead of $-2 < \sfrwun < -1$. The \gls{ms} is poorly recovered in the \gls{EWS}, while better though suboptimal results are observed for the \gls{EDF}, suggesting the need for better-suited, ad hoc priors to adopt for template-fitting.

As we want to evaluate the \gls{ml} methods with what \Euclid will realistically yield, their \gls{EWS} and \gls{EDF} performance are measured training on \phosphoros\ recovered labels and testing on true values. We also checked the metrics improvement when training on true values -- inaccessible in a real scenario, as the ground truth is unknown. Moreover, since we simulated four different versions of the mock catalogs (\gls{EWS}, \gls{EDF}, and two auxilliary fields named C16 and C25 from the number of expected ROS), for the \gls{EWS} we test how a model trained on features and labels coming from deeper photometry fare on the test \gls{EWS} ground truth values.

We found different results by employing two approaches: in the paired labels approach, we are training the models on a rightly coupled set of features and labels coming respectively from the four catalogs and testing on the \gls{EWS} ground truth. With this approach, we notice a well-known pattern in \gls{ml} applications: as there is a mismatch between the training and test features (in terms of noise), it is not guaranteed that a deeper set of features (and labels) would lead to better performance. This is particularly true for stellar masses, as the NMADs and fraction of outliers do not improve significantly (or become even worse, see Table\,\ref{tab:WideRec}) from training and testing on the \gls{EWS} to training on deeper photometries and testing on the \gls{EWS}, with the notable exception of \gls{EDF} training as it benefits from the addition of two IRAC filters.

Better performance is obtained if we employ a mixed labels approach. In this case, the labels are still the \phosphoros\ results on photometry coming from the four catalogs, but the training features are always the ones from the \gls{EWS}. Despite the reduction in the training sample size (as only \gls{EWS} detected sources are employed in training) from a few million to $\sim 500$\,k sources, this approach avoids the data mismatch issue, as the features always come from the \gls{EWS} simulated catalog. More importantly, it acts as a prior on the features-labels association, as the models are able to better distinguish between similar cases near degenerate regions of the feature space, for instance, the ones that generate the outlier cloud of close and less massive galaxies mistaken for far-away and more massive ones.

In fact, those outliers are significantly reduced (or disappear altogether) with this approach. We observed a significant improvement in the performance metrics, matching or even surpassing those obtained with \phosphoros: for \nnpz, $z_{\rm phot}$ NMAD decreases from $0.063$ to $0.055$, and $f_{\rm out}$ is reduced from $22\%$ to $15\%$ passing from Wide training labels to Deep training labels, not so distant from the ideal scenario when the true labels are known ($0.047$ and $15\%$ respectively). The same goes for stellar masses (for the \gls{CCR}, NMAD falls from $0.24$ to $0.19$, outliers from $28\%$ to $11\%$, while they are $0.14$ and $10\%$ in the ideal scenario) and star formation rates (\nnpz\ reaches NMADs of $0.62$ starting from $0.71$, and outliers of $23\%$ from $30\%$, lower limits of $0.43$ and $11\%$). The same is true for the \gls{ms}. 

The full \gls{EDF} would not be finalized until February 2031. In the meantime, the main scientific results will rely on training samples obtained from multiple ROS of the auxiliary fields. We observed how the metrics did not degrade by more than a few percentage points between the C16 and C25 training samples (though here we only use the 9 previously cited bands to infer the results, thus the real ones will benefit from better estimated labels) and the \gls{EDF} one. Moreover, we also checked how a COSMOS-like reference sample does not significantly impact the model performance, as a reduction to $\sim 230$\,k galaxies in the training sample is not enough to degrade those by more than $1$--$2\%$. As a final test, we removed the $u$ band filter from the test sample, as we already know that, for DR1, those observations of the southern sky will not yet be available. However, that should not compromise the scientific outcome, as we only notice a small performance degradation, on the order of $\sim3\%$.

As expected, the \gls{EDF} results are the ones with the best results, where both labels and features come from the deepest available observations (and \phosphoros\ run on). The parameter recovery is optimal, especially the \gls{ms}, once considered how the \gls{EDF} extends the object detection to significantly higher redshifts (one order of magnitude more $z > 4$ galaxies) and lower stellar masses and SFRs, with one order of magnitude increase for $\Mstarwun < 8$ and $\sfrwun < 0$. 

There are some caveats to keep in mind to properly gauge all those results in the right framework. In our simulated catalogs, we are only considering galaxies, thus neglecting the potential impact of misclassified stars, AGN and QSOs, local contaminants, and photometric defects. These will undoubtedly cause a degradation in the considered metrics, at least by a few percent, even though a precise quantitative estimation is nontrivial and outside the scope of this work. The simulated \gls{EWS} and \gls{EDF} are built assuming BC03 models; thus, whatever discrepancy might come from what an actual galaxy emission is (or the chosen IMF) will impact the overall performance, again, in a nontrivial way. Also, we are evaluating performance on the \gls{EWS} and \gls{EDF} simulated catalogs with nine filters, the ground-based $ugriz$ and four \Euclid ones. This is the bare minimum that will be released by \Euclid, but we know that, especially for the $\sim 53\,{\rm deg}^2$ of the \gls{EDF} given their importance in the galaxy formation and evolution context, multiple ancillary or forthcoming multiwavelength data will be available to complement the \Euclid releases.

As such, we can consider our results as a kind of best-case -- as we are not accounting for defects in the catalog, the contamination by AGNs, and the fact that galaxy emission could differ from BC03 models -- of the worst-case scenario, as adding more filters, especially in the mid-IR to far-IR, will undeniably improve the performance. These results highlight \Euclid’s vast potential in assessing galaxy formation and evolution and could serve as a benchmark for all the upcoming large-area surveys in the next decade.

\begin{acknowledgements}
\AckEC

AE, MB, LP, SQ, MT, GDL, VA, JB, SF, MS acknowledge support from the ELSA project. "ELSA: Euclid Legacy Science Advanced analysis tools" (Grant Agreement no. 101135203) is funded by the European Union. Views and opinions expressed are however those of the author(s) only and do not necessarily reflect those of the European Union or Innovate UK. Neither the European Union nor the granting authority can be held responsible for them. UK participation is funded through the UK HORIZON guarantee scheme under Innovate UK grant 10093177.

We acknowledge the support from grant PRIN MIUR 2017-20173ML3WW$\_$001.

We acknowledge the CINECA award under the ISCRA initiative, for the availability of high performance computing resources and support.

M.S. acknowledges support by the Polish National Agency for Academic Exchange (Bekker grant BPN/BEK/2021/1/00298/DEC/1), the State Research Agency of the Spanish Ministry of Science and Innovation under the grants 'Galaxy Evolution with Artificial Intelligence' (PGC2018-100852-A-I00) and 'BASALT' (PID2021-126838NB-I00). This work was partially supported by the European Union's Horizon 2020 Research and Innovation program under the Maria Sklodowska-Curie grant agreement (No. 754510).

\phosphoros\ filters are taken from the SVO Filter Profile Service \citep{2012ivoa.rept.1015R, 2020sea..confE.182R}. This research has made use of the Spanish Virtual Observatory (https://svo.cab.inta-csic.es) project funded by MCIN/AEI/10.13039/501100011033/ through grant PID2020-112949GB-I00.

In preparation for this work, we used the following codes for Python: \texttt{Numpy} \citep{2020Natur.585..357H}, \texttt{Scipy} \citep{2020NatMe..17..261V}, \texttt{Scikit-Learn} \citep{2011JMLR...12.2825P}, \texttt{Pandas} \citep{mckinney-proc-scipy-2010}, \catboost\ \citep{NEURIPS2018_14491b75}, \texttt{Tensorflow} \citep{tensorflow2015-whitepaper}, \nnpz\ \citep{2018PASJ...70S...9T}, \phosphoros\ (Paltani et al. in prep).

\end{acknowledgements}

\bibliography{mybiblio, Euclid}

\begin{appendix} 

\section{Feature importances}\label{app:featimp}
\begin{figure}
    \centering
    \includegraphics[width=.7\hsize]{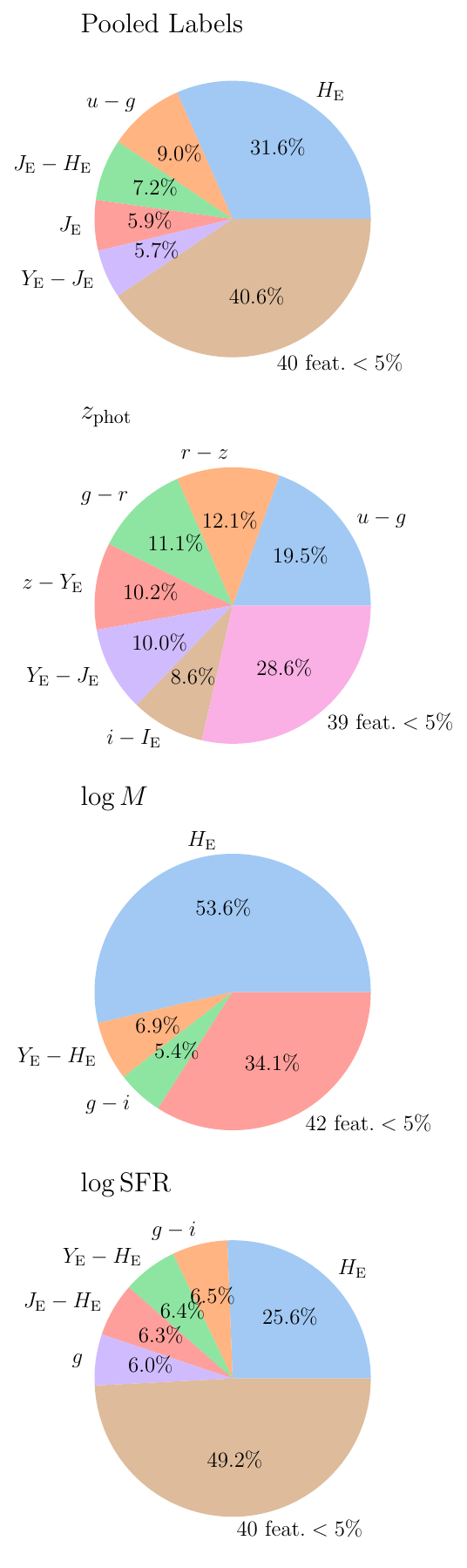}
    \caption{Pie chart highlighting the most important features in recovering the pooled labels with \gls{CSMR}. The feature importance weights (in percentage) how much a single feature influences the final prediction. In the Pooled Labels and $\sfr$ only cases, $45$ features enter the model (magnitudes and all possible color permutations), only five have an importance over $5\%$.}
    \label{fig:featimp}
\end{figure}

In this appendix, we present the feature importance for models evaluated with \gls{CSMR}. In standard \gls{ml} terminology, feature importance refers to the quantification of the impact each feature has on the model's predictions. This is measured by considering the number of times a feature is used for splits across all trees in the ensemble and the corresponding improvement in the model's performance. The higher the number of times a feature is used and the greater the improvement, the more important the feature is considered, giving insights into the relative significance of different features in the data, for example, if and what certain filter or color is more important in correctly assessing a galaxy redshift, stellar mass, or star formation rate.

These findings are reported in Fig.\,\ref{fig:featimp}, where we show the feature importance for four different models, each one specifically trained to recover a single label, whether is $z_{\rm phot}$, $\Mstarwun$, or $\sfrwun$, and a model trained on a set of pooled labels. In each case, to remove altogether every source of noise skewing the results, we are performing the training on the unperturbed catalog (see Sect.\,\ref{sec:res_unp} for further details).

As expected, considering that most of the training galaxies are at $0 < z < 1.5$, the $\HE$ band is by far the most important one in determining a galaxy's stellar mass (more than half importance); similarly, for photometric redshifts, the colors hold the most importance, more than single filter observations. Things are more mixed for star formation rates, as with the exception of $\HE$ ($\sim 25\%$), all the other features hold similar importance (between $3\%$ and $6\%$). When considering all the labels together, we observe a mix between the previous results, with $\HE$ being still the most considered feature at $\sim 25\%$ importance, followed by the colors.

\section{\phosphoros\ results for the calibration fields}\label{app:phospcalib}
\begin{figure*}
    \centering
    \includegraphics[width=\hsize]{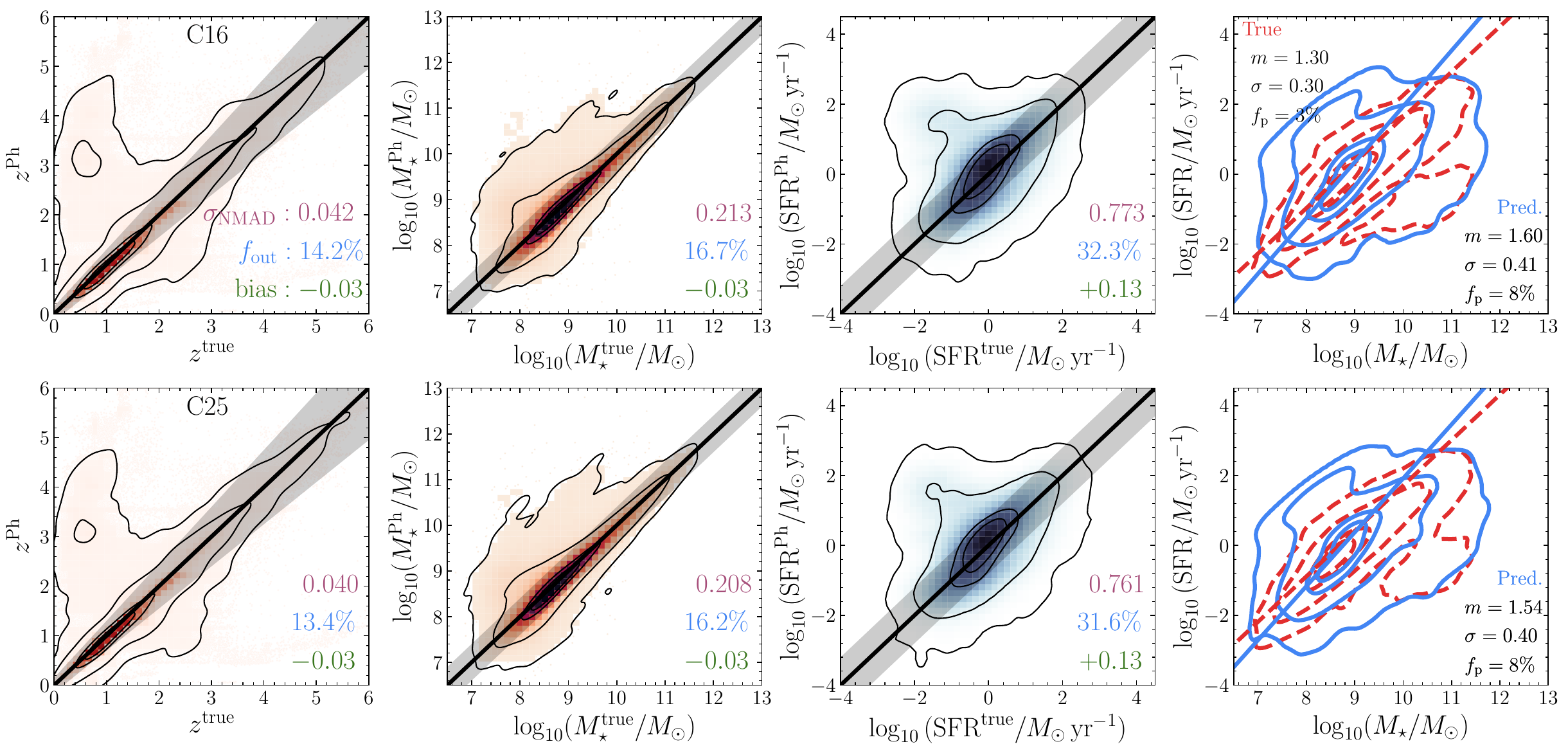}
    \caption{\phosphoros\ results on two simulated \Euclid auxiliary fields, with 16 (C16, bottom panel) and 25 (C25, bottom panel) ROS each. The ground truth is plotted against the \phosphoros\ recovered parameters. The black line is the 1:1 relation; the shaded area is the region beyond which a prediction is an outlier. In every plot, the four contours are the area containing $98\%$, $86\%$, $39\%$ (corresponding to the $3\sigma$, $2\sigma$ and $1\sigma$ levels for a 2D histogram) and $20\%$ of the sample. For \gls{ms} the true distribution is reported in red (dashed), the predicted one in blue (solid). The lines are the \gls{odr} best-fit to the (passive-removed) distribution. The reported metrics are NMAD (purple), the outlier fraction $f_{\rm out}$ (blue) and the bias (green) for the photometric redshifts and physical parameters, as well as the slope $m$, scatter $\sigma$ and fraction of passive galaxies $f_{\rm p}$ for the \gls{ms}, all defined in Sect.\,\ref{sec:metrics}.}
    \label{fig:phosph_calibs}
\end{figure*}
In Sect.\,\ref{sec:res_phosp}, we reported the template-fitting results with \phosphoros\ to the \gls{EWS} and \gls{EDF}. In Fig.\,\ref{fig:phosph_calibs}, we report the corresponding \phosphoros\ run to the auxiliary fields at 16 and 25 ROS each. As expected, the performance is intermediate between the \gls{EDF} and the \gls{EWS}, though closer to the former. Anyway, it should always be kept in mind that the deeper the observations, the more distant and/or less massive galaxies will enter the catalogs, whose photometric redshifts and physical parameters are harder to properly assess, therefore reducing the expected metrics improvement.

\section{Results with the paired labels approach}
In Tabs.\,\ref{tab:WideRec}--\ref{tab:WideRecMS}, we report the results for the \gls{EWS} with the paired labels approach. As described in Sect.\,\ref{sec:res_wide}, with this approach, we train each model with a set of features and labels obtained from the \phosphoros\ run to the corresponding depth (e.g., C16 features, labels from \phosphoros\ run to the C16 photometry), and test on Wide features and ground truth values obtained from the simulation.

\begin{table*}[]
    \caption{Metrics for the \gls{EWS}, with the paired labels approach.}\label{tab:WideRec}
    \centering
    \begin{tabular}{llccc|ccc|ccc|ccc}
    \hline
                                  &     & \multicolumn{3}{c}{\gls{CSMR}}   & \multicolumn{3}{c}{\gls{CCR}}     & \multicolumn{3}{c}{DLNN} & \multicolumn{3}{c}{\nnpz} \\
                                  &     &  NMAD &  $f_{\rm out}$ &  bias &  NMAD &  $f_{\rm out}$ &  bias &  NMAD &  $f_{\rm out}$ &  bias &  NMAD &  $f_{\rm out}$ &  bias\\
    \hline
    \noalign{\vskip 1pt}
    \multirow{3}{*}{C16} & $z$      & $0.07$ & $24\%$ & $-0.03$ & $0.06$ & $20\%$ & $-0.03$ & $0.07$ & $22\%$ & $-0.03$ & $0.06$ & $18\%$ & $-0.04$ \\
                                  & $M_\star$ & $0.25$ & $20\%$ & $-0.10$ & $0.24$ & $20\%$ & $-0.08$ & $0.25$ & $21\%$ & $-0.09$ & $0.23$ & $17\%$ & $-0.01$ \\
                                  & ${\rm SFR}$  & $0.82$ & $34\%$ & $-0.05$ & $0.80$ & $33\%$ & $-0.04$ & $0.83$ & $35\%$ & $-0.11$ & $0.70$ & $28\%$ & $\phantom{+}0.16$ \\
                                  \hline
    \multirow{3}{*}{C25} & $z$      & $0.07$ & $24\%$ & $-0.03$ & $0.06$ & $20\%$ & $-0.03$ & $0.07$ & $25\%$ & $-0.03$ & $0.06$ & $18\%$ & $-0.04$ \\
                                  & $M_\star$ & $0.25$ & $21\%$ & $-0.11$ & $0.24$ & $21\%$ & $-0.08$ & $0.25$ & $21\%$ & $-0.09$ & $0.22$ & $17\%$ & $-0.05$ \\
                                  & ${\rm SFR}$   & $0.82$ & $34\%$ & $-0.08$ & $0.80$ & $33\%$ & $-0.07$ & $0.82$ & $34\%$ & $\phantom{+}0.03$ & $0.69$ & $28\%$ & $\phantom{+}0.14$ \\
                                  \hline
    \multirow{3}{*}{\gls{EDF}}    & $z$      & $0.06$ & $21\%$ & $-0.01$ & $0.05$ & $17\%$ & $-0.01$ & $0.07$ & $22\%$ & $-0.03$ & $0.05$ & $16\%$ & $-0.02$ \\
                                  & $M_\star$ & $0.21$ & $22\%$ & $-0.03$ & $0.19$ & $20\%$ & $-0.05$ & $0.22$ & $23\%$ & $-0.08$ & $0.19$ & $17\%$ & $\phantom{+}0.01$ \\
                                  & ${\rm SFR}$  & $0.79$ & $30\%$ & $-0.06$ & $0.75$ & $28\%$ & $-0.09$ & $0.81$ & $31\%$ & $-0.12$ & $0.65$ & $25\%$ & $-0.09$ \\
    \hline
    \end{tabular}
    \tablefoot{The leftmost column refers to the training (reference) sample, i.e., C25 means a model trained with features and labels from the C25 simulated auxiliary field. All the models are then tested on features from the \gls{EWS} simulation and ground truth labels. The reported metrics are the ones presented in Sect.\,\ref{sec:metrics}. We are not reporting the paired labels \gls{EWS} row as it is the same as the mixed labels T.lab. Wide-Wide case in Figs.\,\ref{fig:WIDE_zphot}--\ref{fig:WIDE_SFR} and Table\,\ref{tab:WideRec_ml}. $M_\star$ refers to $\Mstarwun$, SFR to $\sfrwun$.}
\end{table*}

\begin{table*}[]
    \caption{Metrics for the recovered \gls{ms} in the \gls{EWS}, with the paired labels approach.}\label{tab:WideRecMS}
    \centering
    \begin{tabular}{llccc|ccc|ccc|ccc}
    \hline
                                  &     & \multicolumn{3}{c}{\gls{CSMR}}   & \multicolumn{3}{c}{\gls{CCR}}     & \multicolumn{3}{c}{DLNN} & \multicolumn{3}{c}{\nnpz} \\
                                  &     &  $m$ &  $\sigma$ &  $f_{\rm p}$ &  $m$ &  $\sigma$ &  $f_{\rm p}$ & $m$ &  $\sigma$ &  $f_{\rm p}$ & $m$ &  $\sigma$ &  $f_{\rm p}$ \\
    \hline
    \noalign{\vskip 1pt}
    C16 &   & $1.29$ & $0.41$ & $0.14$ &  $1.32$ & $0.39$ & $0.14$ & $1.35$ & $0.39$ & $0.16$ & $1.27$ & $0.39$ & $0.08$ \\
    C25 &   & $1.28$ & $0.40$ & $0.14$ &  $1.28$ & $0.40$ & $0.14$ & $1.32$ & $0.42$ & $0.13$ & $1.28$ & $0.38$ & $0.08$ \\
    \gls{EDF}  &   & $1.27$ & $0.39$ & $0.14$ &  $1.23$ & $0.39$ & $0.14$ & $1.31$ & $0.42$ & $0.17$ & $1.23$ & $0.36$ & $0.14$ \\
    \hline
    \end{tabular}
    \tablefoot{The reported metrics are the one presented in Sect.\,\ref{sec:metrics}.}
\end{table*}

\end{appendix}

\end{document}